\documentclass{aa}  

\usepackage{graphicx}
\usepackage{txfonts}

\usepackage{natbib}
\bibpunct{(}{)}{;}{a}{}{,} 

\def\epi{E$_{\rm pi}$}

\def\eiso{E$_{\rm iso}$}

\def\eer{\epi\ -- \eiso\ {\rm relation}}
\def\eep{\epi\ -- \eiso\ {\rm plane}}

\def\t9{T$_{90}$}

\def\nallswift{85}
\def\nall{126}

\def\ngrb{71}
\def\ngrbnozall{42}
\def\ngrbnoz{14}
\def\nabove{38}
\def\nbelow{40}

\begin{document}

   \title{ Connecting Prompt and Afterglow GRB emission
   }
   \subtitle{I. Investigating the impact of optical selection effects in the \eep}

   \author{
      D. Turpin\inst{1,2}\fnmsep\thanks{D. Turpin used funds provided by the LabEx OCEVU.}
         \and
      V. Heussaff
      \inst{1,2}
         \and
      J.-P. Dezalay \inst{1,2}
         \and
      J-L. Atteia\inst{1,2}
         \and
      A. Klotz \inst{1,2}
         \and
      D. Dornic \inst{3}
   }

   \institute{
      Universit\'e de Toulouse; UPS-OMP; IRAP; Toulouse, France -- \email{damien.turpin@irap.omp.eu}
         \and
      CNRS; IRAP; 14, avenue Edouard Belin, F-31400 Toulouse, France
         \and
      Aix Marseille Université, CNRS/IN2P3, CPPM UMR 7346, 13288 Marseille, France
          \\
   }

   \date{Received september 5, 2014; accepted ...}

 
  \abstract
   {Measuring GRB properties in their rest-frame is crucial to understand the physics at work in gamma-ray bursts. 
   This can only be done for GRBs with known redshift. Since redshifts are usually measured from the optical spectrum of the afterglow, 
   correlations between prompt and afterglow emissions may introduce subtle biases in the distribution of rest-frame properties of the prompt emission, 
   especially considering that we measure the redshift of only one third of {\textit Swift} GRBs.}
   {In this paper we study the brightness of optical GRB afterglows and the role of optical selection effects in the distribution of various intrinsic properties of GRBs and on the \eer\ discovered by Amati et al. (2002, A\&A, 390, 81). 
   }
   {
   Our analysis is based on a sample of \nallswift\ GRBs with good optical follow-up and well measured prompt emission. 71 of them have a measure of redshift and 14 have no redshift.
   We discuss the connection between the location of GRBs in the \eep\ and their optical brightness measured two hours after the trigger in the GRB rest frame.
   }
   {
We show that the brightness of GRBs in our sample is mainly driven by their intrinsic luminosity and depends only slightly on their redshift.
We also show that GRBs with faint afterglows are preferentially located in the upper part of the \eep.  This optical selection effect favors the detection of GRBs with bright afterglows located below the best fit \eer\ whose redshift is easily measurable. 
}
   {The distributions of prompt GRB properties in the rest frame undergo selection effects due to the need to measure the redshift
from the optical afterglow emission. These biases put significant uncertanties when interpreting the statistical studies of GRB properties in the rest frame. We show that the \eer\ is not immune to these selection effects. The difficulty to measure the redshifts of GRBs located far above the best fit \eer\ may partly explain the observed lack of GRBs with large Epi and low Eiso. However, we observe that bright GRBs may indeed follow an \eer. Studying these selection effects could allow us to better understand the properties of GRBs in their rest frame and the connection between the prompt and the afterglow properties.
   }


   \maketitle


\section{Introduction}
\label{sec_intro}

Gamma-ray bursts (GRBs) are cataclysmic explosions resulting from the collapse of massive stars or the merging of two compact objects, see \cite{Kumar2014} and references therein. During these events an ultra relativistic jet is produced accompanied by an intense gamma-ray flash which can be seen at very high redshifts (up to z=9 for GRB090429B, see \cite{Cucchiara2011a}. The entire gamma-ray emission is produced in only few seconds (prompt emission) with a total energy released up to E$\sim$$10^{54}$ erg assuming an isotropic emission. This prompt emission is followed by a long lasting and fading multi-wavelength emission from X-rays to radio (afterglow emission) attributed to the external shock between the relativistic ejecta and the interstellar medium (ISM). The origin of the prompt $\gamma$-ray emission has been suggested to be due to internal shocks between shells moving at different Lorentz factors, $\Gamma$, inside the jet, see \citep[][]{Rees1994,Paczynski1994,Piran1999,Kumar2014}. However, the physical processes at work in these shocks are still not well understood due to strong uncertainties on the physical conditions (baryon loading, energy dissipation in the jet, acceleration mecanism), see the review by \cite{Kumar2014}. On the contrary, the afterglow emission is better understood and can be explained by the synchrotron emission from the accelerated electrons in front of the shock between the relativistic ejecta and the ISM (forward shock), (see \cite{Sari1998} $\&$ \cite{Granot2002}). An optical/radio flash is also expected and sometimes observed from a reverse shock propagating into the relativistic ejecta, see \citep[][]{Meszaros1993, Kobayashi2000}. In this framework, the dynamics of the afterglow is determined by the microphysics of the shocked material, the ISM properties and the kinetic energy of the blast wave, $E_k$ which depends on the prompt emission properties (the isotropic gamma-ray energy released, $E_{iso}$ and the gamma-ray radiative efficiency, $\eta$). Moreover, studying the optical flash radiation from the reverse shock is a very interesting opportunity to bring constraints on the magnetization parameter of the ejecta \citep[][]{Zhang2005, Narayan2011}, the Lorentz factor of the jet, and also on the jet composition \citep[][]{McMahon2006, Nakar2004}. Understanding of the connection between the prompt and afterglow properties would help us to better constrain the physics of GRBs and relativistic jets. Thus many authors discussed correlation between the afterglow luminosity and the prompt energetics in the rest frame. 

\smallskip 

Correlations between the afterglow optical luminosity and prompt isotropic energy have been found by \cite{Kann2010}, \cite{Nysewander2009} and also between the afterglow X-ray emission and the isotropic energy by \cite{Kaneko2007} and \cite{Margutti2013}. However, it is difficult to assess wether these relations have their origin in the physics of the GRB since some studies have shown that they could undergo strong selection effects. Indeed, recently \cite{Coward2014} detected a strong Malmquist bias in the correlation $E_{iso}-L_{opt,X}$ as we preferentialy detect the brightest part of the GRB population. Other studies lead by \cite{Heussaff2013} and \cite{Shahmoradi2013} have shown that gamma-ray selection effects strongly biased rest frame prompt properties correlations and particularly the \eer\ discovered by \cite{Amati2002}. While it is clear that gamma-ray selection effects can bias statistical studies of prompt GRB properties, the impact of optical selection effects is rarely assessed. This paper is dedicated to the study of the impact of afterglow optical brightness on the \eer\ which is one of the most robust and tighest rest frame prompt correlation.
To check the impact of optical selection effects, we construct a sample of GRBs with good optical and gamma-ray data in section \ref{sec_sample}. In section \ref{sec_analyse_opt} we describe the optical brightness distribution of GRB in our sample and the potential parameters (extrinsic and intrinsic) that could bias it. Then, we compare the optical brightness of GRBs located at different positions in the \eep\ in section \ref{sec_prompt}. In section \ref{discussion}, we briefly discuss the consequences of our findings.
Our conclusions are given in section \ref{sec_conclusion}.


\section{GRB sample and data}
\label{sec_sample}

\subsection{Optical data}
\label{sub_opt}
We collected the afterglow optical lightcurves of \nall\ GRBs with a redshift and \ngrbnozall\ GRBs without a redshift. We specifically choose the R band because it concentrates the largest number of optical measurements. These R band photometric measurements are issued from published articles and GCN Circulars\footnote{http://gcn.gsfc.nasa.gov}.

\begin{figure*}[t]
\centering
\includegraphics[trim = 30 200 0 200,clip=true,width=0.8\textwidth]{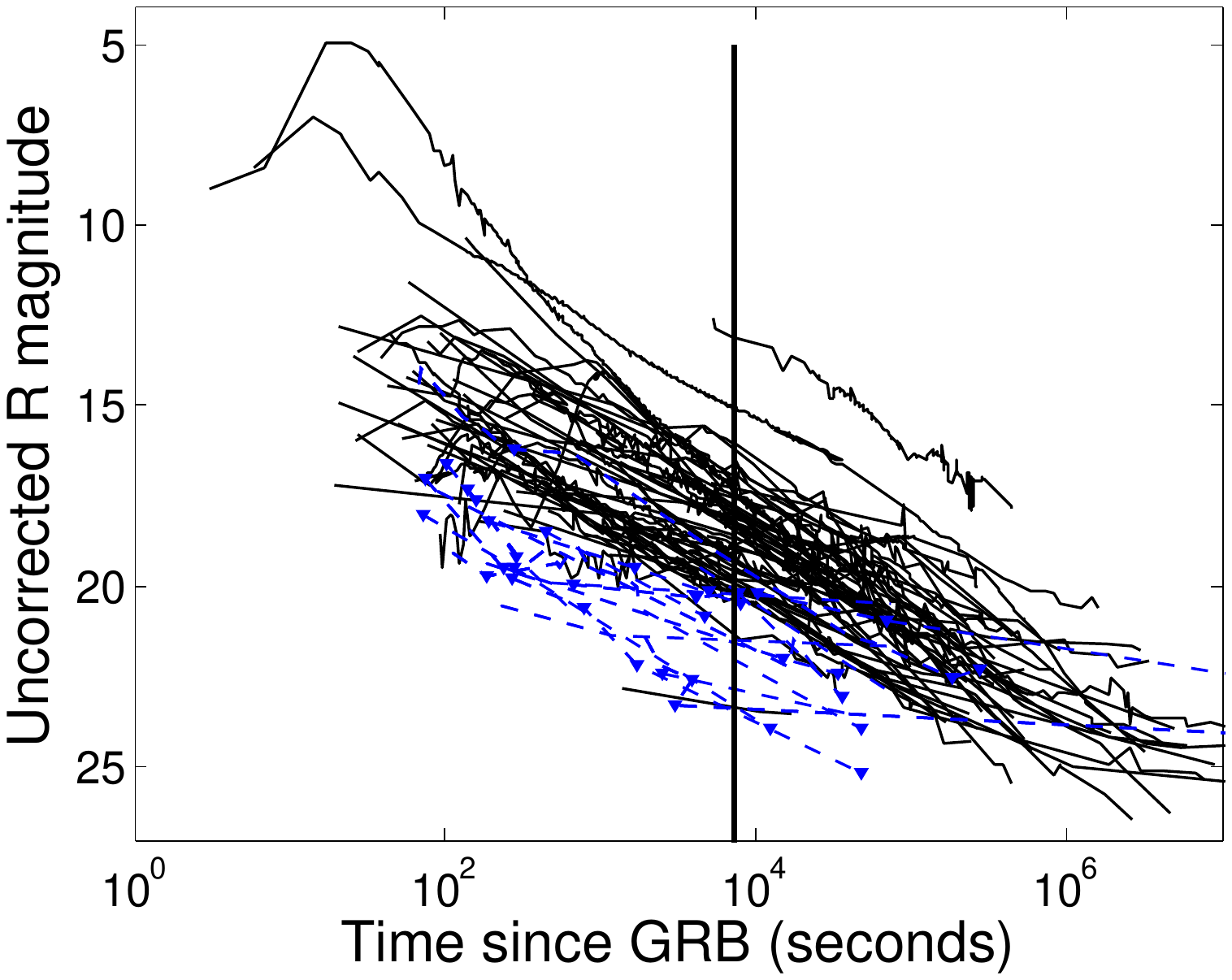}
\caption{R band optical lightcurves in the observer frame of the afterglows of \ngrb\ GRBs with a redshift (black solid line) and \ngrbnoz\ GRBs without a redshift (blue dashed line with triangle for upper limits) considered in this study. The magnitudes are not corrected from the galactic and host extinctions. The vertical solid line represents the time (2 hours) at which we estimate the uncorrected R magnitude.
}
\label{fig_afterglow}
\end{figure*}

Then, we used the apparent R magnitude measured 2 hours after the burst without any extinction correction as a proxy for the observed brightness of the optical afterglow. The R magnitude is directly interpolated from the available measurements. 
To do so, we required that GRBs in our sample have good optical follow up during the first hours after the burst to accurately measure the optical flux of the afterglow. The afterglow light curves of our complete sample of GRBs can be seen in figure \ref{fig_afterglow}.

The choice of the time (2 hours after the trigger) at which we measure the optical brightness results from various constraints: 
\begin{itemize}
\item[$\circ$] We want the afterglow to be in its classical slow cooling and decaying regime, yet to be bright enough to permit reliable measurements of the magnitude. However, we removed few GRBs, like GRB060206 \cite[][]{Wozniak2006, Monfardini2006}, which exhibit strong optical flaring at that time. 
\item[$\circ$] Above all, we want to measure the optical brightness at a time comparable with the time at which the vast majority of GRB redshifts are measured {\it i.e.} in the first few hours after the trigger.
\end{itemize}

We also decided to exclude few GRBs with high visual extinction such that $A_V^{tot}=A_V^{Gal}+A_V^{Host}>1.2$. Indeed, GRB afterglows which are strongly absorbed by dust do not bring any information on their true optical brightness. This cut off is a good trade-off to optimize the number of GRBs for which afterglow brightness isn't so much polluted by strong external effects (dust absorption). The galactic extinction $A_V^{Gal}$ has been calculated from the dust map of \cite{Schlegel1998} and the host extinctions $A_V^{Host}$ are issued from varying sources. However, for some bursts we didn't have access to the host extinction. In order to have a rough estimate of it we performed a simple linear fit between $A_V^{Host}$ and the intrinsic X-ray absorption $NH_{X,i}$ derived from the Swift XRT catalogue\footnote{$http://www.swift.ac.uk/xrt\_live\_cat/$}. For this fit we had 114 GRBs available. The best fit gives us the following relation : $A_V^{Host} = 3.9\times 10^{-23}\times NH_{X,i} + 0.06$ with a standard deviation of $\sigma\sim~0.34$ magnitude that we considered as an acceptable uncertainty in our $A_V^H$ estimates.

At last, for GRB afterglows with only upper limit of detection, we required that their optical upper limit are deeply constrained by large telescopes (at least one 2.0 meter telescope). Moreover, to be selected these upper limits of detection must have been measured close to 2 hours after the burst.

After passing the optical selection criteria we finally ended with \ngrb\ GRBs with a redshift (69 detections and 2 upper limits) and \ngrbnoz\ GRBs without one (3 detections and 11 upper limits). These 85 GRBs constitute our full sample which is summarized in Table \ref{tab_GRB} for GRBs with a redshift and Table \ref{tab_GRB_sansz} for GRBs without a redshift. This sample covers about 15 years of pre-Swift and Swift GRB observations (from 1999 to 2014).

\subsection{Gamma-ray data}
\label{sub_HE}
For our selected GRBs we also collected the observed gamma-ray properties of the prompt emission : the spectral indexes $\alpha$ and $\beta$, the time for which 90\% of the energy is released, $T_{90}$, the gamma-ray fluence $S_\gamma$ and the observed peak energy of the spectral energy distribution $E_{po}$.
Our gamma-ray selection criteria followed a similar procedure as \cite{Heussaff2013}.
We selected events with well measured spectral parameters from the GCN Circulars $\&$ \citep[][]{Pelangeon2008, Gruber2014}. We parametrize GRB spectra with the Band function, \cite{Band1993}, that consists of two smoothly connected power laws. Following standard naming, we call $\alpha$ the photon spectral index of the low-energy power law ($\alpha >-2$), and $\beta$ the photon spectral index of the high-energy power law ($\beta < -2$). The $\nu F\nu$ spectrum peaks at the energy $E_{po}$, near the junction of the two power laws. The selection was made with the application of the following cuts:

\begin{itemize}
\item[-]First, we made a selection on the duration. We considered GRBs with $T_{90}$ between 2 and 1000 s. This criterion excludes short GRBs ($T_{90}<2s$), and very long GRBs that are superimposed on a varying background and whose $E_{peak}$ is difficult to measure accurately.
\item[-]Second, we require accurate spectral parameters. We excluded GRBs with one or more spectral parameters missing. We excluded GRBs with an error on alpha (the low-energy index of the Band function) larger than 0.5. We excluded a few GRBs with $\alpha <-2.0$ and GRBs with $\beta>\alpha$ because such values suggest a confusion between fitting parameters. We excluded GRBs with large errors on $E_{po}$, defined by a ratio of the $90\%$ upper limit to the $90\%$ lower limit larger than 3.5. We have less stringent constraints on beta (the high-energy index of
the Band function) since we have checked that the position of GRBs in the \eep\ changes very little with beta. When the error on $\beta$ in the catalog is lacking or larger than 1.0, we assigned to $\beta$ the classical value of –2.3 and we give no error. In a few cases, the high energy spectral index in the Fermi catalog is incompatible with being <-2.0 at 2 $\sigma$ level, and the catalog gives the energy of a spectral break that is not $E_{po}$. In these cases we look for $E_{po}$ in the GCN Circulars, and if we cannot find it, we simply remove the burst from the sample.
\end{itemize}


\section{Afterglow optical brighness and instrinsic luminosity}
\label{sec_analyse_opt}

\begin{figure}[t]
\centering
\includegraphics[trim = 30 200 0 200,clip=true,width=0.55\textwidth]{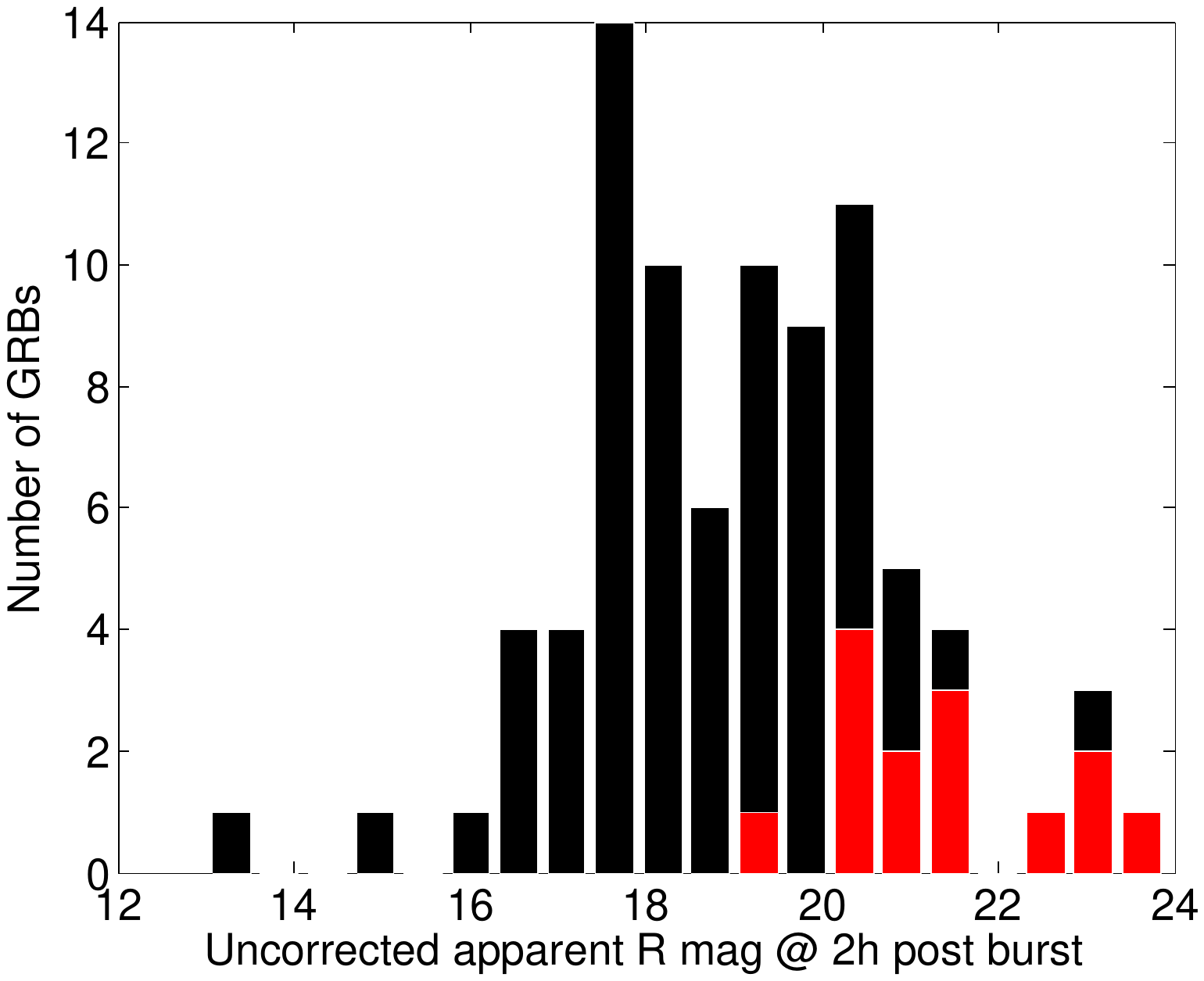}
\caption{Distribution of the afterglow optical brightness taken 2 hours after the burst (uncorrected R magnitude). The 71 GRBs with a redshift are shown in black and the 14 GRBs without a redshift are in red.
}
\label{Rmag_dist}
\end{figure}

The afterglows of our 85 GRBs span a large range of optical brightness from mag R$^{2h}$ = 13.02 to mag R$^{2h}$ = 23.9, see figure \ref{Rmag_dist}. We noticed that GRBs without a redshift have faint optical counterparts which can't be due to high visual extinction since they pass our optical selection criteria. These GRBs are probably high-z GRBs or they could have intrinsically sub-luminous afterglows. Our GRBs with a redshift are also distributed in a wide range of redshift from z=0.168 to z=8.26 and we need to understand if the distribution of the R magnitudes is dominated by the redshift distribution of our GRBs. To verify this hypothesis we separate our 71 GRBs with a redshift into three classes of afterglow brightness equally populated. 

\begin{enumerate}
\item The class of {\it bright} GRBs is composed of 24 GRBs with afterglow optical brightness brighter than R=17.9
\item The class of {\it intermediate brightness} GRBs is composed of 23 GRBs with afterglow optical brightness in the range $17.9<R\leq19.1$
\item The class of {\it faint} GRBs is composed of 24 GRBs with afterglow optical brightness fainter than R=19.1
\end{enumerate}

We will refer to these classes all along this paper.

\subsection{Impact of the redshift on the afterglow optical brightness distribution}
\label{redshift_dist}

\begin{figure}[t]
\centering
\includegraphics[trim = 30 200 0 200,clip=true,width=0.5\textwidth]{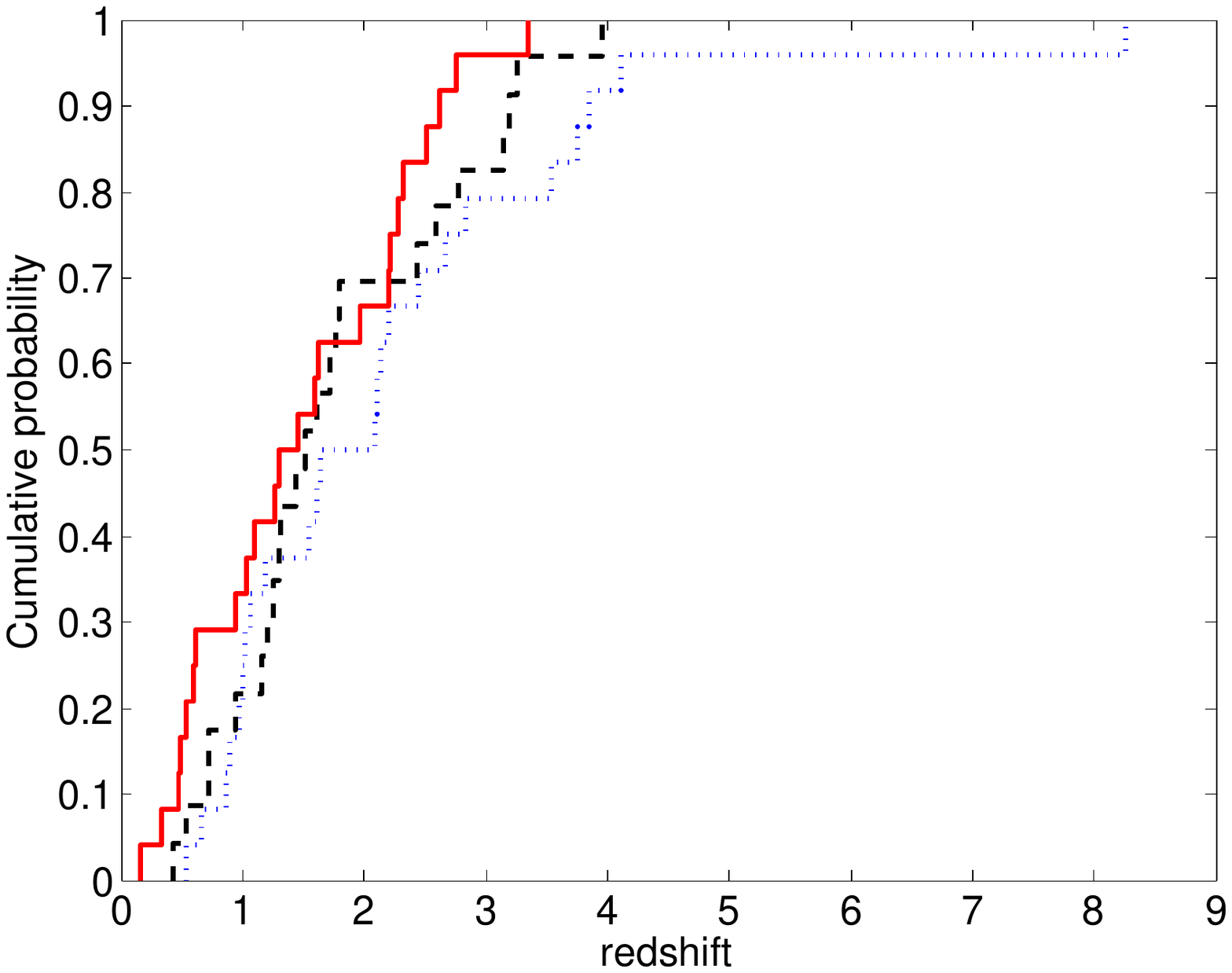}
\caption{Redshift Cumulative distribution function for our three classes of GRB. The bright, intermediate and faint GRB afterglows are indicated with the solid red, dashed black and dotted blue lines, respectively. 
}
\label{z_dist}
\end{figure}

We find that The median redshift for the three classes of GRBs are $z_{med}^{bright}=1.38$, $z_{med}^{int}=1.52$ and $z_{med}^{faint}=1.87$.
We then compare the redshift distribution of the three classes of GRBs with a Kolomogorov-Smirnov statistical test (KS test). The KS test clearly reveals that the redshift distribution of the three classes are similar. The p-value\footnote{The probability of observing a test statistic as extreme as, or more extreme than, the observed value under the null hypothesis} of each test are 0.656 between bright and intermediate GRBs, 0.387 between bright and faint GRBs and 0.711 between intermediate and faint GRBs. This indicates that the redshift should not bias the optical brightness distribution, see figure \ref{z_dist}. 
We perform an additional simple test that consists in boostrapping n-times the redshifts between the bright and faint GRBs and calculating the median redshift difference $\Delta z_{med}$ distribution. Then the probability of obtaining an absolute difference $|\Delta z_{med}|$ at least as strong as we observed with $N_{tot}$ random simulations is given by : 

\begin{equation}
P(X>|\Delta z_{obs}|) = \frac{N(X>|\Delta z_{obs}|)}{N_{tot}},here~ N_{tot}=10^5
\end{equation}
According to this boostrapping test, such a redshift difference between bright and faint afterglows is unsignificant. This result confirms that the redshift doesn't drive the observed optical flux. The results of the statistical tests are summed up in table \ref{tab_stat1}.

\subsection{Impact of the visual extinction in the afterglow optical brightness distribution}
\label{Av_dist}
\begin{figure}[t]
\centering
\includegraphics[trim = 30 200 0 200,clip=true,width=0.55\textwidth]{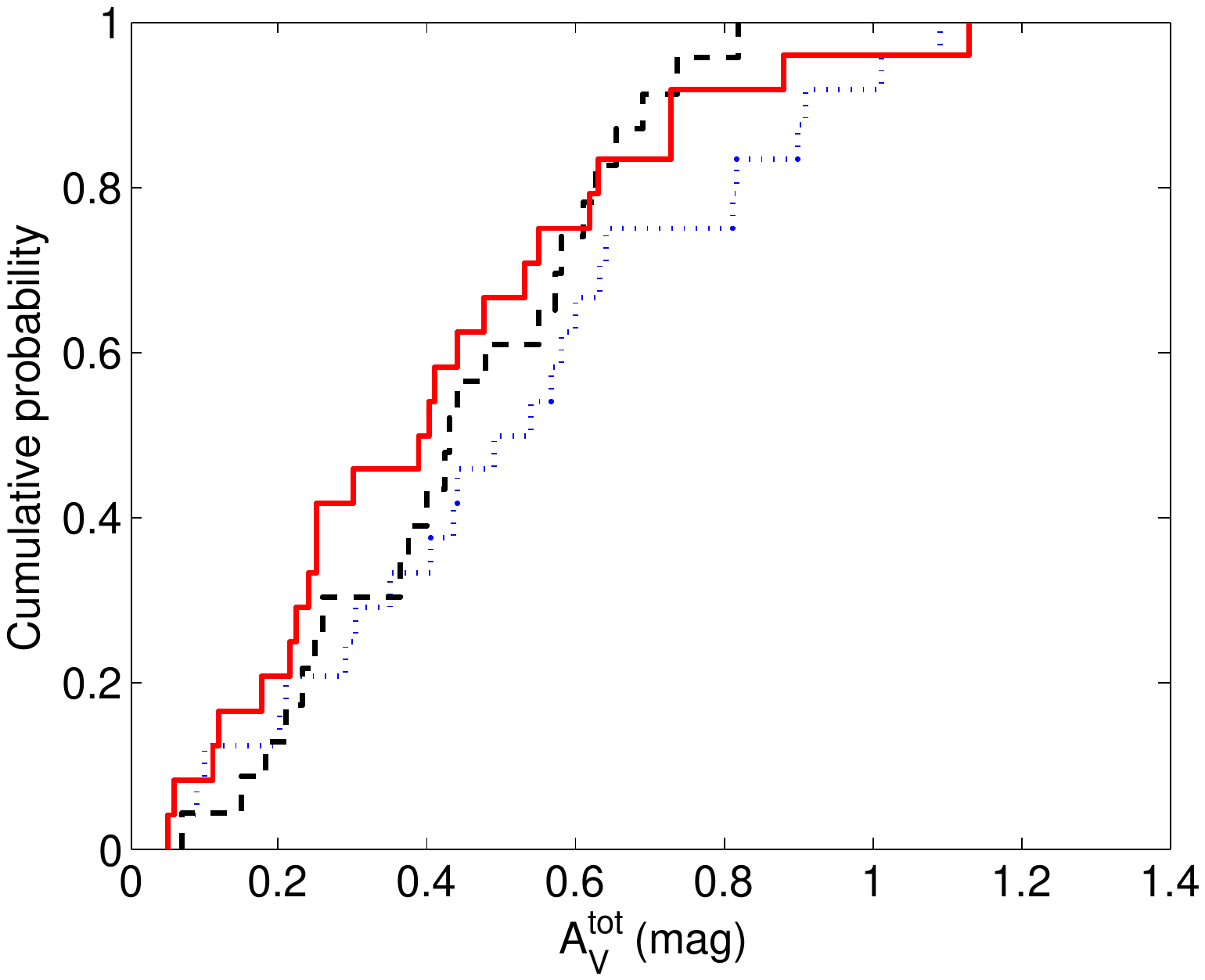}
\caption{Cumulative distribution function of $A_V^{tot}$ for our three classes of GRB. The bright, intermediate and faint GRB afterglows are indicated with the solid red, dashed black and dotted blue lines, respectively. 
}
\label{AV_dist}
\end{figure}

Albeit we have selected GRBs with relatively low total visual extinction ($A_V^{tot}<1.2$) we also checked if this parameter could bias our afterglow brightness distribution, i.e wether faint afterglows are more obscured by dust. We found $median[Av^{tot}_{bright}] = 0.39$ for bright GRBs, $median[Av^{tot}_{int}] = 0.43$ for intermediate GRBs and $median[Av^{tot}_{faint}] = 0.54$ for faint GRBs. We performed a KS-test to compare the $A_V^{tot}$ distribution of our three classes of GRB afterglow brightness, see figure \ref{AV_dist}. We found that the three populations of GRBs are drawn from the same underlying distribution. For example, a p-value as high as 0.622 is found between bright and faint GRBs. We also performed a boostrapping test to compare the total visual extinction of bright and faint GRB afterglows. This test is completely compatible with the result of the KS test, see table \ref{tab_stat1} for the complete results of ours statistical tests. We conclude that our selected GRBs with faint afterglows are as obscured as the bright ones and thus visual extinction doesn't drive our afterglow optical distribution.

\subsection{Afterglow intrinsinc optical luminosity}
\label{Lopt_dist}
As the extrinsic factors (redshift, visual extinction) don't seem to play a major role in the observed optical brightness distribution, we investigated the impact of the intrinsic optical luminosity of the afterglows.
We calculated the optical luminosity density (in units of erg/s/Hz) taken two hours {\it in the rest frame} using the formula:

\begin{equation}
L_R(t_{rest})=\frac{4\pi D_L(z)^2}{(1+z)^{1-\beta_o+\alpha_o}}\times F_{R}(t_{obs})(\frac{\nu_R}{\nu_{obs}})^{-\beta_o}
\end{equation}

where $F_R$ is the optical flux density corrected from the Galactic and host extinction measured at $t_{obs}=2h$ after the burst in the observer frame, z is the GRB redshift, $D_L(z)$ is the luminosity distance, $\beta_o$ is the optical spectral index and $\alpha_o$ is the optical temporal index, $\nu_R$ is the typical frequency of the R band (Vega system) and $\nu_{obs}$ is the observed frequency (here $\nu_{obs}=\nu_R$). 
For GRBs with no optical detection we used their optical upper limits to compute the $F_R$ and the value of $\alpha_o^{med}$ and $\beta_o^{med}$ to estimate an upper limit on their optical luminosity density. 

\begin{figure}[t]
\centering
\includegraphics[trim = 30 200 0 200,clip=true,width=0.55\textwidth]{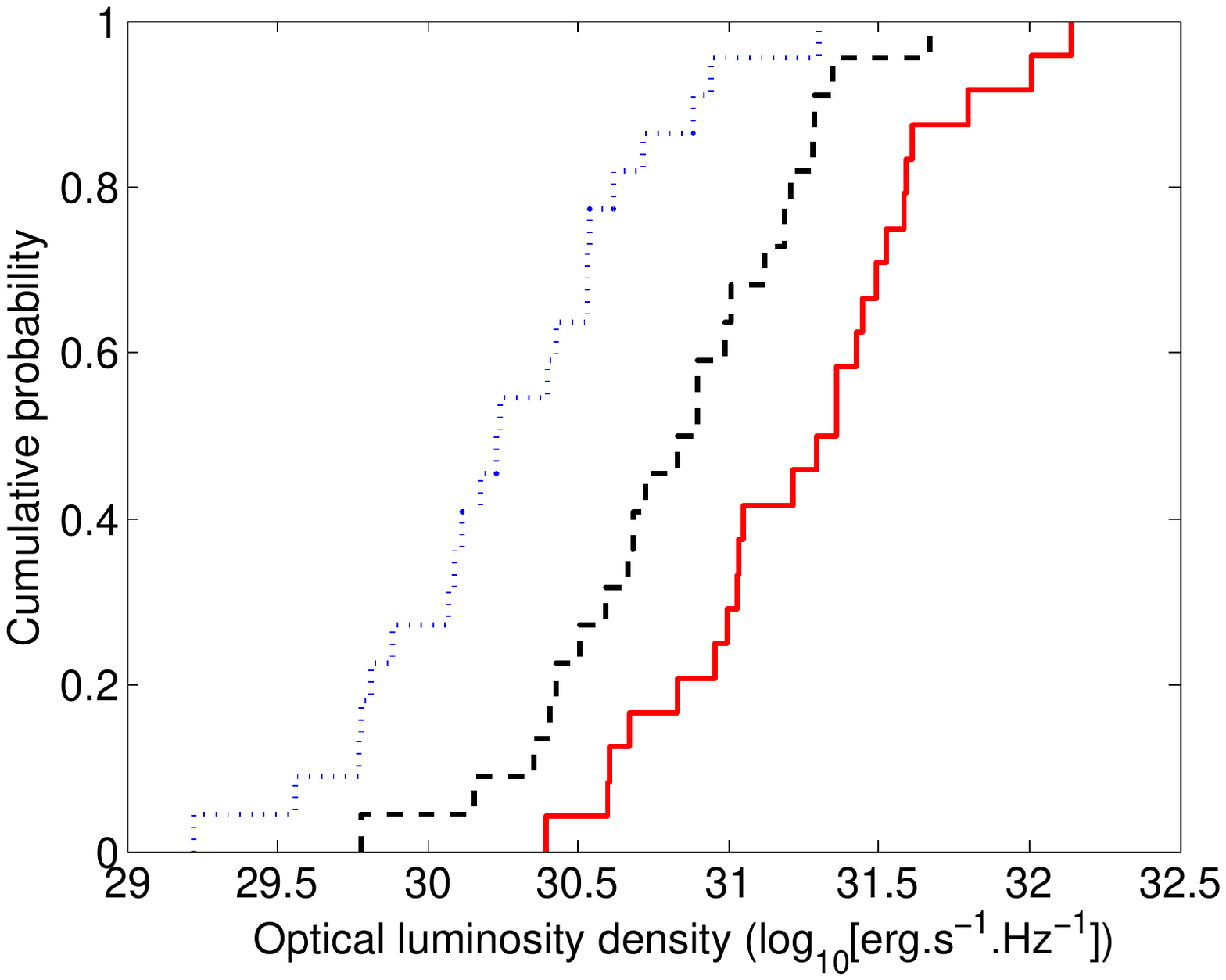}
\caption{Cumulative distribution function of the optical luminosity density (taken 2h after the burst in the rest frame) for our three classes of GRB. The bright, intermediate and faint GRB afterglows are indicated with the solid red, dashed black and dotted blue lines, respectively. 
}
\label{CDF_LR}
\end{figure}

Then, we compared the optical luminosity densities in the rest frame of the three classes of GRB brightness ({\it bright}, {\it intermediate} and {\it faint}), see figure \ref{CDF_LR}. To do so, we apply the same statistical tests than previously. The KS test reveals that :
\begin{itemize}
\item[$\circ$]the population of bright and faint GRBs have very different optical luminosity density distribution with a the p-value of $1.73\times 10^{-6}$ ($\sim 5\sigma$ confidence level).
\item[$\circ$]the population of bright and intermediate GRBs have marginally different optical luminosity density distribution with a the p-value of $1.126\times 10^{-2}$ ($\sim 2.5\sigma$ confidence level).
\item[$\circ$]the population of intermediate and faint GRBs have different optical luminosity density distribution with a the p-value of $4.861\times 10^{-3}$ ($\sim 3\sigma$ confidence level).
\end{itemize}

\begin{figure*}[t]
\centering
\includegraphics[trim = 30 200 0 200,clip=true,width=0.9\textwidth]{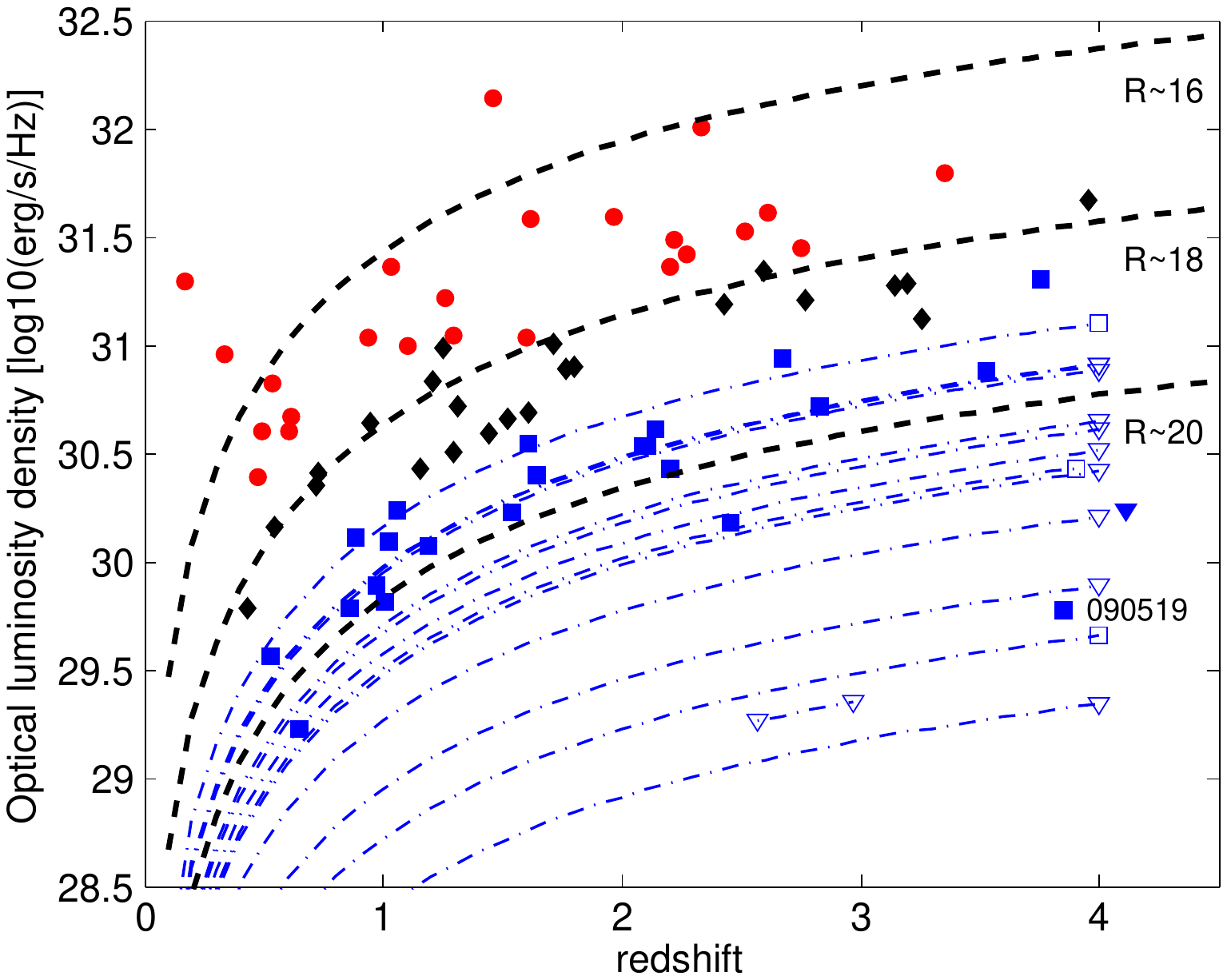}
\caption{Intrinsic optical luminosity density as a function of redshift for 71 GRBs with a redshift and 14 GRBs without a redshift (blue dash-dotted lines). Upper limit are plotted as downward triangles. The red circles represent bright GRBs in the observer frame, the black diamonds represent intermediate GRBs in the observer frame and faint GRBs in the observer frame are plotted as blue squares.
}
\label{z_vs_LR}
\end{figure*}

The median optical luminosity densities (in log scale) for the three classes of GRBs are $LR_{med}^{bright}$=31.33$~erg.s^{-1}.Hz^{-1}$, $LR_{med}^{int}$=30.86$~erg.s^{-1}.Hz^{-1}$ and $LR_{med}^{faint}$=30.23$~erg.s^{-1}.Hz^{-1}$. We observe that faint GRBs are one order of magnitude less luminous on average than bright GRBs ($\Delta L_R = 1.1~erg.s^{-1}.Hz^{-1}$).The bootstrapping simulations of the optical luminosity density between the class of bright and faint GRBs show that such observed $\Delta L_R$ is never seen in the total random cases ($10^5$ simulations) making this difference of luminosity highly significant. The results of the statistical tests are summed up in table \ref{tab_stat1}. We conclude that our afterglow optical brightness distribution is strongly shaped by the intrinsic optical luminosity densities of the GRB afterglows. Moreover, the GRBs without a redshift also follow this trend. Indeed, they have faint afterglows in the observer frame (see figure \ref{Rmag_dist}) and at any redshift, most of them would have had intrinsically under luminous afterglows, see figure \ref{z_vs_LR}. Thus, we conclude that a large population of GRBs with intrinsic faint afterglows may escape detection in the optical domain creating a strong bias in the afterglow intrinsic luminosity distribution. We particularly remarked that GRBs with $L_{R}<30.0 ~log_{10}(erg.s^{-1}.Hz^{-1})$ are systematically not beyond $z\sim1$, except for GRB090519.

\subsection{The case of GRB 090519}

\label{090519}
GRB 090519 is one exception of a very faint GRB with a redshift measurement that illustrates the optical selection effects that a large population of GRBs undergo. GRB090519 was both detected by Fermi and Swift on 2009 May 19 at 21:08:56 UT, see \cite{Perri2009}. Swift-XRT and Swift-UVOT rapidly observed the field of GRB090519 (<130s after the the burst). An X-ray counterpart was clearly identified by Swift-XRT allowing a refined localization of the burst at RA(J2000) = $ 09^h29^m06.85^s$ and Dec(J2000) = $+00^d10'48.6"$. However, no significant optical counterpart was detected by Swift-UVOT and a corresponding 3-sigma upper limit of 19.6 mag (white filter) was estimated at $\sim200$s after the burst. On the ground, fast robotic telescopes rapidly respond to the GCN notice as TAROT at Calern observatory in France, \cite{Klotz2009a}, FARO at Chante Perdrix Observatory in France, \cite{Klotz2009b}, and BOOTES-1B in Spain \cite{Jelinek2009}. No optical counterpart was detected and a limiting magnitude of R>18.5 at t$\sim$230s after the burst was estimated from TAROT Calern observations. The low galactic extinction ($A_V^{Gal}=0.13$) suggested that GRB090519 is a high-z GRB or is embedded in a very dusty environment. In the next hours, NOT, see \cite{Thoene2009a}, and GROND telescope, see \cite{Rossi2009a}, detected a new fading optical source in the XRT field of view which was associated to the afterglow emission from GRB090519. The magnitude measured by NOT at t$\sim 0.33$h after the burst was R$\sim22.8$ reavealing that the afterglow of GRB090519 is among the faintest afterglow ever observed, see figure \ref{fig_afterglow}. The faintness of the afterglow was due to a combination of a high redshift value and an intrinsic weak luminosity of the afterglow, see figure \ref{z_vs_LR}. We estimate from it's high $NH_{X,i}$ measurement a high $A_v^{Host} = 0.96$, however \cite{Greiner2011} found $AV_{Host}\sim0.01$ by fitting GROND/Swift-XRT broadband spectrum. So, we exclude a very dusty environment surrounding GRB 090519 and this burst should be even intrinsically fainter than the value we report here with a $A_v^{Host} = 0.96$. Albeit faint, the redshift of GRB090519 was determined by the VLT, see \cite{Thoene2009b}, thanks to a fast response to the GCN notice ($\sim$4 h after the burst). We can reasonably believe that a delay of a few additionnal hours in the VLT observation schedule would have made the afterglow of GRB090519 unreachable for a redshift measurement and useless for GRBs studies.

\begin{table*}[ht]
\caption{Result of the different statistical tests (Kolomogorov-Smirnov, bootstrap). The probabilities indicated correspond to the p-values, i.e, the probability of observing a test statistic as extreme as, or more extreme than, the observed value under the null hypothesis
}
\begin{center}
\hspace*{-1.0cm}
\begin{tabular}{|c|c|c|c|}

\hline
\hline
{\tiny Median value}&{\tiny Tested parameter}&{\tiny KS test (p-value)} &{\tiny Bootstrap test (p-value)}\\
\hline
\hline
{\tiny } & {\tiny } & {\tiny } & {\tiny }\\
{\tiny $z^{bright}=1.38$} & {\tiny $\Delta~z^{bright}_{int}$} & {\tiny 0.66} & {\tiny 0.73 }\\
{\tiny } & {\tiny } & {\tiny } & {\tiny }\\
{\tiny $z^{int}=1.52$} & {\tiny $\Delta~z^{bright}_{faint}$} & {\tiny 0.38} & {\tiny 0.34 }\\
{\tiny } & {\tiny } & {\tiny } & {\tiny }\\
{\tiny $z^{faint}=1.87$} & {\tiny $\Delta~z^{int}_{faint}$} & {\tiny 0.71} & {\tiny 0.31 }\\
{\tiny } & {\tiny } & {\tiny } & {\tiny }\\
\hline
{\tiny } & {\tiny } & {\tiny } & {\tiny }\\
{\tiny $A_V^{bright}=0.39$} & {\tiny $\Delta~Av^{bright}_{int}$} & {\tiny 0.92} & {\tiny 0.56 }\\
{\tiny } & {\tiny } & {\tiny } & {\tiny }\\
{\tiny $A_V^{int}=0.43$} & {\tiny $\Delta~Av^{bright}_{faint}$} & {\tiny 0.62} & {\tiny 0.24 }\\
{\tiny } & {\tiny } & {\tiny } & {\tiny }\\
{\tiny $A_V^{faint}=0.54$} & {\tiny $\Delta~Av^{int}_{faint}$} & {\tiny 0.65} & {\tiny 0.47 }\\
{\tiny } & {\tiny } & {\tiny } & {\tiny }\\
\hline
{\tiny } & {\tiny } & {\tiny } & {\tiny }\\
{\tiny $L_R^{bright}=31.33~erg.s^{-1}.Hz^{-1}$} & {\tiny $\Delta~LR^{bright}_{int}$} & {\tiny $1.26\times 10^{-2}$} & {\tiny 0.005 }\\
{\tiny } & {\tiny } & {\tiny } & {\tiny }\\
{\tiny $L_R^{int}=30.86~erg.s^{-1}.Hz^{-1}$} & {\tiny\bf $\Delta~LR^{bright}_{faint}$} & {\tiny $1.73\times 10^{-6}$} & {\tiny $<10^{-5}$ }\\
{\tiny } & {\tiny } & {\tiny } & {\tiny }\\
{\tiny $L_R^{faint}=30.23~erg.s^{-1}.Hz^{-1}$} & {\tiny $\Delta~LR^{int}_{faint}$} & {\tiny 0.005} & {\tiny $2.3\times 10^{-4}$ }\\
{\tiny } & {\tiny } & {\tiny } & {\tiny }\\
\hline
\end{tabular}
\end{center}
\label{tab_stat1}
\end{table*}

\section{Optical selection effect in the \eep}
\label{sec_prompt}
With the first measurement of a GRB redshift in 1997, came the possibility to measure the intrinsic properties of GRBs. 
Few years later, \cite{Amati2002} discovered a strong correlation  linking the energy of the maximum of the prompt emission, \epi, and \eiso , the isotropic energy of GRBs. 
Since \eiso\ depends on the cosmology, and \epi\ is cosmology free, this relation has been used by various authors to standardize GRBs,
and eventually to constrain the parameters of cosmological models \citep[e.g.][]{Schaefer2007a,Amati2014}. 
It was also used as a tool to infer GRB redshifts \citep[][]{Atteia2003,Xiao2009}, to constrain the physics of the prompt emission \cite[][]{Eichler2004, Rees2005} and the geometry of GRB jets \citep[][]{Lamb2005,Toma2005}.
However, few years after this discovery, 
other studies showed that this relation could be affected by various selection effects, raising a debate about the reality of the \eer\ \cite[e.g.][]{Band2005, Ghirlanda2005, Nakar2005, Sakamoto2006, Butler2007, Cabrera2007, Schaefer2007a, Butler2009, Firmani2009, Krimm2009, Butler2010, Shahmoradi2011, Collazzi2012a, Kocevski2012, Goldstein2012b}. 
\cite{Heussaff2013} proposed to explain these conflicting results with the combination of two effects: a true lack of GRBs with large \eiso\ and low \epi\ and selection effects preventing the detection of GRBs with low \eiso\ and large \epi , which have comparatively less photons.
While they mainly investigated the selection effects related with the detection of the prompt gamma-ray photons, \cite{Heussaff2013}  also mentioned the possible impact of optical selection effects on the \eer, without investigating this possibility in detail. 
This issue is discussed in this section. 

\subsection{Our GRB sample in the \eep}
\begin{figure*}[t]
\begin{minipage}{0.5\linewidth}
\hspace{-1.3cm}
\includegraphics[trim = 30 200 0 200,clip=true,width=1.25\textwidth]{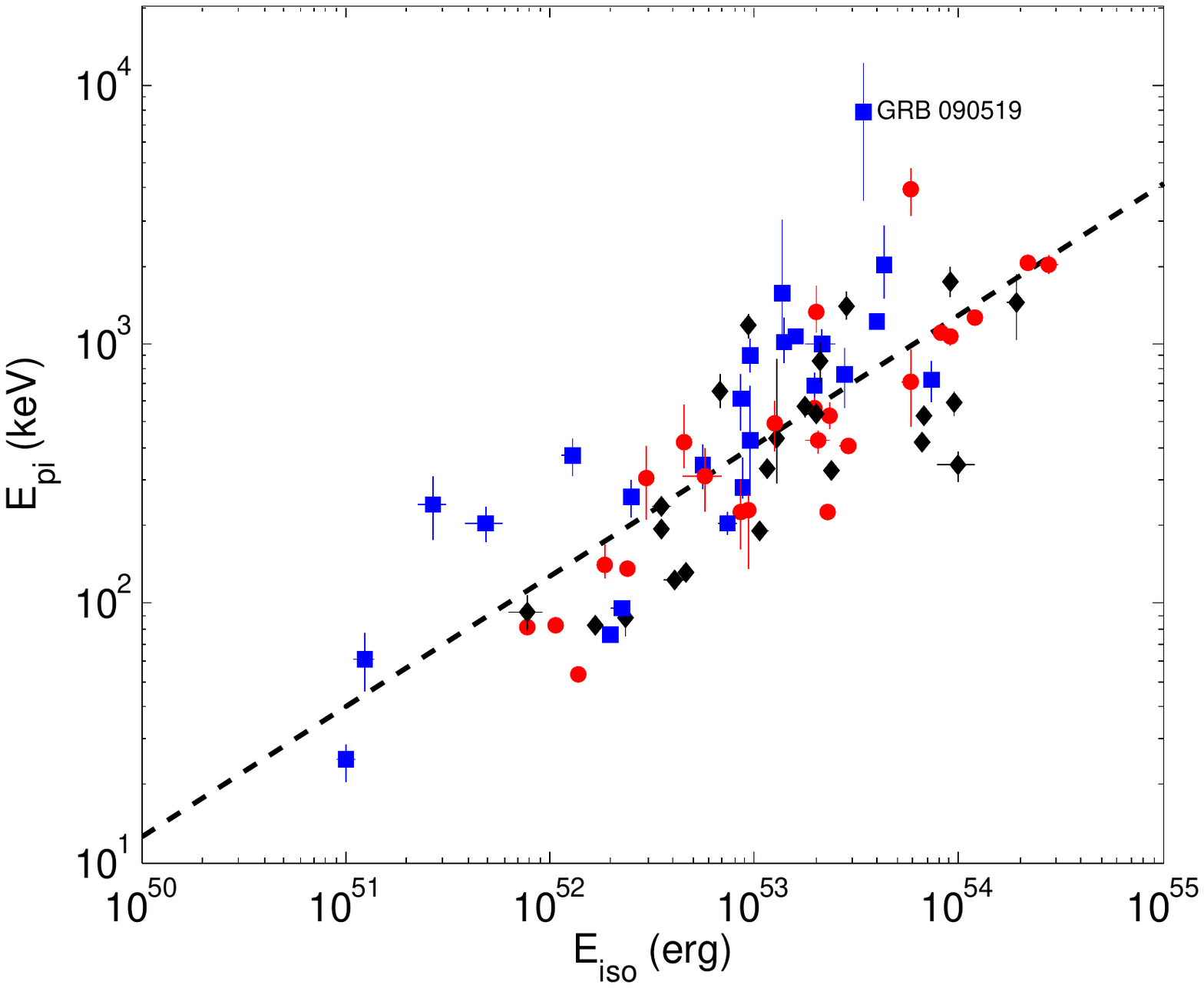}
\end{minipage}
\begin{minipage}{0.5\linewidth}
\hspace{-0.8cm}
\includegraphics[trim = 30 200 0 200,clip=true,width=1.35\textwidth]{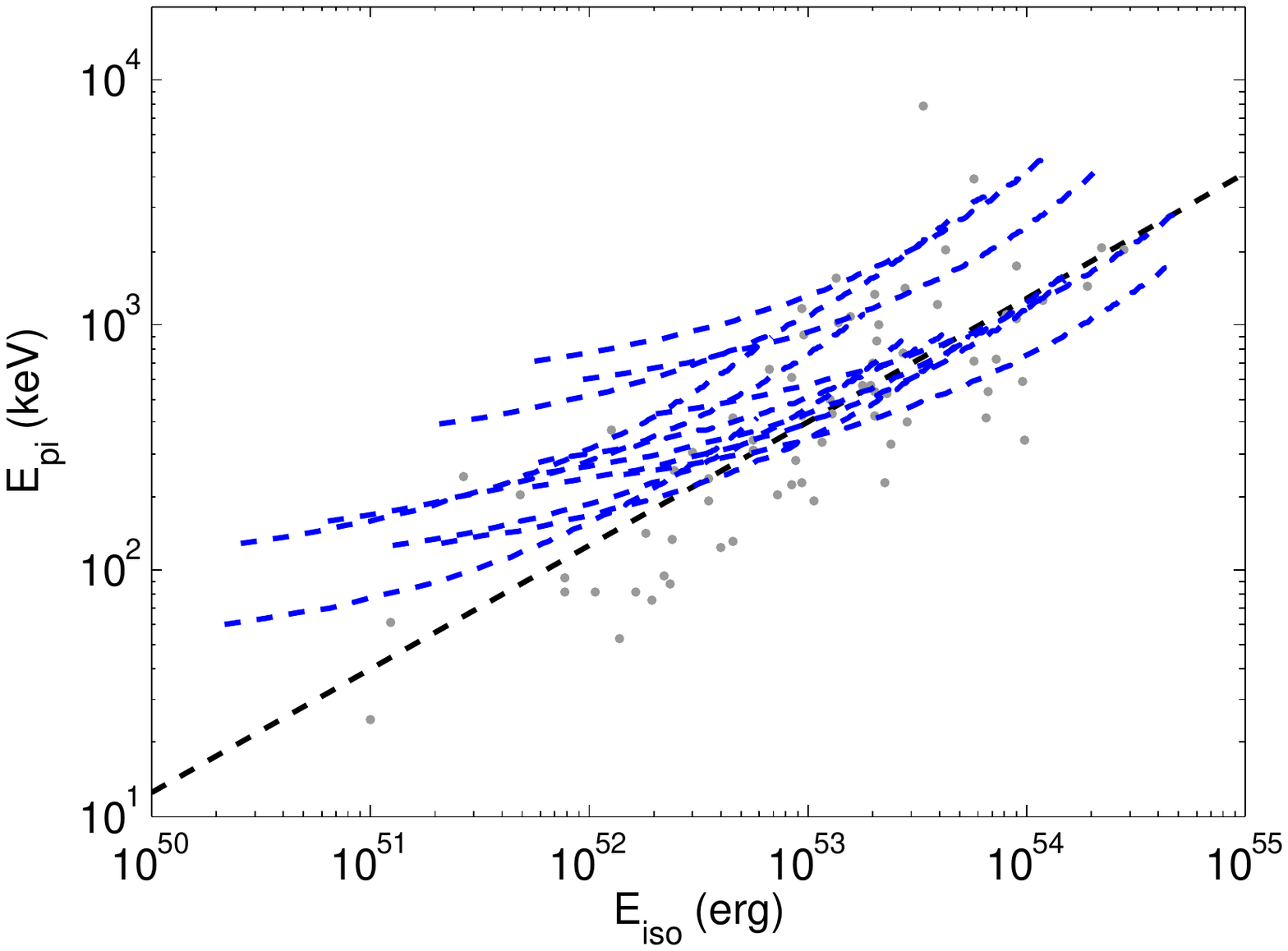}

\end{minipage}
\caption{Distribution of the \ngrb\ GRBs with a redshift (left panel) and the 14 GRBs without a redshift (right panel) in the \eep. {\sf Left panel} : The red circles corresponds to the population of GRBs with bright afterglows, the black diamonds corresponds to the GRB with intermediate brightness and the blue squares corresponds to the faint GRB afterglows. {\sf Right panel} : The dashed lines represents the possible location of the GRBs witout a redshift in the \eep\ and their color shows their class of afterlow brightness (red = bright, black = intermediate, blue = faint). The grey dots indicate the location of our 71 GRBs with a redshift in the \eep. {\sf Both panels} : The best fit \eer\ is represented as the black dashed solid line, which is fully compatible with other recent studies \citep[e.g.][]{Nava2012,Gruber2012}.
}
\label{fig_amati}
\end{figure*}

Our selected \ngrb\ GRBs with a redshift follow a standard \eer, as shown in figure \ref{fig_amati}. 
The best fit \eer\ for this sample is ${\rm E}_{\rm pi} = 126~{\rm E}_{52}^{0.504}~ {\rm keV}$, where E$_{52}$ is the GRB isotropic energy in units of $10^{52}$ ergs (see figure \ref{fig_amati}). 
This best fit relation is fully compatible with the \eer\ found by other authors \cite[e.g.][] {Nava2012,Gruber2012}, showing that our sample is not significantly biased for what concerns the distribution of GRBs with a redshift  in the \eep . The dispersion of the points around the best fit relation along the vertical axis, $\sigma = 0.29$, is also comparable with the values found by \cite{Nava2012} and \cite{Gruber2012}, ($\sigma = 0.34$). 

\subsection{Afterglow optical brightness in the \eep}
In order to investigate the distribution of the afterglow optical brightnesses in the \eep, we did two complementary analysis. One comparing the vertical distance to the best fit \eer\ for the three classes of GRB afterglow brightness and a second comparing the afterglow optical brighness of GRBs located above and below the best fit \eer.
\subsubsection{Comparing the vertical distances}
\label{brightness_eep}
\begin{figure*}[t]
\centering
\includegraphics[trim = 30 200 0 200,clip=true,width=0.8\textwidth]{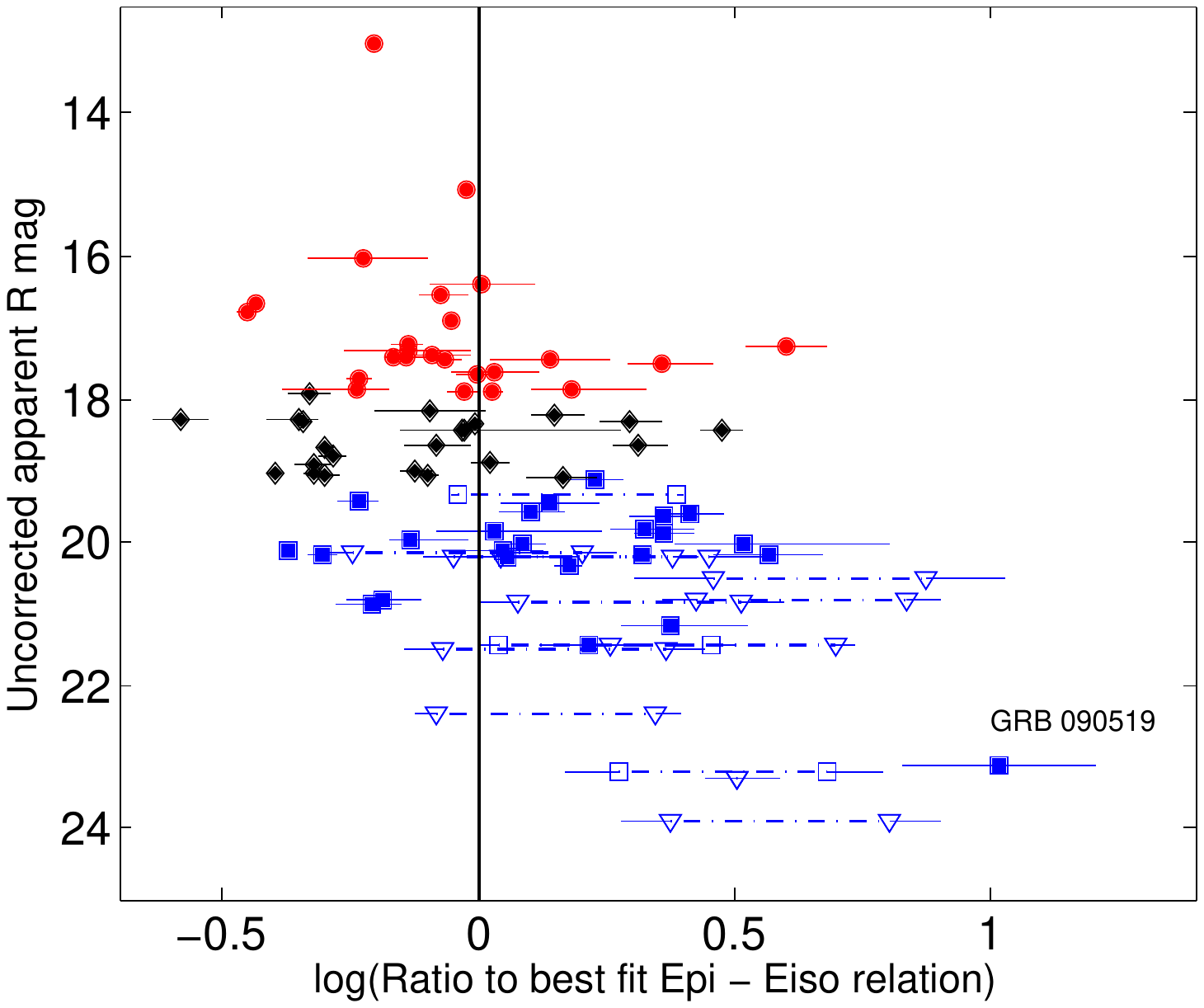}
\caption{Distribution of the afterglow optical brightness as a function of their vertical distance to the \eer. The black solid line separates the population of GRBs located above ($dist>0$) and below ($dist<0$) the best fit \eer. The red circles represent the class of bright GRBs, the black diamonds represent the GRBs of intermediate brightness and the blue squares represent the faint GRBs. The upper limit of detections are plotted as downward trinagles. The GRBs without a redshift are represented with dash dotted line showing their maximum and minimum vertical distance to the \eer.
}
\label{dist_Rmag}
\end{figure*}

\begin{figure}[t]
\centering
\includegraphics[trim = 30 200 0 200,clip=true,width=0.55\textwidth]{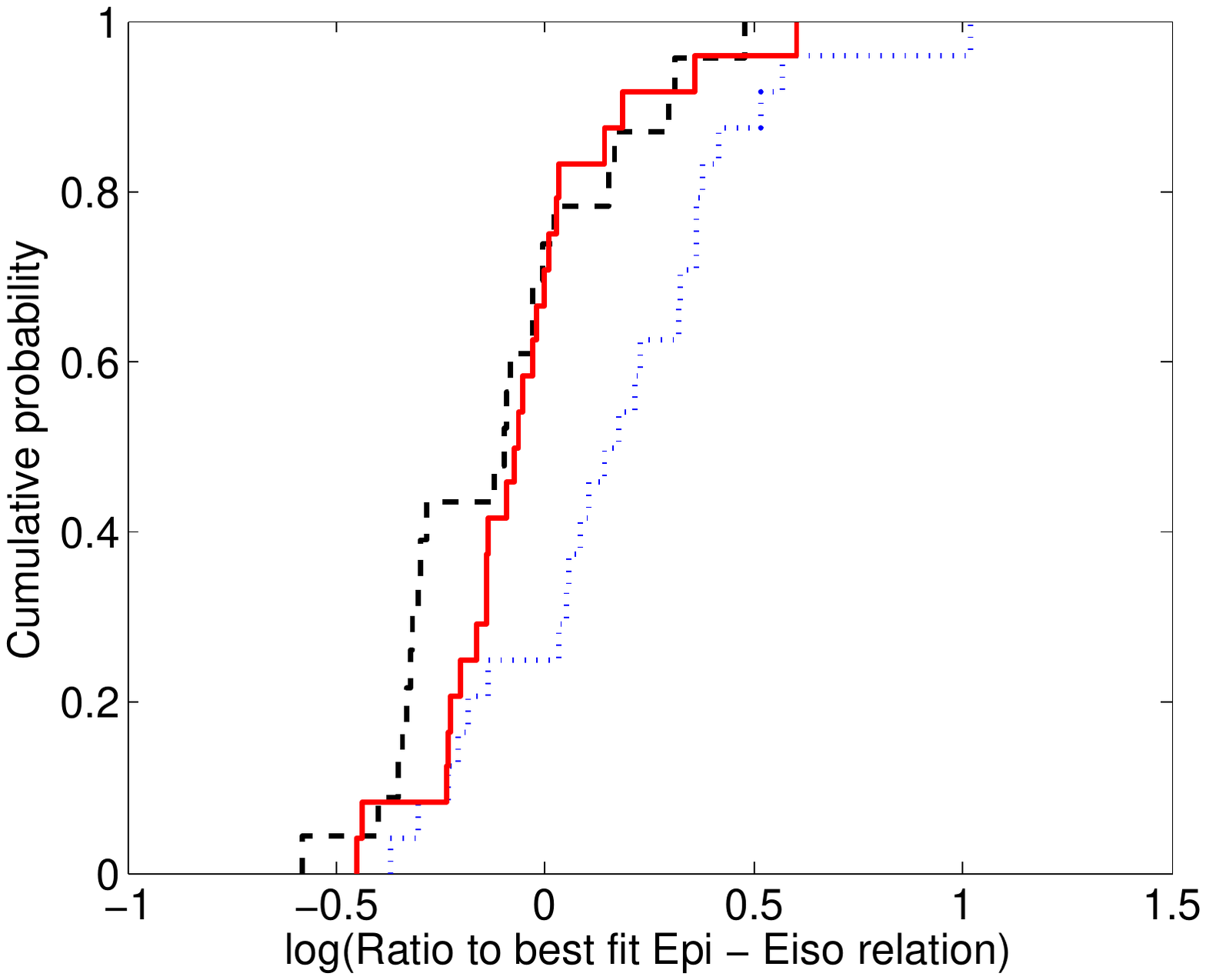}
\caption{Cumulative distribution function of the vertical distance to the best fit \eer\ for our three classes of GRB. The bright, intermediate and faint GRB afterglows are indicated with the solid red, dashed black and dotted blue lines, respectively. 
}
\label{CDF_dist}
\end{figure}

The vertical distance is defined as $log_{10}(E_{pi}) - log_{10}[best fit~(E_{pi}-E_{iso}~relation)]$ and we show in figure \ref {dist_Rmag} our GRB sample in a plane combining the vertical distances to the best fit \eer\ and the afterglow optical brightness. We found that the median vertical distance of the bright GRBs is $dist_{med}^{bright} = -0.07 $, $dist_{med}^{int} = -0.09 $ for the intermediate GRBs and $dist_{med}^{faint} = 0.16 $ for the faint GRBs. We performed a KS test to compare the vertical distance distributions for our three class of GRBs. The KS test reveals that the two populations of bright and faint GRBs are drawn from two different underlying distributions with a p-value =$9.31\times 10^{-4}$, see figure \ref{CDF_dist}. We also calculated the KS-test between the bright and intermediate GRBs and between the intermediate and faint GRBs. These tests reveal that intermediate and bright GRBs obey to the same distribution while faint GRBs clearly differ from the two other group of GRBs. To verify the validity of the KS-test we perform a booststrapping simulation considering the vertical distances and we also find that faint GRBs differ from the other group of GRBs. The results of statistical tests are summed up in table \ref{tab_stat2}.

\smallskip

These two statistical tests confirm that GRBs with faint afterglows are preferentially located above the best fit \eer\ (18/24 faint GRBs are "above") whereas bright GRBs are generally located below the best fit \eer\ (17/24 bright GRBs are "below"). This particular distribution of the afterglow optical brightnesses in the \eep\ is significant at the level >3$\sigma$.

\subsubsection{Comparing the afterglow optical brightness with respect to the best fit \eer}

We performed a complementary analysis, looking if we can find a significant difference in optical brightness between GRBs located above and below the best fit \eer, see figure \ref{dist_Rmag}. We compared the optical brightness (R magnitude) of the two groups of GRBs. In order to include GRBs without a redshift, we calculated their minimum and maximum vertical distance to the best fit \eer. Then we only kept those which are strictly above the best fit \eer\ or strictly below it at all redshifts considered here ($0.34<z<3.72$). 7 GRBs without a redshift were selected and all of them were strictly located above the best fit \eer. So, we finally used 78 GRBs in the analysis.

\begin{figure}[t]
\centering
\includegraphics[trim = 30 200 0 200,clip=true,width=0.55\textwidth]{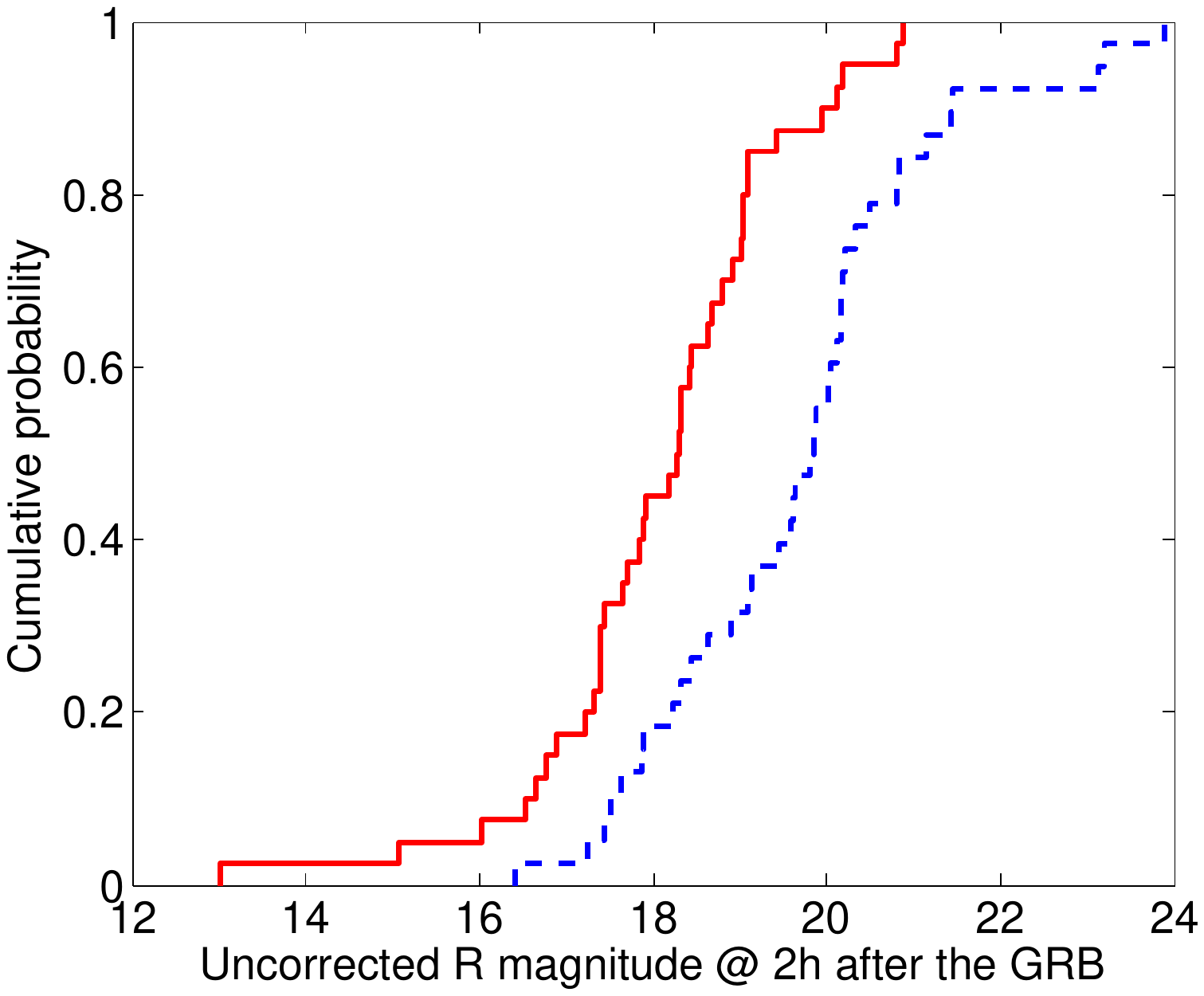}
\caption{Cumulative distribution function of the afterglow R magnitudes. The \nabove\ GRBs located above the best fit \eer\ are indicated with the blue dashed line while the \nbelow\ GRBs located below the best fit \eer\ are represented with the red solid line.
}
\label{significance}
\end{figure}

First, we compared the afterglow optical brightness distribution of GRBs above and below the best fit \eer\ applying a KS-test. The statistical test confirms that the two distributions clearly differ with a p-value $p = 1.41 \times 10^{-5}$. 
We found that the mean R magnitude of the GRBs located above the best fit \eer\ (38 GRBs) is R$_{\rm mean}^{above} = 19.69$, while the GRBs below the best fit \eer\ (40 GRBs) have R$_{\rm mean}^{below} = 18.1$. We again performed bootstrapping simulations (with the R magnitudes) to evaluate the significance of such brightness difference ($\Delta R_{\rm mean} = 1.59 $) between GRBs "above \& below". An R magnitude difference as strong as we observe is obtained in only 0.003\% of the cases, indicating that it is significant at the level of 99.997\% ($>\sim4\sigma$ confidence level), see figure \ref{significance}. The high significance of these tests confirm that GRBs located below the best fit \eer\ have brighter afterglows than those of GRBs located above it. The results of the statistical tests are summed up in table \ref{tab_stat2}. 

\begin{table*}[t]
\caption{Result of the different statistical tests (Kolomogorov-Smirnov, bootstrap). The probabilities indicated correspond to the p-values, i.e, the probability of observing a test statistic as extreme as, or more extreme than, the observed value under the null hypothesis
}
\begin{center}
\hspace*{-1.0cm}
\begin{tabular}{|c|c|c|c|}
\hline
\hline
{\tiny Median value}&{\tiny Tested parameter}&{\tiny KS test (p-value)} &{\tiny Bootstrap test (p-value)}\\
\hline
\hline
{\tiny } & {\tiny } & {\tiny } & {\tiny }\\
{\tiny $dist^{bright}= -0.07$} & {\tiny $\Delta~dist^{bright}_{int}$} & {\tiny 0.084} & {\tiny 0.62 }\\
{\tiny } & {\tiny } & {\tiny } & {\tiny }\\
{\tiny $dist^{int}= -0.09$} & {\tiny\bf $\Delta~dist^{bright}_{faint}$} & {\tiny $9.31\times 10^{-4}$} & {\tiny $9.0\times 10^{-3}$ }\\
{\tiny } & {\tiny } & {\tiny } & {\tiny }\\
{\tiny $dist^{faint}= 0.16$} & {\tiny $\Delta~dist^{int}_{faint}$} & {\tiny $1.40\times 10^{-3}$} & {\tiny 0.026 }\\
{\tiny } & {\tiny } & {\tiny } & {\tiny }\\
\hline
{\tiny } & {\tiny } & {\tiny } & {\tiny }\\
{\tiny $Rmag~^{above}_{below} = ^{19.69}_{18.10}$} & {\tiny $\Delta~Rmag^{above}_{below}$} & {\tiny $1.41\times 10^{-5}$} & {\tiny $3.0\times 10^{-5}$ }\\
{\tiny } & {\tiny } & {\tiny } & {\tiny }\\
\hline
\end{tabular}
\end{center}
\label{tab_stat2}
\end{table*}

\subsection{Consequence for the \eer\ and other GRB correlations}

According to these results, we conclude that there is a significant correlation between the afterglow optical brightness of our GRB sample and their location in the \eep. Faint GRBs preferentially fill in the region above the best fit \eer\ and conversely for the bright GRBs which fill preferentially the region below the best fit \eer. This effect could prevent many GRBs in the upper part of the \eep\ from having a redshift measurement. Indeed, we noticed that the majority of GRBs without a redshift, which have very faint afterglows (fainter than R=20 mag at 2h after the burst) are mainly located in the upper part of the \eep. Thus, we confirm that the \eer\ is not immune to significant optical selection effets that could jointly act with the gamma-ray selection effect against the detection of a wider population of GRBs in the upper part of the \eep. In addition, as the redshift is a crucial ingredient to study GRB rest frame properties, it's clear that optical selection effects would also apply to other GRB rest frame correlations.

\subsection{Using the \eer\ for cosmology}
Despite the existence of significant selection effects in the \eep\ we observed that GRBs brighter than Rmag$\sim$ 19.7 are rather symetrically distributed around the best fit \eer. Below that optical brightness, GRBs are clearly concentrated towards the upper part of the \eep, see figure \ref{dist_Rmag}. First, this means that bright GRBs may indeed follow a standard \eer\ and could be suitable for cosmological studies based on this correlation. Second, faint and dark GRBs cannot be used for cosmological studies in the same way as bright GRBs, especially since they are not compatible with the \eer. In any case the selection effects due to the measure of the redshift should be taken into account when discussing the \eer\ as a genuine rest frame prompt property of long GRBs and when attempting to use it for cosmology. Moreover, as shown in the figure \ref{z_vs_LR}, GRBs that can be calibrated with supernovae Ia (at redshifts z < 1.5), are not representative of the GRB population at higher redshift (fig. \ref{z_vs_LR}). We have shown in section \ref{Lopt_dist} that high-z GRBs are mostly optically intrinsically bright and that a large population of faint GRBs remains undetected, while these faint GRBs represent a significant fraction of the population detected at lower redshifts. According to this feature, we think that cosmological studies using high-z GRBs may undergo large uncertainties.

\section{Discussion}
\label{discussion}

\subsection{Impact of the \eiso\ on the observed afterglow brightness}

\begin{figure}[t]
\centering
\includegraphics[trim = 30 200 0 200,clip=true,width=0.55\textwidth]{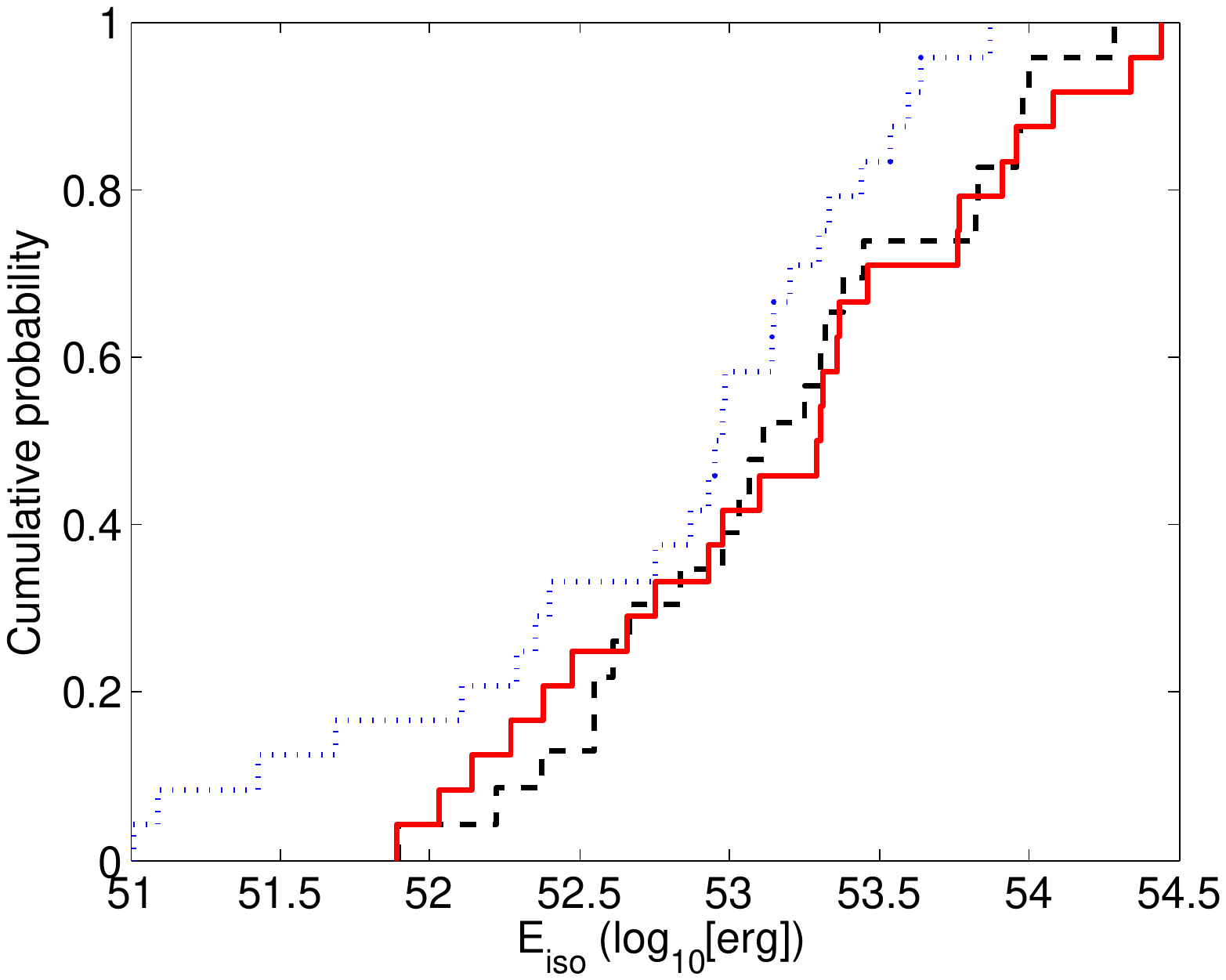}
\caption{Cumulative distribution function of \eiso\ for our three classes of GRBs. The bright, intermediate and faint GRB afterglows are indicated with the solid red, dashed black and dotted blue lines, respectively. 
}
\label{CDF_Eiso}
\end{figure}

In the framework of the fireball model, the intensity of the afterglow emission is partly determined by the internal kinetic energy of the jet $E_k^{52}= E_{iso}^{52}\times\frac{1-\eta}{\eta}$ where $\eta$ is the gamma-ray radiative efficiency. Thus, the \eiso\ is an important parameter to power optical GRB afterglows . We decided to check if the \eiso\ distribution is biased by the optical brightness distribution of GRB's afterglow. First, we calculated the median \eiso\ value $Eiso_{med}^{bright}=53.3~log_{10}(erg)$ for the bright GRBs, $Eiso_{med}^{int}=53.1~log_{10}(erg)$ for the bright GRBs and $Eiso_{med}^{faint}=52.9~log_{10}(erg)$ for the faint GRBs. We still applied the same statistical method to compare the \eiso\ distribution of our three class of afterglow brightness, see figure \ref{CDF_Eiso}. The KS test reveals that :
\begin{itemize}
\item[$\circ$]the population of bright and faint GRBs have similar \eiso\ distributions with a the p-value of 0.387.
\item[$\circ$]the population of bright and intermediate GRBs have similar \eiso\ distributions with a the p-value of 0.993.
\item[$\circ$]the population of intermediate and faint GRBs have similar \eiso\ distributions with a the p-value of 0.569.
\end{itemize}

To compare the \eiso\ distribution between the bright and faint GRBs, we also performed a simulation based on booststrapping. According to the these two statistical tests, we conclude that the afterglow optical brightness distribution does not bias the observed \eiso\ distribution of the GRBs. The results of the statistical tests are summed up in table \ref{tab_stat3}.

\subsection{Impact of the \epi\ on the observed afterglow brightness}
\begin{figure}[t]
\centering
\includegraphics[trim = 30 200 0 200,clip=true,width=0.55\textwidth]{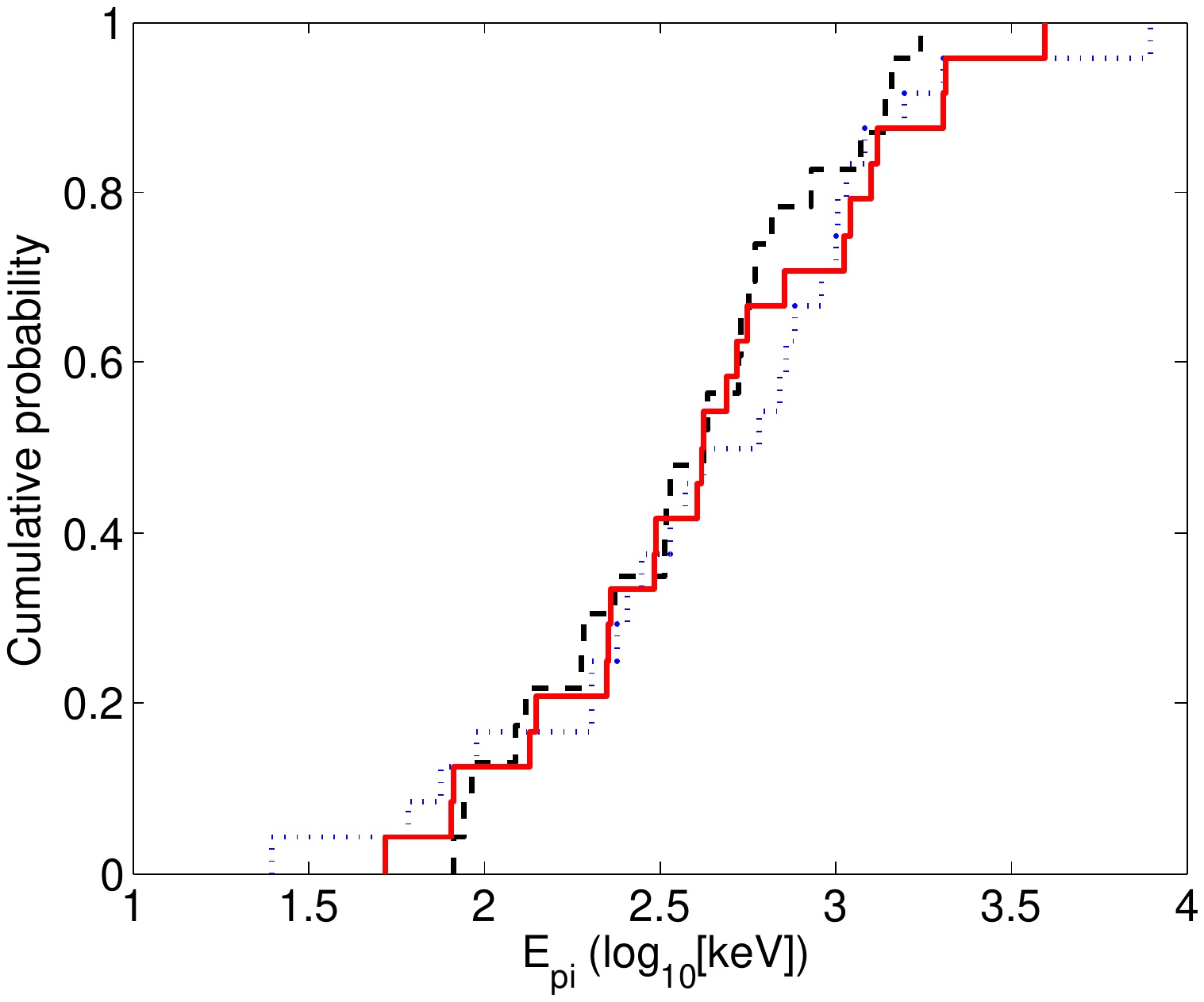}
\caption{Cumulative distribution function of \epi\ for our three classes of GRBs. The bright, intermediate and faint GRB afterglows are indicated with the solid red, dashed black and dotted blue lines, respectively. 
}
\label{CDF_epi}
\end{figure}

We performed the same statistical test for \epi. We calculated the median $Epi_{med}^{bright}=420.2~keV$ for the bright GRBs, $Epi_{med}^{int}=421.0~keV$ for the intermediate GRBs and $Epi_{med}^{faint}=519.6~keV$ for the faint ones. Then we applied a KS test to compare the \epi\ distributions of our three class of GRB's afterglow brightness. This test reveals that the three populations of GRB have clearly the same \epi\ distributions, see figure \ref{CDF_epi}. We noted the lack of significance difference between the bright and the faint afterlows measured with the KS test (p-value = 0.861) indicating that optical afterglow brighness does not strongly bias the observed \epi\ distribution. As for the previously tested parameter, we performed additional simulations based on bootstrapping the \epi\ between the bright and faint GRBs. We found that a \epi\ difference as we observed ($\Delta E_{pi} = 100 ~keV$) is not significant. The statistical test results are summarized in table \ref{tab_stat3}.

\begin{table*}
\caption{Result of the different statistical tests (Kolomogorov-Smirnov, bootstrap). The probabilities indicated correspond to the p-values, i.e, the probability of observing a test statistic as extreme as, or more extreme than, the observed value under the null hypothesis
}
\begin{center}
\hspace*{-1.0cm}
\begin{tabular}{|c|c|c|c|}
\hline
\hline
{\tiny Median value}&{\tiny Tested parameter}&{\tiny KS test (p-value)} &{\tiny Bootstrap test (p-value)}\\
\hline
\hline
{\tiny } & {\tiny } & {\tiny } & {\tiny }\\
{\tiny $E_{iso}^{bright} = 53.3~log_{10}(erg)$ } & {\tiny $\Delta~Eiso^{bright}_{int}$} & {\tiny 0.99} & {\tiny 0.66 }\\
{\tiny } & {\tiny } & {\tiny } & {\tiny }\\
{\tiny $E_{iso}^{int} = 53.1~log_{10}(erg)$ } & {\tiny $\Delta~Eiso^{bright}_{faint}$} & {\tiny 0.39} & {\tiny 0.16 }\\
{\tiny } & {\tiny } & {\tiny } & {\tiny }\\
{\tiny $E_{iso}^{faint} = 52.9~log_{10}(erg)$ } & {\tiny $\Delta~Eiso^{int}_{faint}$} & {\tiny 0.57} & {\tiny 0.48 }\\
{\tiny } & {\tiny } & {\tiny } & {\tiny }\\
\hline
{\tiny } & {\tiny } & {\tiny } & {\tiny }\\
{\tiny $E_{pi}^{bright} = 420~keV$ } & {\tiny $\Delta~Epi^{bright}_{int}$} & {\tiny 0.99} & {\tiny $\sim$ 1.0 }\\
{\tiny } & {\tiny } & {\tiny } & {\tiny }\\
{\tiny $E_{pi}^{int} = 421~keV$ } & {\tiny $\Delta~Epi^{bright}_{faint}$} & {\tiny 0.86} & {\tiny 0.62 }\\
{\tiny } & {\tiny } & {\tiny } & {\tiny }\\
{\tiny $E_{pi}^{faint} = 520~keV$ } & {\tiny $\Delta~Epi^{int}_{faint}$} & {\tiny 0.45} & {\tiny 0.69 }\\
{\tiny } & {\tiny } & {\tiny } & {\tiny }\\
\hline
\end{tabular}
\end{center}
\label{tab_stat3}
\end{table*}

\subsection{Redshift measurement}

It has been shown by \cite{Fynbo2009} that GRBs with a redshift measured from the spectroscopy of their optical afterglow lack of dark GRBs and have also brighter high energy emission from the prompt phase than GRBs for which redshift was not determined from the afterglow spectroscopy. These authors also showed that GRBs with a spectroscopic redshift have significantly less X-ray absorption than GRBs with non spectroscopic redshift. They concluded that these GRBs easily detectable are non-representative of the whole GRB population. As GRBs with a spectroscopic redshift represent the majority of the Swift GRBs with a redshift and that the whole population of Swift GRBs is composed of only one third of GRBs with a redshift, this selection effect introduces in GRBs studies (rest frame prompt properties, star forming region of GRBs, standardization of GRBs, etc.) that should not be ignored. 

\smallskip

The exceptional case of GRB090519 shows us that it is possible to detect and measure the redshift from faint optical afterglows as soon as large telescopes fastly respond to the GCN notice (few hours after the burst at most) when a very early non detection of an optical counterpart is assessed by fast slewing robotic telescope. It is obvious that a joint observational strategy between fast robotic telescope and very large telescope is needed to measure the redshift of such optically optically faint GRB population.

\smallskip

A way to reduce this bias from spectroscopic redshift measurement is to look for the host galaxy of GRBs. If it is sufficiently bright and well identified, a redshift can be determined from the spectroscopy of the host galaxy emission line. This work has already been done by \cite{Hjorth2012} with the TOUGH survey realized with the VLT instruments (FORS1, FORS2, ISAAC and X-shooter). They were able to determine 15 new redshifts over 23 host galaxy spectra increasing the redshift completeness of their total sample from 55\% to 77\%. They also were able to determine with their own TOUGH program 17 redshift larger than z=2. Such method of redshift measurement should be encouraged in order to enrich the GRB samples and unbias statistical study of GRBs and host properties.

\subsection{The optical afterglow of Fermi GRBs}
Our GRB sample is mainly composed of pre-Swift and Swift GRBs and so our study does not take into account the recent GRBs only detected by Fermi-GBM or Fermi-LAT. This could potentially create a bias in our sample since we missed a part of the GRBs population.
However most of Fermi GRBs have no redshift measurements due to the difficulty to locate them with sufficiently high precision. This prevents astronomers from performing optical follow-up with ground based telescopes and spectroscopy of the afterglow optical counterpart. It's clear that this lack of redshift measurements participate to hide numerous Fermi GRBs in the \eep\ and so avoid to test the \eer\ in the context of Fermi GRBs. Some studies, like \citep[][]{Gruber2012, Nava2012, Heussaff2013} particularly showed that Fermi GRBs with a redshift follow a broader \eer\ than pre-Fermi GRBs. Moreover, some outliers (12\% of the Fermi GRBs) have been identified by \cite{Heussaff2013}.
A need for the redshift measurement of Fermi GRBs would be important for GRB studies. 

\smallskip

\begin{figure*}[t]
\begin{minipage}{0.5\linewidth}
\hspace{-1.3cm}
\includegraphics[trim = 30 200 0 200,clip=true,width=1.25\textwidth]{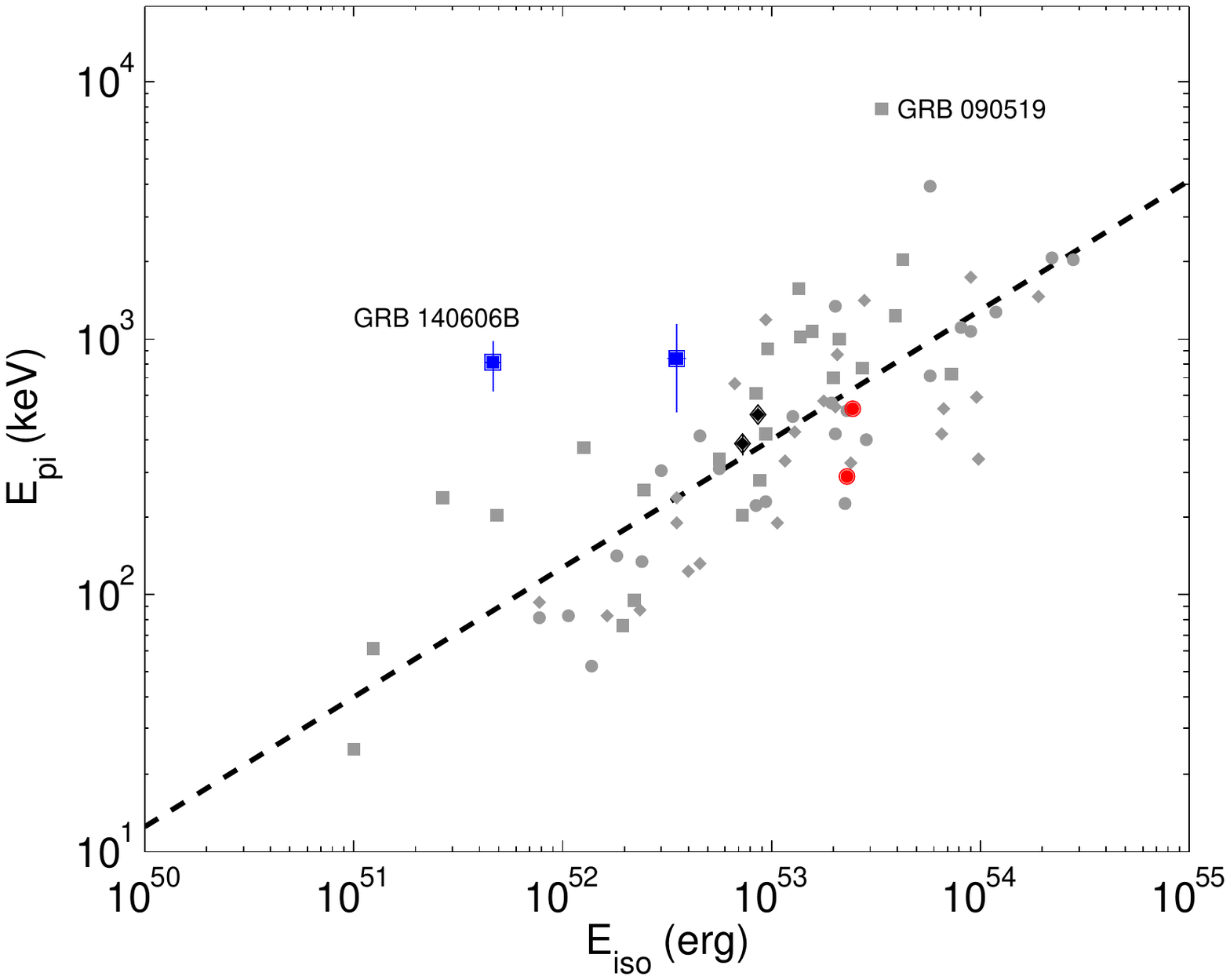}
\end{minipage}
\begin{minipage}{0.5\linewidth}
\includegraphics[trim = 30 200 0 200,clip=true,width=1.31\textwidth]{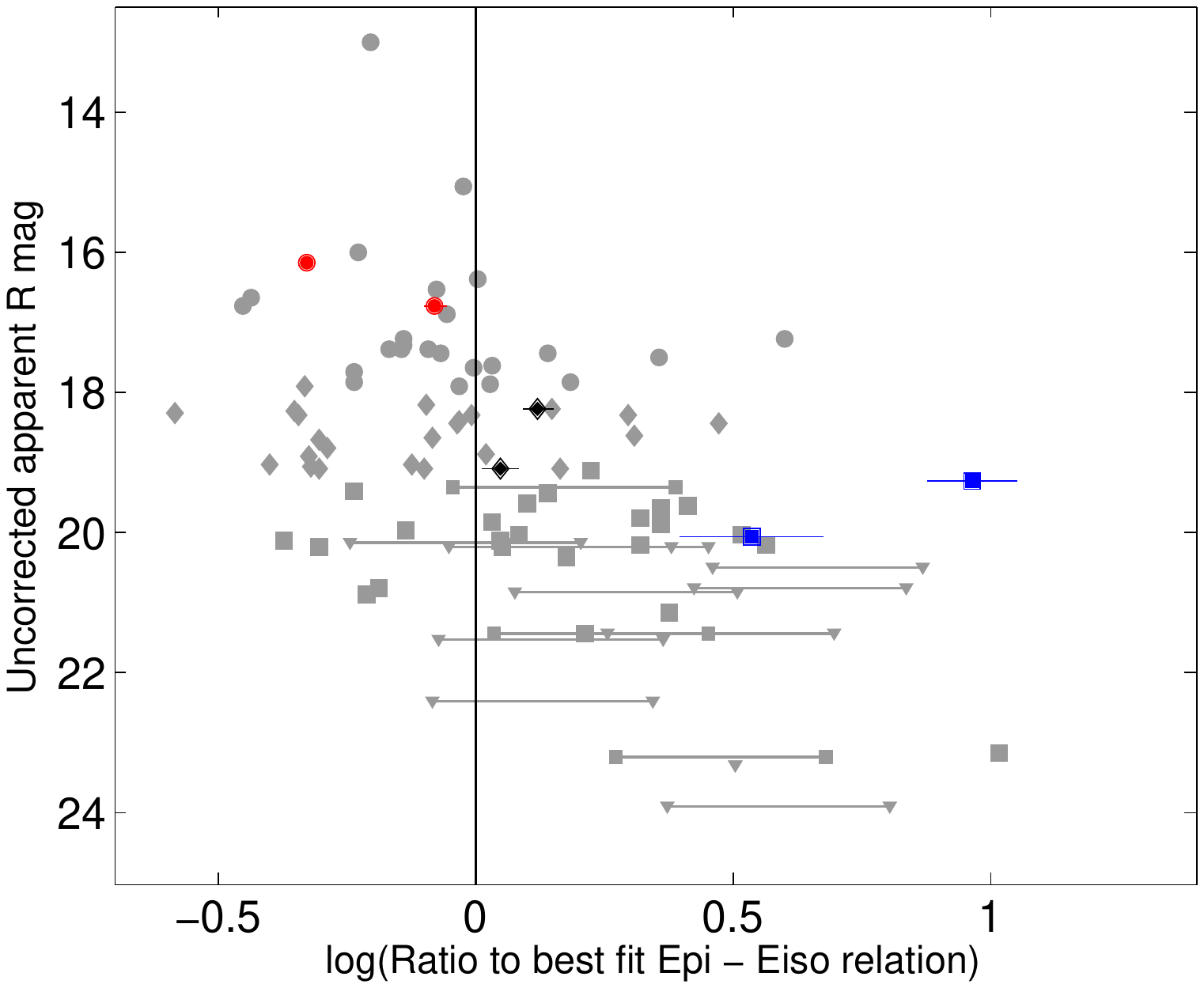}

\end{minipage}
\caption{{\sf Left panel} : 6 Fermi GRBs from \cite{Singer2015} in the \eep. {\sf Right panel} : The Fermi GRBs in the vertical distances/afterglow brightness plane. The colors and symbols show the class of the afterglow optical brightness (blue squares : faint afterglows, black diamonds : afterglow with intermediate brightness and red circles : bright afterglows).
}
\label{fig_Fermi}
\end{figure*}

Recently, \cite{Singer2015} reported the results of their intermediate Palomar Transient Factory observational program. This program consists in searching the optical counterparts of Fermi GRBs to perform broadband follow up and spectroscopy (from X-ray to radio). They present broadband observations of 8 Fermi GRBs for which they could derive spectroscopic redshift. We note that they found a new outlier (140606B/iPTF14bfu) to the \eer\ located at z=0.384. This kind of study is interesting to test our correlation between the afterglow optical brightness and the GRBs location in the \eep\ in the context of Fermi GRBs. According to their optical and gamma-ray data, we have included 6/8 of their GRBs to our analysis (see figure \ref{fig_Fermi}) and found that they confirm it, slightly improving the significance of the magnitude difference between GRBs located above and below the best fit \eer\ (see table \ref{tab_stat4}).

\begin{table*}
\caption{Result of the different statistical tests (Kolomogorov-Smirnov, bootstrap) when the 6 Fermi GRBs are taken into account in the analysis. The probabilities indicated correspond to the p-values, i.e, the probability of observing a test statistic as extreme as, or more extreme than, the observed value under the null hypothesis
}
\begin{center}
\hspace*{-1.0cm}
\begin{tabular}{|c|c|c|}
\hline
\hline
{\tiny Tested parameter}&{\tiny KS test (p-value)} &{\tiny Bootstrap test (p-value)}\\
\hline
{\tiny } & {\tiny } & {\tiny }\\
{\tiny $\Delta~Rmag^{above}_{below}$} & {\tiny $5.70\times 10^{-6}$} & {\tiny $<10^{-5}$ }\\
{\tiny } & {\tiny } & {\tiny }\\
\hline
\hline
{\tiny } & {\tiny } & {\tiny }\\
{\tiny $\Delta~dist^{bright}_{faint}$} & {\tiny $1.73\times 10^{-3}$} & {\tiny 0.0021 }\\
{\tiny } & {\tiny } & {\tiny }\\
\hline
\end{tabular}
\end{center}
\label{tab_stat4}
\end{table*}

   
\section{Conclusions}
\label{sec_conclusion}
We have studied the optical afterglows of 85 GRBs and have determined that the distribution of their optical brightness is mainly shaped by the intrinsic optical luminosity function. The extrinsic factors (redshift and visual extinction) seem to only behave as perturbations in this distribution. We also showed that optical selection effects act against the redshift measurement of faint GRBs in the rest frame, particularly for GRBs with $L_R<30.0 log_{10}(erg.s^{-1}.Hz^{-1}$) and z>1. These non optical detections prevent many GRBs from a redshift measurement.
We also found a strong correlation between the location of the GRBs with respect to the \eer\ and their afterglow optical brighnesses.
This optical bias acts jointly with selection effects in the hard X-ray range against the detection of GRBs located well above the best fit \eer.
Thus we conclude that the need to measure the redshift to obtain GRB rest-frame properties introduces significant biases in the observed distribution of GRBs in the \eep . 

The complete understanding of the distribution of GRBs in the \eep\ and the connection with their afterglow emission will require detailed simulations of the GRB afterglow physics, of their luminosity function and their redshift distribution. The first point will be discussed in a future paper with the aim of providing clues about the connection between the prompt and afterglow physics of GRBs.


\begin{table*}
\caption{Parameters of 71 GRBs with a redshift that make our sample. (1) GRB name, (2) redshift, (3) rest frame peak energy of the gamma-ray spectral energy distribution, (4) isotropic gamma-ray energy, (5) $R_{mag}$ is the apparent R magnitude 2 hours after the burst not corrected from the galactic and host extinctions. (6) $L_R^{rest}$ is the optical luminosity density taken 2 hours in the rest frame, (7) galactic extinction from \cite{Schlegel1998}, (8) host extinction (* indicates that the host extinction is estimated from the $NH_{X,i}$ measurement), (9) vertical distance to the best fit \eer\ (dist<0 if the GRB is below the best fit \eer\ and dist>0 if it is above).}
\begin{center}
\begin{tabular}{ccccccccccc}

\hline
\hline
{\tiny GRB }& {\tiny z} & {\tiny $E_{pi}$} & {\tiny $E_{iso}$}& {\tiny $R_{mag}$} &{\tiny $log_{10}(L_R^{rest})$}& {\tiny $A_V^{Gal}$}& {\tiny $A_V^{Host}$} & {\tiny dist}& {\tiny Réf. z}& {\tiny Réf. $R_{mag}$}\\
{\tiny }& {\tiny } & {\tiny (keV)} & {\tiny ($10^{52} erg$)}& {\tiny (t$_{\rm obs}$=2h)} &{\tiny (erg/s/Hz)}&{\tiny (mag)}&{\tiny (mag)}&{\tiny }& {\tiny } & {\tiny }\\
{\tiny (1)}& {\tiny (2)} & {\tiny (3)} & {\tiny (4)}& {\tiny (5)} &{\tiny (6)}&{\tiny (7) }& {\tiny (8)}& {\tiny (9)}& {\tiny (10)}& {\tiny (11)}\\
\hline
\hline
{\tiny } & {\tiny } & {\tiny  }&{\tiny }& {\tiny } & {\tiny }&{\tiny }&{\tiny }&{\tiny }&{\tiny  }&{\tiny }\\
{\tiny 990123} & {\tiny 1.60} & {\tiny $2030.0^{+161.0}_{-161.0}$ }&{\tiny $278.0^{+31.5}_{-31.5}$ }& {\tiny $17.90$ }&{\tiny 31.03}&{\tiny0.05}&{\tiny$\sim 0$$~^{a)}$}&{\tiny-0.029}&{\tiny A.}&{\tiny (1)}\\
{\tiny 990510} & {\tiny 1.619} & {\tiny $423.0^{+42.0}_{-42.0}$ }&{\tiny $20.6^{+2.9}_{-2.9}$ }& {\tiny $17.40$ }&{\tiny 31.58}&{\tiny0.66}&{\tiny0.22$~^{a)}$}&{\tiny-0.1401}&{\tiny B.}&{\tiny (2)}\\
{\tiny 990712} & {\tiny 0.434} & {\tiny $93.0^{+15.0}_{-15.0}$ }&{\tiny $0.78^{+0.15}_{-0.15}$ }& {\tiny $18.64$ }&{\tiny 29.78}&{\tiny0.11}&{\tiny0.15$~^{b)}$}&{\tiny-0.0810}&{\tiny C.}&{\tiny (3)}\\
{\tiny 020124} & {\tiny 3.198} & {\tiny $339.6^{+44.0}_{-44.0}$ }&{\tiny $99.8^{+21.0}_{-21.0}$ }& {\tiny $18.29$ }&{\tiny 31.29}&{\tiny0.16}&{\tiny0.28$~^{a)}$}&{\tiny-0.5810}&{\tiny D.}&{\tiny (4) (5)}\\
{\tiny 020813} & {\tiny 1.25} & {\tiny $592.9^{+60.2}_{-60.2}$ }&{\tiny $95.9^{+3.6}_{-3.6}$ }& {\tiny $17.91$ }&{\tiny 30.99}&{\tiny0.36}&{\tiny0.12$~^{a)}$}&{\tiny-0.3302}&{\tiny E.}&{\tiny (6) (7) (8) (9) (10)}\\
{\tiny} & {\tiny } & {\tiny }&{\tiny }& {\tiny }& {\tiny }&{\tiny}&{\tiny}&{\tiny}&{\tiny }&{\tiny(11) (12) (13) (14)}\\
{\tiny 021004} & {\tiny 2.33} & {\tiny $310.5^{+84.0}_{-84.0}$ }&{\tiny $5.68^{+1.26}_{-1.26}$ }& {\tiny $16.40$ }&{\tiny 32.01}&{\tiny0.20}&{\tiny0.26$~^{a)}$}&{\tiny0.0078}&{\tiny F.}&{\tiny (15) (16) (17) (18) (19)}\\
{\tiny 021211} & {\tiny 1.01} & {\tiny $94.8^{+6.8}_{-6.8}$ }&{\tiny $2.23^{+0.23}_{-0.23}$ }& {\tiny $20.19$ }&{\tiny 29.81}&{\tiny0.09}&{\tiny$\sim 0$$~^{a)}$}&{\tiny-0.3027}&{\tiny G.}&{\tiny (20) (21) (22) (23) (24) (25)}\\
{\tiny 030328} & {\tiny 1.52} & {\tiny $327.3^{+22.6}_{-22.6}$ }&{\tiny $24.0^{+0.73}_{-0.73}$ }& {\tiny $18.79$ }&{\tiny 30.66}&{\tiny0.15}&{\tiny$\sim 0$$~^{a)}$}&{\tiny-0.2849}&{\tiny H.}&{\tiny (26) (27) (28) (29) (30) (31)}\\
{\tiny 030329} & {\tiny 0.168} & {\tiny $82.2^{+1.5}_{-1.5}$ }&{\tiny $1.07^{+0.02}_{-0.02}$ }& {\tiny $13.02$ }&{\tiny 31.29}&{\tiny0.08}&{\tiny0.54$~^{a)}$}&{\tiny-0.2038}&{\tiny I.}&{\tiny (32) (33)}\\
{\tiny 040924} & {\tiny 0.859} & {\tiny $75.9^{+2.5}_{-2.5}$ }&{\tiny $1.96^{+0.14}_{-0.14}$ }& {\tiny $20.12$ }&{\tiny 29.78}&{\tiny0.19}&{\tiny0.16$~^{a)}$}&{\tiny-0.3710}&{\tiny J.}&{\tiny (34) (35) (36)}\\
{\tiny 041006} & {\tiny 0.716} & {\tiny $82.2^{+3.1}_{-3.1}$ }&{\tiny $1.66^{+0.05}_{-0.05}$ }& {\tiny $18.68$ }&{\tiny 30.35}&{\tiny0.71}&{\tiny0.11$~^{a)}$}&{\tiny-0.3000}&{\tiny K.}&{\tiny (37) (38) (39)}\\
{\tiny 050416A} & {\tiny 0.6535} & {\tiny $24.8^{+3.8}_{-4.5}$ }&{\tiny $0.10^{+0.01}_{-0.01}$ }& {\tiny $20.87$ }&{\tiny 29.22}&{\tiny0.10}&{\tiny0.19$~^{a)}$}&{\tiny-0.2087}&{\tiny L.}&{\tiny (40)}\\
{\tiny 050525A} & {\tiny 0.606} & {\tiny $135.1^{+2.7}_{-2.7}$ }&{\tiny $2.41^{+0.04}_{-0.04}$ }& {\tiny $17.40$ }& {\tiny 30.60}&{\tiny0.32}&{\tiny0.26$~^{c)}$}&{\tiny-0.1656}&{\tiny M.}&{\tiny (41) (42) (43) (44) (45)}\\
{\tiny} & {\tiny } & {\tiny }&{\tiny }& {\tiny }& {\tiny }&{\tiny}&{\tiny}&{\tiny}&{\tiny }&{\tiny(46) (47) (48) (49) (50) (51)}\\
{\tiny 050820A} & {\tiny 2.612} & {\tiny $1326.7^{+343.4}_{-224.1}$ }&{\tiny $20.3^{+1.43}_{-1.43}$ }& {\tiny $17.51$ }&{\tiny 31.61}&{\tiny0.15}&{\tiny0.065$~^{d)}$}&{\tiny0.3595}&{\tiny N.}&{\tiny (52) (53)}\\
{\tiny 050922C} & {\tiny 2.198} & {\tiny $417.5^{+162.8}_{-85.7}$ }&{\tiny $4.56^{+0.15}_{-0.15}$ }& {\tiny $17.87$ }&{\tiny 31.36}&{\tiny0.34}&{\tiny0.07$~^{e)}$}&{\tiny0.1844}&{\tiny O.}&{\tiny (54)}\\ 
{\tiny 060115} & {\tiny 3.53} & {\tiny $280.9^{+86.1}_{-27.2}$ }&{\tiny $8.87^{+0.78}_{-0.78}$ }& {\tiny $19.97$ }&{\tiny 30.88}&{\tiny0.44}&{\tiny$\sim 0^*$}&{\tiny-0.1333}&{\tiny P.}&{\tiny (55) (56) (57)}\\
{\tiny 060908} & {\tiny 1.1884} & {\tiny $425.3^{+264.1}_{-130.3}$ }&{\tiny $9.48^{+0.38}_{-0.38}$ }& {\tiny $19.85$ }&{\tiny 30.08}&{\tiny0.12}&{\tiny0.09$~^{a)}$}&{\tiny0.0324}&{\tiny Q.}&{\tiny (58)}\\ 
{\tiny 061007} & {\tiny 1.261} & {\tiny $1065.4^{+81.4}_{-81.4}$ }&{\tiny $91.41^{+1.16}_{-1.16}$ }& {\tiny $17.44$ }&{\tiny 31.21}&{\tiny0.07}&{\tiny0.66$~^{f)}$}&{\tiny-0.0652}&{\tiny R.}&{\tiny (59)}\\ 
{\tiny 061121} & {\tiny 1.314} & {\tiny $1402.3^{+208.3}_{-166.6}$ }&{\tiny $28.21^{+0.41}_{-0.41}$ }& {\tiny $18.63$ }&{\tiny 30.72}&{\tiny0.15}&{\tiny0.28$~^{a)}$}&{\tiny0.3116}&{\tiny S.}&{\tiny (60) (61) (62) (63)}\\  
{\tiny} & {\tiny } & {\tiny }&{\tiny }& {\tiny }& {\tiny }&{\tiny}&{\tiny}&{\tiny}&{\tiny }&{\tiny(64) (65) (66)}\\
{\tiny 070612A} & {\tiny 0.617} & {\tiny $305.6^{+95.4}_{-95.4}$ }&{\tiny $3.0^{+0.17}_{-0.17}$ }& {\tiny $17.45$ }&{\tiny 30.67}&{\tiny0.17}&{\tiny0.46$^*$}&{\tiny$0.1409^*$}&{\tiny T.}&{\tiny (67) (68) (69) (70) (71)}\\ 
{\tiny 071010B} & {\tiny 0.947} & {\tiny $87.6^{+7.8}_{-1.4}$ }&{\tiny $2.35^{+0.05}_{-0.05}$ }& {\tiny $18.28$ }&{\tiny --}&{\tiny0.03}&{\tiny$0.18^*$}&{\tiny-0.3488}&{\tiny U.}&{\tiny (72) (73) (74) (75) (76)}\\
{\tiny 071020} & {\tiny 2.145} & {\tiny $1014.3^{+252.0}_{-167.0}$ }&{\tiny $14.06^{+0.61}_{-0.61}$ }& {\tiny $19.81$ }&{\tiny 30.61}&{\tiny0.21}&{\tiny0.28$~^{a)}$}&{\tiny-0.3235}&{\tiny V.}&{\tiny (77) (78) (79) (80) (81)}\\ 
{\tiny 080319B} & {\tiny 0.937} & {\tiny $1261.0^{+27.1}_{-25.2}$ }&{\tiny $120.36^{+1.49}_{-1.49}$ }& {\tiny $16.89$ }&{\tiny 31.03}&{\tiny0.04}&{\tiny0.07$~^{a)}$}&{\tiny-0.0523}&{\tiny W.}&{\tiny (82) (83)}\\ 
{\tiny 080411} & {\tiny 1.03} & {\tiny $525.8^{+71.1}_{-54.8}$ }&{\tiny $23.24^{+0.09}_{-0.09}$ }& {\tiny $16.54$ }&{\tiny 31.36}&{\tiny0.11}&{\tiny$0.28^*$}&{\tiny-0.0720}&{\tiny X.}&{\tiny (84) (85)}\\
{\tiny 080413A} & {\tiny 2.433} & {\tiny $432.6^{+449.7}_{-144.2}$ }&{\tiny $12.97^{+0.37}_{-0.37}$ }& {\tiny $18.43$ }&{\tiny 31.19}&{\tiny0.51}&{\tiny$\sim 0$$~^{d)}$}&{\tiny-0.0290}&{\tiny Y.}&{\tiny (86) (87) (88) (89) (90)}\\ 
{\tiny 080413B} & {\tiny 1.10} & {\tiny $140.7^{+27.3}_{-16.8}$ }&{\tiny $1.85^{+0.06}_{-0.06}$ }& {\tiny $17.39$ }&{\tiny 31.00}&{\tiny0.12}&{\tiny$\sim 0$$~^{a)}$}&{\tiny-0.0906}&{\tiny Z.}&{\tiny (91) (92)}\\ 
{\tiny 080605} & {\tiny 1.64} & {\tiny $768.2^{+198.0}_{-198.0}$ }&{\tiny $27.60^{+0.41}_{-0.41}$ }& {\tiny $20.20$ }&{\tiny 30.40}&{\tiny0.44}&{\tiny0.47$~^{g)}$}&{\tiny0.0551}&{\tiny AA.}&{\tiny (93) (94) (95) (96)}\\ 
{\tiny} & {\tiny } & {\tiny }&{\tiny }& {\tiny }& {\tiny }&{\tiny}&{\tiny}&{\tiny}&{\tiny }&{\tiny(97) (98) (99)}\\
{\tiny 080721} & {\tiny 2.591} & {\tiny $1747.0^{+241.3}_{-212.5}$ }&{\tiny $91.00^{+7.58}_{-7.58}$ }& {\tiny $18.24$ }&{\tiny 31.34}&{\tiny0.34}&{\tiny0.35$~^{d)}$}&{\tiny0.1506}&{\tiny AB.}&{\tiny (100) (101) (102)}\\ 
{\tiny 080804} & {\tiny 2.2045} & {\tiny $697.2^{+78.4}_{-78.4}$ }&{\tiny $19.80^{+1.01}_{-1.01}$ }& {\tiny $20.03$ }&{\tiny 30.43}&{\tiny0.05}&{\tiny0.06$~^{g)}$}&{\tiny0.0857}&{\tiny AC.}&{\tiny (103) (104) (105)}\\ 
{\tiny 080810} & {\tiny 3.355} & {\tiny $3957.3^{+801.7}_{-801.7}$ }&{\tiny $58.58^{+2.55}_{-2.55}$ }& {\tiny $17.25$ }&{\tiny 31.79}&{\tiny0.09}&{\tiny0.16$~^{d)}$}&{\tiny0.6021}&{\tiny AD.}&{\tiny (106) (107) (108)}\\ 
{\tiny 081007} & {\tiny 0.5295} & {\tiny $61.2^{+15.3}_{-15.3}$ }&{\tiny $0.12^{+0.01}_{-0.01}$ }& {\tiny $19.45$ }&{\tiny 29.56}&{\tiny0.05}&{\tiny$\sim 0$$~^{a)}$}&{\tiny0.1403}&{\tiny AE.}&{\tiny (109)}\\ 
{\tiny 081008} & {\tiny 1.9685} & {\tiny $493.3^{+107.7}_{-107.7}$ }&{\tiny $12.70^{+0.59}_{-0.59}$ }& {\tiny $17.62$ }&{\tiny 31.59}&{\tiny0.31}&{\tiny0.46$~^{a)}$}&{\tiny0.0326}&{\tiny AF.}&{\tiny (110) (111) (112) (113)}\\ 
{\tiny 081121} & {\tiny 2.512} & {\tiny $564.6^{+58.1}_{-58.1}$ }&{\tiny $19.57^{+1.43}_{-1.43}$ }& {\tiny $17.65$ }&{\tiny 31.52}&{\tiny0.17}&{\tiny$0.13^*$}&{\tiny-0.0034}&{\tiny AG.}&{\tiny (114) (115) (116)}\\ 
{\tiny 081222} & {\tiny 2.77} & {\tiny $538.1^{+36.1}_{-36.1}$ }&{\tiny $20.28^{+0.42}_{-0.42}$ }& {\tiny $18.45$ }&{\tiny 31.21}&{\tiny0.07}&{\tiny$\sim 0$$~^{a)}$}&{\tiny-0.321}&{\tiny AH.}&{\tiny (117) (118)(119) (120)}\\ 
{\tiny} & {\tiny } & {\tiny }&{\tiny }& {\tiny }& {\tiny }&{\tiny}&{\tiny}&{\tiny}&{\tiny }&{\tiny(121) (122)}\\
{\tiny 090102} & {\tiny 1.543} & {\tiny $1073.8^{+45.9}_{-45.9}$ }&{\tiny $15.96^{+0.70}_{-0.70}$ }& {\tiny $20.17$ }&{\tiny 30.22}&{\tiny0.15}&{\tiny0.45$~^{a)}$}&{\tiny0.3204}&{\tiny AI.}&{\tiny (123)}\\ 
{\tiny 090418A} & {\tiny 1.608} & {\tiny $1567.4^{+1444.8}_{-560.7}$ }&{\tiny $13.75^{+0.60}_{-0.60}$ }& {\tiny $20.03$ }&{\tiny 30.54}&{\tiny0.15}&{\tiny$0.67^*$}&{\tiny0.5174}&{\tiny AJ.}&{\tiny (124) (125) (126) (127) (128)}\\ 
{\tiny 090423} & {\tiny 8.26} & {\tiny $613.9^{+154.8}_{-154.8}$ }&{\tiny $8.51^{+0.58}_{-0.58}$ }& {\tiny $21.45^{*}$ }&{\tiny --}&{\tiny0.10}&{\tiny$\sim 0$$~^{a)}$}&{\tiny0.2153}&{\tiny AK.}&{\tiny (129) (130) (131) (132) (133)}\\  
{\tiny 090424} & {\tiny 0.544} & {\tiny $237.1^{+5.9}_{-5.9}$ }&{\tiny $3.54^{+0.35}_{-0.35}$ }& {\tiny $18.33$ }&{\tiny 30.15}&{\tiny0.08}&{\tiny0.50$~^{d)}$}&{\tiny-0.0055}&{\tiny AL.}&{\tiny (134) (135) (136) (137) (138)}\\ 
{\tiny} & {\tiny } & {\tiny }&{\tiny }& {\tiny }& {\tiny }&{\tiny}&{\tiny}&{\tiny}&{\tiny }&{\tiny(139) (140) (141) (142) (143) (144)}\\
{\tiny 090516A} & {\tiny 4.11} & {\tiny $725.9^{+135.2}_{-135.2}$ }&{\tiny $73.95^{+4.93}_{-4.93}$ }& {\tiny $20.88^{*}$ }&{\tiny --}&{\tiny0.17}&{\tiny$0.84^*$}&{\tiny-0.1854}&{\tiny AM.}&{\tiny (145) (146) (147) (148)}\\
{\tiny} & {\tiny } & {\tiny }&{\tiny }& {\tiny }& {\tiny }&{\tiny}&{\tiny}&{\tiny}&{\tiny }&{\tiny(149) (150)}\\  
{\tiny 090519} & {\tiny 3.85} & {\tiny $7848.2^{+4282.6}_{-4282.6}$ }&{\tiny $34.30^{+2.86}_{-2.86}$ }& {\tiny $23.14$ }&{\tiny 29.77}&{\tiny0.13}&{\tiny0.01$~^{g)}$}&{\tiny1.0167}&{\tiny AN.}&{\tiny (151) (152) (153) (154)}\\  
{\tiny} & {\tiny } & {\tiny }&{\tiny }& {\tiny }& {\tiny }&{\tiny}&{\tiny}&{\tiny}&{\tiny }&{\tiny(155) (156)}\\  
{\tiny 090618} & {\tiny 0.54} & {\tiny $226.2^{+5.6}_{-5.6}$ }&{\tiny $22.95^{+0.22}_{-0.22}$ }& {\tiny $16.65$ }&{\tiny 30.83}&{\tiny0.28}&{\tiny0.25$~^{a)}$}&{\tiny-0.4356}&{\tiny AO.}&{\tiny (157) (158) (159) (160) (161)}\\  
{\tiny} & {\tiny } & {\tiny }&{\tiny }& {\tiny }& {\tiny }&{\tiny}&{\tiny}&{\tiny}&{\tiny }&{\tiny(162) (163) (164) (165) (166)}\\ 
{\tiny} & {\tiny } & {\tiny }&{\tiny }& {\tiny }& {\tiny }&{\tiny}&{\tiny}&{\tiny}&{\tiny }&{\tiny(167) (168) (169) (170)}\\   
{\tiny 090812} & {\tiny 2.452} & {\tiny $2022.9^{+838.8}_{-524.7}$ }&{\tiny $43.26^{+1.49}_{-1.49}$ }& {\tiny $21.15$ }&{\tiny 30.18}&{\tiny0.08}&{\tiny0.41$~^{g)}$}&{\tiny0.3771}&{\tiny AP.}&{\tiny (171) (172) (173) (174)}\\  
{\tiny 091020} & {\tiny 1.71} & {\tiny $661.9^{+99.5}_{-99.5}$ }&{\tiny $6.81^{+0.18}_{-0.18}$ }& {\tiny $18.31$ }&{\tiny 31.01}&{\tiny0.06}&{\tiny$0.36^*$}&{\tiny0.2969}&{\tiny AQ.}&{\tiny (175) (176) (177) (178) (179)}\\ 
{\tiny} & {\tiny } & {\tiny }&{\tiny }& {\tiny }& {\tiny }&{\tiny}&{\tiny}&{\tiny}&{\tiny }&{\tiny (180) (181)}\\    
{\tiny } & {\tiny } & {\tiny  }&{\tiny }& {\tiny } & {\tiny }&{\tiny }&{\tiny }&{\tiny }&{\tiny  }&{\tiny }\\  
\end{tabular}
\end{center}
\label{tab_GRB}
\end{table*}

\begin{table*}
\addtocounter{table}{-1}
\caption{continued}
\begin{center}

\begin{tabular}{ccccccccccc}
\hline
\hline
{\tiny GRB }& {\tiny z} & {\tiny $E_{pi}$} & {\tiny $E_{iso}$}& {\tiny $R_{mag}$} &{\tiny $log_{10}(L_R^{rest})$}& {\tiny $A_V^{Gal}$}& {\tiny $A_V^{Host}$} & {\tiny dist}& {\tiny Ref. z}& {\tiny Ref. $R_{mag}$}\\
{\tiny }& {\tiny } & {\tiny (keV)} & {\tiny ($10^{52} erg$)}& {\tiny (t$_{\rm obs}$=2h)} &{\tiny (erg/s/Hz)}&{\tiny (mag)}&{\tiny (mag)}&{\tiny }& {\tiny}& {\tiny }\\
{\tiny (1)}& {\tiny (2)} & {\tiny (3)} & {\tiny (4)}& {\tiny (5)} &{\tiny (6)}&{\tiny (7) }& {\tiny (8)}& {\tiny (9)}& {\tiny (10)}& {\tiny (11)}\\
\hline
\hline
{\tiny } & {\tiny } & {\tiny  }&{\tiny }& {\tiny } & {\tiny }&{\tiny }&{\tiny }&{\tiny }&{\tiny  }&{\tiny }\\
{\tiny 091029} & {\tiny 2.752} & {\tiny $229.7^{+33.8}_{-94.5}$ }&{\tiny $9.46^{+0.39}_{-0.39}$ }& {\tiny $17.85$ }&{\tiny 31.44}&{\tiny0.06}&{\tiny$\sim 0$$~^{g)}$}&{\tiny-0.2348}&{\tiny AR.}&{\tiny (182) (183) (184) (185)}\\
{\tiny 091127} & {\tiny 0.49} & {\tiny $52.9^{+2.3}_{-2.3}$ }&{\tiny $1.39^{+0.05}_{-0.05}$ }& {\tiny $16.77$ }&{\tiny 30.60}&{\tiny0.13}&{\tiny0.11$~^{a)}$}&{\tiny-0.4520}&{\tiny AS.}&{\tiny (186) (187) (188) (189) (190) }\\ 
{\tiny} & {\tiny } & {\tiny }&{\tiny }& {\tiny }& {\tiny }&{\tiny}&{\tiny}&{\tiny}&{\tiny }&{\tiny (191) (192) (193) (194) (195) (196)}\\     
{\tiny 091208B} & {\tiny 1.063} & {\tiny $255.4^{+41.4}_{-40.0}$ }&{\tiny $2.50^{+0.15}_{-0.15}$ }& {\tiny $19.59$ }&{\tiny 30.24}&{\tiny0.18}&{\tiny0.40$~^{a)}$}&{\tiny0.1029}&{\tiny AT.}&{\tiny (197) (198) (199) (200) (201)}\\  
{\tiny} & {\tiny } & {\tiny }&{\tiny }& {\tiny }& {\tiny }&{\tiny}&{\tiny}&{\tiny}&{\tiny }&{\tiny (202) (203) (204)}\\     
{\tiny 100728B} & {\tiny 2.106} & {\tiny $341.2^{+68.5}_{-68.5}$ }&{\tiny $5.66^{+0.33}_{-0.33}$ }& {\tiny $20.11$ }&{\tiny 30.53}&{\tiny0.22}&{\tiny$0.35^*$}&{\tiny0.0495}&{\tiny AU.}&{\tiny (205) (206) (207) (208) (209)}\\ 
{\tiny 100814A} & {\tiny 1.44} & {\tiny $330.9^{+25.5}_{-25.5}$ }&{\tiny $11.65^{+0.26}_{-0.26}$ }& {\tiny $19.02$ }&{\tiny 30.59}&{\tiny0.07}&{\tiny$0.11^*$}&{\tiny-0.1220}&{\tiny AV.}&{\tiny (210)}\\ 
{\tiny 110205A} & {\tiny 2.22} & {\tiny $714.8^{+238.3}_{-238.3}$ }&{\tiny $57.82^{+5.78}_{-5.78}$ }& {\tiny $17.33$ } & {\tiny 31.49}&{\tiny0.05}&{\tiny0.20$~^{a)}$}&{\tiny-0.1382}&{\tiny AW.}&{\tiny (211) (212) (213)}\\ 
{\tiny 110213A} & {\tiny 1.46} & {\tiny $223.9^{+76.3}_{-64.0}$ }&{\tiny $8.57^{+0.58}_{-0.58}$ }& {\tiny $16.02$ } & {\tiny 32.14}&{\tiny1.06}&{\tiny$0.07^*$}&{\tiny-0.2244}&{\tiny AX.}&{\tiny (214) (215) (216) (217) (218) }\\ 
{\tiny} & {\tiny } & {\tiny }&{\tiny }& {\tiny }& {\tiny }&{\tiny}&{\tiny}&{\tiny}&{\tiny }&{\tiny (219) (220) (221) (222)}\\     
{\tiny 110422A} & {\tiny 1.77} & {\tiny $421.0^{+13.9}_{-13.9}$ }&{\tiny $66.04^{+1.61}_{-1.61}$ }& {\tiny $19.04$ } & {\tiny 30.89}&{\tiny0.09}&{\tiny$0.65^*$}&{\tiny-0.3972}&{\tiny AY.}&{\tiny (223) (224) (225) (226) (227) }\\ 
{\tiny} & {\tiny } & {\tiny }&{\tiny }& {\tiny }& {\tiny }&{\tiny}&{\tiny}&{\tiny}&{\tiny }&{\tiny (228) (229) (230) (231) (232)}\\     
{\tiny 110503A} & {\tiny 1.613} & {\tiny $572.3^{+52.3}_{-49.3}$ }&{\tiny $17.84^{+0.71}_{-0.71}$ }& {\tiny $18.89$ } & {\tiny 30.68}&{\tiny0.08}&{\tiny$0.15^*$}&{\tiny0.0227}&{\tiny AZ.}&{\tiny (233) (234) (235) (236) (237)}\\ 
{\tiny} & {\tiny } & {\tiny }&{\tiny }& {\tiny }& {\tiny }&{\tiny}&{\tiny}&{\tiny}&{\tiny }&{\tiny (238) (239) (240) (241)}\\     
{\tiny 110731A} & {\tiny 2.83} & {\tiny $1223.0^{+73.4}_{-73.4}$ }&{\tiny $39.84^{+0.66}_{-0.66}$ }& {\tiny $20.36$ } & {\tiny 30.72}&{\tiny0.57}&{\tiny0.24$~^{a)}$}&{\tiny0.1766}&{\tiny BA.}&{\tiny (242) (243) (244) (245)}\\ 
{\tiny} & {\tiny } & {\tiny }&{\tiny }& {\tiny }& {\tiny }&{\tiny}&{\tiny}&{\tiny}&{\tiny }&{\tiny (246) (247)}\\     
{\tiny 120326A} & {\tiny 1.798} & {\tiny $122.9^{+10.8}_{-10.8}$ }&{\tiny $4.08^{+0.47}_{-0.47}$ }& {\tiny $18.91$ } & {\tiny 30.90}&{\tiny0.17}&{\tiny$0.23^*$}&{\tiny-0.3222}&{\tiny BB.}&{\tiny (248) (249) (250) (251) (252)}\\ 
{\tiny} & {\tiny } & {\tiny }&{\tiny }& {\tiny }& {\tiny }&{\tiny}&{\tiny}&{\tiny}&{\tiny }&{\tiny (253) (254) (255) (256) (257)}\\     
{\tiny 120811C} & {\tiny 2.671} & {\tiny $204.0^{+19.6}_{-19.6}$ }&{\tiny $7.42^{+0.74}_{-0.74}$ }& {\tiny $19.42$ } & {\tiny 30.94}&{\tiny0.11}&{\tiny$0.53^*$}&{\tiny-0.2331}&{\tiny BC.}&{\tiny (258) (259) (260) (261)}\\ 
{\tiny 120907A} & {\tiny 0.97} & {\tiny $241.2^{+67.3}_{-67.3}$ }&{\tiny $0.27^{+0.04}_{-0.04}$ }& {\tiny $20.17$ } & {\tiny 29.88}&{\tiny0.31}&{\tiny$0.13^*$}&{\tiny0.5665}&{\tiny BD.}&{\tiny (262)}\\ 
{\tiny 121211A} & {\tiny 1.023} & {\tiny $202.8^{+32.0}_{-32.0}$ }&{\tiny $0.49^{+0.10}_{-0.10}$ }& {\tiny $19.65$ } & {\tiny 30.09}&{\tiny0.03}&{\tiny$0.37^*$}&{\tiny0.3606}&{\tiny BE.}&{\tiny (263) (264) (265) (266)}\\ 
{\tiny 130408A} & {\tiny 3.758} & {\tiny $1003.9^{+138.0}_{-138.0}$ }&{\tiny $21.39^{+3.72}_{-3.72}$ }& {\tiny $19.13$ } & {\tiny 31.30}&{\tiny0.84}&{\tiny$0.06^*$}&{\tiny0.2271}&{\tiny BF.}&{\tiny (267) (268) (269) (270)}\\ 
{\tiny 130420A} & {\tiny 1.297} & {\tiny $131.6^{+7.2}_{-7.2}$ }&{\tiny $4.61^{+0.19}_{-0.19}$ }& {\tiny $19.06$ } & {\tiny 30.51}&{\tiny0.04}&{\tiny$0.21^*$}&{\tiny-0.3196}&{\tiny BG.}&{\tiny (271) (272) (273) (274) (275)}\\
{\tiny} & {\tiny } & {\tiny }&{\tiny }& {\tiny }& {\tiny }&{\tiny}&{\tiny}&{\tiny}&{\tiny }&{\tiny (276) (277) (278) (279)}\\      
{\tiny 130427A} & {\tiny 0.3399} & {\tiny $1112.1^{+6.7}_{-6.7}$ }&{\tiny $81.93^{+0.79}_{-0.79}$ }& {\tiny $15.08$ } & {\tiny 30.96}&{\tiny0.07}&{\tiny$0.11^*$}&{\tiny-0.0226}&{\tiny BH.}&{\tiny (280) (281)}\\ 
{\tiny 130505A} & {\tiny 2.27} & {\tiny $2063.4^{+101.4}_{-101.4}$ }&{\tiny $220.26^{+10.49}_{-10.49}$ }& {\tiny $17.89$ }&{\tiny 31.42}&{\tiny0.13}&{\tiny$0.35^*$}&{\tiny0.0293}&{\tiny BI.}&{\tiny (282) (283) (284) (285)}\\
{\tiny} & {\tiny } & {\tiny }&{\tiny }& {\tiny }& {\tiny }&{\tiny}&{\tiny}&{\tiny}&{\tiny }&{\tiny (286) (287)}\\       
{\tiny 130610A} & {\tiny 2.092} & {\tiny $911.8^{+132.7}_{-132.7}$ }&{\tiny $9.59^{+0.38}_{-0.38}$ }& {\tiny $19.88$ } & {\tiny 30.53}&{\tiny0.07}&{\tiny$0.23^*$}&{\tiny0.3609}&{\tiny BJ.}&{\tiny (289) (290) (291) (292) (293)}\\ 
{\tiny} & {\tiny } & {\tiny }&{\tiny }& {\tiny }& {\tiny }&{\tiny}&{\tiny}&{\tiny}&{\tiny }&{\tiny (294) (295)}\\       
{\tiny 130701A} & {\tiny 1.155} & {\tiny $191.8^{+8.6}_{-8.6}$ }&{\tiny $3.52^{+0.08}_{-0.08}$ }& {\tiny $19.08$ } & {\tiny 30.43}&{\tiny0.28}&{\tiny$0.09^*$}&{\tiny-0.0967}&{\tiny BK.}&{\tiny (296) (297) (298)}\\ 
{\tiny 130831A} & {\tiny 0.4791} & {\tiny $81.4^{+5.9}_{-5.9}$ }&{\tiny $0.775^{+0.002}_{-0.002}$ }& {\tiny $17.23$ } & {\tiny 30.39}&{\tiny0.15}&{\tiny$0.07^*$}&{\tiny-0.1376}&{\tiny BL.}&{\tiny (299) (300) (301) (302) (303)}\\
{\tiny} & {\tiny } & {\tiny }&{\tiny }& {\tiny }& {\tiny }&{\tiny}&{\tiny}&{\tiny}&{\tiny }&{\tiny (304) (305) (306) (307) (308)}\\   
{\tiny} & {\tiny } & {\tiny }&{\tiny }& {\tiny }& {\tiny }&{\tiny}&{\tiny}&{\tiny}&{\tiny }&{\tiny (309) (310) (311) (312) (313)}\\   
{\tiny} & {\tiny } & {\tiny }&{\tiny }& {\tiny }& {\tiny }&{\tiny}&{\tiny}&{\tiny}&{\tiny }&{\tiny (314) (315)}\\                   
{\tiny 131030A} & {\tiny 1.293} & {\tiny $405.9^{+22.9}_{-22.9}$ }&{\tiny $28.91^{+2.89}_{-2.89}$ }& {\tiny $17.70$ } & {\tiny 31.04}&{\tiny0.19}&{\tiny$0.21^*$}&{\tiny-0.2323}&{\tiny BM.}&{\tiny (316) (317) (318) (319) (320)}\\ 
{\tiny} & {\tiny } & {\tiny }&{\tiny }& {\tiny }& {\tiny }&{\tiny}&{\tiny}&{\tiny}&{\tiny }&{\tiny (321) (322) (323) (324) (325)}\\
{\tiny} & {\tiny } & {\tiny }&{\tiny }& {\tiny }& {\tiny }&{\tiny}&{\tiny}&{\tiny}&{\tiny }&{\tiny (326) (327) (328) (329) (330)}\\              
{\tiny} & {\tiny } & {\tiny }&{\tiny }& {\tiny }& {\tiny }&{\tiny}&{\tiny}&{\tiny}&{\tiny }&{\tiny (331)}\\
{\tiny 140213A} & {\tiny 1.2076} & {\tiny $191.2^{+7.8}_{-7.8}$ }&{\tiny $10.75^{+1.07}_{-1.07}$}& {\tiny $18.32$ } & {\tiny 30.83}&{\tiny0.49}&{\tiny$0.06^*$}&{\tiny-0.3423}&{\tiny BN.}&{\tiny (332) (333)}\\ 
{\tiny 140419A} & {\tiny 3.956} & {\tiny $1452.1^{+416.3}_{-416.3}$ }&{\tiny $191.66^{+19.17}_{-19.17}$ }& {\tiny $18.17$ } & {\tiny 31.67}&{\tiny0.10}&{\tiny$0.47^*$}&{\tiny-0.0928}&{\tiny BO.}&{\tiny (334) (335) (336) (337) (338)}\\
{\tiny} & {\tiny } & {\tiny }&{\tiny }& {\tiny }& {\tiny }&{\tiny}&{\tiny}&{\tiny}&{\tiny }&{\tiny (339) (340) (341) (342) (343)}\\ 
{\tiny} & {\tiny } & {\tiny }&{\tiny }& {\tiny }& {\tiny }&{\tiny}&{\tiny}&{\tiny}&{\tiny }&{\tiny (344) (345) (346)}\\ 
{\tiny 140423A} & {\tiny 3.26} & {\tiny $532.5^{+38.3}_{-38.3}$ }&{\tiny $67.97^{+2.17}_{-2.17}$ }& {\tiny $19.08$ } & {\tiny 31.12}&{\tiny0.04}&{\tiny$0.32^*$}&{\tiny-0.3015}&{\tiny BP.}&{\tiny (347) (348) (349) (350) (351)}\\ 
{\tiny} & {\tiny } & {\tiny }&{\tiny }& {\tiny }& {\tiny }&{\tiny}&{\tiny}&{\tiny}&{\tiny }&{\tiny (352) (353) (354) (355) (356)}\\ 
{\tiny} & {\tiny } & {\tiny }&{\tiny }& {\tiny }& {\tiny }&{\tiny}&{\tiny}&{\tiny}&{\tiny }&{\tiny (356) (357) (358) (359) (360)}\\
{\tiny} & {\tiny } & {\tiny }&{\tiny }& {\tiny }& {\tiny }&{\tiny}&{\tiny}&{\tiny}&{\tiny }&{\tiny (361) (362)}\\  
{\tiny 140506A} & {\tiny 0.889} & {\tiny $373.2^{+61.5}_{-61.5}$ }&{\tiny $1.28^{+0.14}_{-0.14}$ }& {\tiny $19.60$ } & {\tiny 30.11}&{\tiny0.31}&{\tiny$0.32^*$}&{\tiny0.4140}&{\tiny BQ.}&{\tiny (363)}\\ 
{\tiny 140512A} & {\tiny 0.725} & {\tiny $1177.8^{+121.3}_{-121.3}$ }&{\tiny $9.44^{+0.20}_{-0.20}$ }& {\tiny $18.44$ } & {\tiny 30.41}&{\tiny0.53}&{\tiny$0.10^*$}&{\tiny0.4755}&{\tiny BR.}&{\tiny (364) (365) (366) (367)}\\ 
{\tiny 140703A} & {\tiny 3.14} & {\tiny $861.3^{+148.3}_{-148.3}$ }&{\tiny $20.96^{+1.61}_{-1.61}$ }& {\tiny $19.10$ } & {\tiny 31.28}&{\tiny0.34}&{\tiny$0.27^*$}&{\tiny0.1650}&{\tiny BS.}&{\tiny (368)}\\
{\tiny } & {\tiny } & {\tiny  }&{\tiny }& {\tiny } & {\tiny }&{\tiny }&{\tiny }&{\tiny }&{\tiny  }&{\tiny }\\
\hline

\end{tabular}
\end{center}
\end{table*}

\begin{table*}
\caption{The sample parameters of GRBs without a redshift. (1) GRB name, (2) and (3) minimum and maximum vertical distance to the best fit \eep\, respectively. (4) and (5) redshift at the minimum and maximum distance to the best fit \eer. For GRB 051008 we used the strong constraint on redshift determined by \cite{Volnova2014f}. (6) and (7) rest frame peak energy of the gamma-ray spectral energy distribution at the minimum and maximum distance to the best fit \eer, respectively. (8) and (9) isotropic gamma-ray energy at the minimum and maximum distance to the best fit \eer, respectively. (10) $R_{mag}$ is the apparent R magnitude 2 hours after the burst not corrected from the galactic and host extinctions ($\dagger$ : calibrated with the V-band magnitude).}
\begin{center}
\hspace*{0.7cm}
\begin{tabular}{ccccccccccc}
\hline
\hline
{\tiny GRB }&{\tiny dmin} &{\tiny dmax} &{\tiny $z_{dmin}$}&{\tiny $z_{dmax}$}& {\tiny $E^{dmin}_{pi}$} & {\tiny $E^{dmax}_{pi}$}&{\tiny $E^{dmin}_{iso}$}&{\tiny $E^{dmax}_{iso}$}& {\tiny $R_{mag}$} & {\tiny Ref. $R_{mag}$}\\
{\tiny }& {\tiny } & {\tiny } & {\tiny }& {\tiny } &{\tiny (keV)}&{\tiny (keV) }& {\tiny ($10^{52} erg$)}& {\tiny ($10^{52} erg$)}& {\tiny (t$_{\rm obs}$=2h)} & {\tiny }\\
{\tiny (1)}& {\tiny (2)} & {\tiny (3)} & {\tiny (4)}& {\tiny (5)} &{\tiny (6)}&{\tiny (7) }& {\tiny (8)}& {\tiny (9)}& {\tiny (10)} & {\tiny (11)}\\
\hline
\hline
{\tiny } & {\tiny } & {\tiny  }&{\tiny }& {\tiny }&{\tiny }&{\tiny }&{\tiny }&{\tiny }&{\tiny } & {\tiny }\\
{\tiny 051008} & {\tiny 0.5051} & {\tiny 0.5056 }&{\tiny 2.84 }& {\tiny 2.62 }&{\tiny 3317.8}&{\tiny 3131.3}&{\tiny 64.33}&{\tiny 0.95}&{\tiny $23.30^{*}$}& {\tiny (372) (373) (374)}\\
{\tiny 060105} & {\tiny -0.0496} & {\tiny 0.3812 }&{\tiny 3.40 }& {\tiny 0.3399 }&{\tiny 1437.3}&{\tiny 438.1}&{\tiny 154.15}&{\tiny 57.09}&{\tiny $20.20^{*}$} & {\tiny (375) (376) (377)}\\
{\tiny 060117} & {\tiny -0.2439} & {\tiny 0.2050 }&{\tiny 3.72 }& {\tiny 0.3399 }&{\tiny 867.9}&{\tiny 246.5}&{\tiny 137.66}&{\tiny 2.04}&{\tiny $20.15^{*}$} & {\tiny (378)}\\
{\tiny 060904} & {\tiny -0.0707} & {\tiny 0.3652 }&{\tiny 3.56 }& {\tiny 0.3399 }&{\tiny 742.6}&{\tiny 218.4}&{\tiny 45.81}&{\tiny 1.46}&{\tiny $21.51^{*}$} & {\tiny (379) (380) (381)}\\
{\tiny 090813} & {\tiny 0.0458} & {\tiny 0.4526 }&{\tiny 3.15 }& {\tiny 0.3399 }&{\tiny 394.3}&{\tiny 127.3}&{\tiny 7.68}&{\tiny 0.55}&{\tiny $20.21^{*}$} & {\tiny (382) (383)}\\
{\tiny 091221} & {\tiny 0.0378} & {\tiny 0.4557 }&{\tiny 3.24 }& {\tiny 0.3399 }&{\tiny 878.6}&{\tiny 277.4}&{\tiny 38.96}&{\tiny 0.13}&{\tiny 21.44} & {\tiny (384) (385) (386)}\\
{\tiny 100413A} & {\tiny 0.4256} & {\tiny 0.8369 }&{\tiny 3.15 }& {\tiny 0.3399 }&{\tiny 1232.1}&{\tiny 397.4}&{\tiny 12.97}&{\tiny 0.59}&{\tiny 23.20} & {\tiny (387) (388) (389) (390)}\\
{\tiny 101011A} & {\tiny 0.2738} & {\tiny 0.6830 }&{\tiny 3.07 }& {\tiny 0.3399 }&{\tiny 1816.9}&{\tiny 597.6}&{\tiny 56.02}&{\tiny 0.21}&{\tiny $20.80^{*\dagger}$} & {\tiny (391) (392) (393) (394)}\\
{\tiny 140102A} & {\tiny -0.0407} & {\tiny 0.3890 }&{\tiny 3.40 }& {\tiny 0.3399 }&{\tiny 817.5}&{\tiny 249.2}&{\tiny 48.35}&{\tiny 0.64}&{\tiny 19.34} & {\tiny (395) (396) (397) (398) (399)}\\
{\tiny } & {\tiny } & {\tiny  }&{\tiny  }& {\tiny }&{\tiny }&{\tiny }&{\tiny }&{\tiny }&{\tiny } & {\tiny (400) (401) (402) (403)}\\
{\tiny 140626A} & {\tiny 0.0796} & {\tiny 0.5119 }&{\tiny 3.47 }& {\tiny 0.3399 }&{\tiny 200.0}&{\tiny 59.9}&{\tiny 1.71}&{\tiny 0.02}&{\tiny $20.84^{*}$} & {\tiny (404) (405)}\\
{\tiny 140709B} & {\tiny 0.4601} & {\tiny 0.8719 }&{\tiny 3.15 }& {\tiny 0.3399 }&{\tiny 2201.7}&{\tiny 710.1}&{\tiny 35.02}&{\tiny 0.57}&{\tiny $20.50^{*}$}& {\tiny (406) (407) (408)}\\
{\tiny 140713A} & {\tiny 0.3737} & {\tiny 0.8044 }&{\tiny 3.48 }& {\tiny 0.3399 }&{\tiny 429.7}&{\tiny 128.6}&{\tiny 2.03}&{\tiny 0.03}&{\tiny $23.90^{*}$} & {\tiny (409) (410)}\\
{\tiny 141005A} & {\tiny 0.2597} & {\tiny 0.6998 }&{\tiny 3.64 }& {\tiny 0.3399 }&{\tiny 551.7}&{\tiny 159.4}&{\tiny 5.62}&{\tiny 0.06}&{\tiny $21.44^{*}$} & {\tiny (411) (412)}\\
{\tiny 141017A} & {\tiny -0.0806} & {\tiny 0.3469 }&{\tiny 3.39 }& {\tiny 0.3399 }&{\tiny 426.3}&{\tiny 130.0}&{\tiny 15.95}&{\tiny 0.22}&{\tiny $22.40^{*}$} & {\tiny (413) (414)}\\
{\tiny } & {\tiny } & {\tiny  }&{\tiny }& {\tiny }&{\tiny }&{\tiny }&{\tiny }&{\tiny }&{\tiny } & {\tiny }\\
\hline
\hline
\end{tabular}
\end{center}
\label{tab_GRB_sansz}
\end{table*}

\newpage

\begin{table*}
\caption{Table of references}
\begin{tabular}{c}
\hline
\hline
{}\\
{\bf References for the afterglow optical light curve}\\
{}\\
\hline
{(1) \cite{Kulkarni1999}, (2) \cite{Harrison1999}, (3) \cite{Sahu2000}, (4) \cite{Berger2002}}\\
{(5) \cite{Hjorth2003}, (6) \cite{Urata2003}, (7) \cite{Laursen2003}, (8) \cite{Li2002a}}\\
{(9) \cite{Li2002b}, (10) \cite{Williams2002}, (11) \cite{Beskin2002}, (12) \cite{Kiziloglu2002}}\\
{(13) \cite{Fiore2002}, (14) \cite{Gorosabel2002}, (15) \cite{Pandey2003a}, (16) \cite{Holland2003}}\\
{(17) \cite{Bersier2003}, (18) \cite{deUgartePostigo2005}, (19) \cite{Garnavich2002}, (20) \cite{Fox2003}}\\
{(21) \cite{Li2003}, (22) \cite{Holland2004}, (23) \cite{Pandey2003b}, (24) \cite{Testa2003}}\\
{(25) \cite{Price2002a}, (26) \cite{Maiorano2006}, (27) \cite{Burenin2003}, (28) \cite{Rumyantsev2003}}\\
{(29) \cite{Fugazza2003}, (30) \cite{Ibrahimov2003}, (31) \cite{Martini2003}, (32) \cite{Gorosabel2006}}\\
{(33) \cite{Lipkin2004}, (34) \cite{Huang2005}, (35) \cite{Soderberg2006}, (36) \cite{Wiersema2008}}\\
{(37) \cite{Stanek2005}, (38)\cite{Urata2007}, (39) \cite{Misra2005}, (40) \cite{Kann2010}}\\ 
{(41) \cite{Resmi2012}, (42) \cite{Klotz2005}, (43) \cite{Rykoff2005}, (44) \cite{Torii2005}}\\
{(45) \cite{Malesani2005}, (46) \cite{Mirabal2005}, (47) \cite{Cobb2005}, (48) \cite{Durig2005}}\\
{(49) \cite{Haislip2005}, (50) \cite{Yanagisawa2005}, (51) \cite{Kann2010} ,(52) \cite{Cenko2006a}}\\
{(53) \cite{Kann2010}, (54)\cite{Kann2010}(55) \cite{Yanagisawa2006}, (56) \cite{Distefano2006}}\\
{(57) \cite{Nysewander2006a}, (58) \cite{Covino2010}, (59) \cite{Mundell2007}(60) \cite{Kann2010}}\\
{(61) \cite{Halpern2006a}, (62) \cite{Halpern2006b}, (63) \cite{Halpern2006c}}\\
{(64) \cite{Yost2006}, (65) \cite{Melandri2006}, (66) \cite{Uemura2006}, (67) \cite{Updike2007a}}\\
{(68) \cite{Updike2007b}, (69) \cite{Updike2007c}, (70) \cite{Mirabal2007}, (71) \cite{Malesani2007}}\\
{(72) \cite{Lee2010}, (73) \cite{Oksanen2008}, (74) \cite{Wang2008}, (75) \cite{Kann2007}}\\
{(76) \cite{Kocevski2007}, (77) \cite{Kann2010}, (78) \cite{Schaefer2007b}, (79) \cite{Ishimura2007}}\\
{(80) \cite{Hentunen2007}, (81) \cite{Im2007}, (82) \cite{Bloom2009}, (83) \cite{Wozniak2009}}\\
{(84) \cite{Kruehler2008a}, (85) \cite{Thoene2008a}, (86) \cite{Kann2010}, (87) \cite{Rykoff2008}}\\
{(88) \cite{Gomboc2008a}, (89) \cite{Cobb2008a}, (90) \cite{Klotz2008}, (91) \cite{Brennan2008}}\\
{(92) \cite{Gomboc2008b}, (93) \cite{Zafar2012}, (94) \cite{Kuvshinov2008}, (95) \cite{Rumyantsev2008}}\\
{(96) \cite{Gomboc2008c}, (97) \cite{Kann2008}, (98) \cite{Jakobsson2008a}, (99) \cite{Yoshida2008a}}\\
{(100) \cite{Starling2009}, (101) \cite{Chen2008}, (102) \cite{Huang2008}, (103) \cite{Rujopakarn2008}}\\
{(104) \cite{Kruehler2008b}, (105) \cite{Guidorzi2008}, (106) \cite{Page2009}, (107) \cite{Kann2010}}\\
{(108) \cite{Yoshida2008b}, (109) \cite{Jin2013}, (110) \cite{Yuan2010}, (111) \cite{Cobb2008b}}\\
{(112) \cite{Kann2010}, (113) \cite{Cucchiara2008b}, (114) \cite{Yuan2008}, (115) \cite{Loew2008}}\\
{(116) \cite{Cobb2008c}, (117) \cite{Covino2013}, (118) \cite{Cwiok2008}, (119) \cite{Sonoda2008}}\\
{(120) \cite{Kuroda2008}, (121) \cite{Melandri2008}, (122) \cite{Roy2008}, (123) \cite{Gendre2010}}\\
{(124) \cite{Yuan2009a}, (125) \cite{Henden2009a}, (126) \cite{Chornock2009a}, (127) \cite{Bikmaev2009}}\\
{(128) \cite{Pavlenko2009}, (129) \cite{Henden2009b}, (130) \cite{Cenko2009a}, (131) \cite{Melandri2009a}}\\
{(132) \cite{Yoshida2009a}, (133) \cite{Olivares2009a}, (134) \cite{Kann2010}, (135) \cite{Jin2013}}\\
{(136) \cite{Urata2009}, (137) \cite{Yuan2009b}, (138) \cite{Mao2009}, (139) \cite{Roy2009}}\\ 
{(140) \cite{Rumyantsev2009a}, (141) \cite{Oksanen2009}, (142) \cite{Gorosabel2009a}, (143) \cite{Olivares2009b}}\\ 
{(144) \cite{Im2009a}(145) \cite{Guidorzi2009}, (146) \cite{Christie2009a}, (147) \cite{Vaalsta2009a}}\\
{(148) \cite{Gorosabel2009b}, (149) \cite{deUgartePostigo2009a}, (150) \cite{Rossi2009b}, (151) \cite{Klotz2009a}}\\ 
{(152) \cite{Klotz2009b}, (153) \cite{Jelinek2009}, (154) \cite{Rossi2009a},(155) \cite{Thoene2009a}}\\
{(156) \cite{Thoene2009b}, (157) \cite{Rujopakarn2009}, (158) \cite{Li2009}, (159) \cite{Cenko2009b}}\\
{(160) \cite{Im2009b}, (161) \cite{Updike2009a}, (162) \cite{Melandri2009b}, (163) \cite{Klunko2009}}\\
{(164) \cite{Anumapa2009}, (165) \cite{Rumyantsev2009b}, (166) \cite{Cano2009a}, (167) \cite{Fernandez-soto2009}}\\
{(168) \cite{Fatkhullin2009}, (169) \cite{Galeev2009}, (170) \cite{Khamitov2009}, (171) \cite{Wren2009}}\\
{(172) \cite{Smith2009a}, (173) \cite{Haislip2009a}, (174) \cite{Updike2009b}, (175) \cite{Gorbovskoy2009}}\\
{(176) \cite{Kann2009a}, (177) \cite{Kann2009b}, (178) \cite{Xu2009a}, (179) \cite{Perley2009a}}\\
{(180) \cite{Perley2009b}, (181) \cite{Xin2009a}, (182) \cite{Filgas2009a}, (183) \cite{Chornock2009c}}\\
{(184) \cite{Christie2009b}, (185) \cite{deUgartePostigo2009c}, (186) \cite{Troja2012}, (187) \cite{Filgas2011}}\\
{(188) \cite{Xu2009b}, (189) \cite{Klotz2009c}, (190) \cite{Cobb2010}, (191) \cite{Kinugasa2009a}}\\
{(192) \cite{Vaalsta2009b}, (193) \cite{Andreev2009a}, (194) \cite{Haislip2009b}, (195) \cite{Haislip2009c}}\\
{(196) \cite{Thoene2009c}, (197) \cite{Nakajima2009}, (198) \cite{Cano2009b}, (199) \cite{Yoshida2009b}}\\
{(200)\cite{Xin2009b}, (201) \cite{Kinugasa2009b}, (202) \cite{Andreev2009b}, (203) \cite{Xu2009c}}\\
{(204) \cite{Updike2009c}, (205) \cite{Perley2010a}, (206) \cite{Perley2010b}, (207) \cite{Elenin2010}}\\
{(208) \cite{Decia2010}, (209) \cite{Olivares2010}, (210) \cite{Nardini2014}, (211) \cite{Zheng2012}}\\
{(212) \cite{Gendre2012}, (213) \cite{Cucchiara2011b}, (214) \cite{Cucchiara2011b}, (215) \cite{Rujopakarn2011}}\\
{(216) \cite{Wren2011}, (217) \cite{Ukwatta2011}, (218) \cite{Hentunen2011a}, (219) \cite{Kuroda2011a}}\\
{(220) \cite{Volnova2011}, (221) \cite{Zhao2011}, (222) \cite{Hentunen2011b}, (223) \cite{Elunko2011}}\\
{(224) \cite{Moskvitin2011a}, (225) \cite{Rumyantsev2011a}, (226) \cite{Hentunen2011c}, (227) \cite{Xu2011a}}\\
{(228) \cite{Melandri2011}, (229) \cite{Jeon2011}, (230) \cite{Kuroda2011b}, (231) \cite{Xu2011b}}\\
{(232) \cite{Rumyantsev2011b}, (233) \cite{Klotz2011}, (234) \cite{Kann2011}, (235) \cite{Leloudas2011}}\\
\end{tabular}
\end{table*}

\begin{table*}
\addtocounter{table}{-1}
\caption{continued}
\begin{tabular}{c}
{(236) \cite{Tasselli2011}, (237) \cite{Broens2011}, (238) \cite{Davanzo2011}, (239) \cite{Pavlenko2011}}\\
{(240) \cite{Im2011}, (241) \cite{Updike2011}, (242) \cite{Bersier2011}, (243) \cite{Kuroda2011c}}\\
{(244) \cite{Moskvitin2011b}, (245) \cite{Malesani2011b}, (246) \cite{Tanvir2011}, (247) \cite{Ackermann2013}}\\
{(248) \cite{Melandri2014}, (249) \cite{Klotz2012}, (250) \cite{Hentunen2012}, (251) \cite{Jang2012}}\\
{(252) \cite{Zhao2012}, (253) \cite{Xin2012a}, (254) \cite{Xin2012b}, (255) \cite{Soulier2012}}\\
{(256) \cite{Quadri2012}, (257) \cite{Sahu2012}, (258) \cite{Elenin2012}, (259) \cite{Litvinenko2012}}\\
{(260) \cite{Galeev2012a}, (261) \cite{Galeev2012b}, (262) \cite{Nevski2012}, (263) \cite{Japelj2012}}\\
{(264) \cite{Kuroda2012}, (265) \cite{Butler2012a}, (266) \cite{Butler2012b}, (267) \cite{Melandri2013a}}\\
{(268) \cite{Sudilovsky2013}, (269) \cite{Trotter2013a}, (270) \cite{Dereli2013}, (271) \cite{Guver2013}}\\
{(272) \cite{Elenin2013a}, (273) \cite{Guidorzi2013a}, (274) \cite{Trotter2013b}, (275) \cite{Hentunen2013a}}\\
{(276) \cite{Butler2013a}, (277) \cite{Watson2013a}, (278) \cite{Butler2013b}, (279) \cite{Zhao2013}}\\
{(280) \cite{Vestrand2014}, (281) \cite{Maselli2014}, (282) \cite{Hentunen2013b}, (283) \cite{Kuroda2013}}\\
{(284) \cite{Xu2013a}, (285) \cite{Xin2013a}, (286) \cite{Kann2013}, (287) \cite{Krugly2013}}\\
{(288) \cite{Watson2013b}, (289) \cite{Klotz2013}, (290) \cite{Melandri2013b}, (291) \cite{Hentunen2013c}}\\
{(292) \cite{Elenin2013b}, (293) \cite{Cenko2013a}, (294) \cite{Im2013a}, (295) \cite{Rumyantsev2013}}\\
{(296) \cite{Leloudas2013}, (297) \cite{Cenko2013b}, (298) \cite{Xu2013b}, (299) \cite{Cano2014a}}\\
{(300) \cite{Yoshii2013}, (301) \cite{Guidorzi2013b}, (302) \cite{Volnova2013a}, (303) \cite{Xu2013c}}\\
{(304) \cite{Izzo2013}, (305) \cite{Xin2013b}, (306) \cite{Sonbas2013}, (307) \cite{Volnova2013b}}\\
{(308) \cite{Volnova2013c}, (309) \cite{Hentunen2013d}, (310) \cite{Leonini2013}, (311) \cite{Khorunzhev2013}}\\
{(312) \cite{Masi2013}, (313) \cite{Butler2013c}, (314) \cite{Watson2013c}, (315) \cite{Lee2013}}\\
{(316) \cite{Moskvitin2013}, (317) \cite{Xu2013d}, (318) \cite{Virgili2013}, (319) \cite{Gorbovskoy2013}}\\
{(320) \cite{Terron2013}, (321) \cite{Im2013b}, (322) \cite{Perley2013}, (323) \cite{Littlejohns2013a}}\\
{(324) \cite{Littlejohns2013b}, (325) \cite{Littlejohns2013c}, (326) \cite{Littlejohns2013d}, (327) \cite{Littlejohns2013e}}\\
{(328) \cite{Hentunen2013e}, (329) \cite{Tanigawa2013}, (330) \cite{Xin2013c}, (331) \cite{Pandey2013}}\\
{(332) \cite{Trotter2014}, (333) \cite{Elliott2014}, (334) \cite{Guver2014}, (335) \cite{Zheng2014}}\\
{(336) \cite{Butler2014a}, (337) \cite{Littlejohns2014a}, (338) \cite{Littlejohns2014b}, (339) \cite{Cenko2014a}}\\
{(340) \cite{Hentunen2014}, (341) \cite{Choi2014b}, (342) \cite{Kuroda2014a}, (343) \cite{Kuroda2014b}}\\
{(344) \cite{Volnova2014a}, (345) \cite{Pandey2014b}, (346) \cite{Xu2014b}, (347) \cite{Ferrante2014}}\\
{(348) \cite{Elenin2014}, (349) \cite{Cenko2014b}, (350)\cite{Harbeck2014a}, (351) \cite{Harbeck2014b}}\\
{(352) \cite{Akitaya2014}, (353) \cite{Kuroda2014c}, (354) \cite{Fujiwara2014}, (355) \cite{Volnova2014b}}\\
{(356) \cite{Sahu2014}, (357) \cite{Cano2014b}, (358) \cite{Volnova2014c}, (359) \cite{Bikmaev2014}}\\
{(360) \cite{Littlejohns2014c}, (361) \cite{Butler2014b}, (362) \cite{Volnova2014d}, (363) \cite{Fynbo2014}}\\
{(364) \cite{Gorbovskoy2014},(365) \cite{Klunko2014}, (366) \cite{deUgartePostigo2014a}, (367) \cite{Graham2014}}\\
{(368) \cite{Ciabattari2014}, (369) \cite{Perley2014b}, (370) \cite{Butler2014c}, (371) \cite{Volnova2014e} }\\
{(372) \cite{Rumyantsev2005}, (373) \cite{Pozanenko2005}, (374) \cite{Volnova2014f}, (375) \cite{Urata2006a} }\\
{(376) \cite{Zimmerman2006}, (377) \cite{Kann2006}, (378) \cite{Nysewander2006b}, (379) \cite{Klotz2006} }\\
{(380) \cite{Cenko2006b}, (381) \cite{Urata2006b}, (382) \cite{Smith2009b}, (383) \cite{Volnova2009} }\\
{(384) \cite{Zheng2009}, (385) \cite{Filgas2009b}, (386) \cite{Haislip2009d}, (387) \cite{Ivanov2010} }\\
{(388) \cite{Guidorzi2010}, (389) \cite{Xin2010}, (390) \cite{Filgas2010}, (391) \cite{Laas-Bourez2010} }\\
{(392) \cite{Nardini2010}, (393) \cite{dePasquale2010}, (394) \cite{Tello2010}, (395) \cite{Yoshii2014} }\\
{(396) \cite{Guziy2014}, (397) \cite{Chen2014}, (398) \cite{Xu2014a}, (399) \cite{Pandey2014a} }\\
{(400) \cite{Tanga2014a}, (401) \cite{Perley2014a}, (402) \cite{Choi2014a}, (403) \cite{Volnova2014g} }\\
{(404) \cite{Klotz2014a}, (405) \cite{Tanga2014b}, (406) \cite{Ivanov2014}, (407) \cite{Volnova2014i} }\\
{(408) \cite{Volnova2014j}, (409) \cite{Xin2014}, (410) \cite{Cano2014b}, (411) \cite{Perley2014c} }\\
{(412) \cite{Schmidl2014}, (413) \cite{Klotz2014b}, (414) \cite{Kann2014}}\\
\end{tabular}
\end{table*}

\begin{table*}
\addtocounter{table}{-1}
\caption{continued}
\begin{tabular}{c}
\hline
\hline
{}\\
{\bf References for the redshift}\\
{}\\
\hline
{A. \cite{Kulkarni1999}, B. \cite{Vreeswijk1999}, C. \cite{Galama1999}, D. \cite{Hjorth2003}}\\
{E. \cite{Price2002b}, F. \cite{Castro-Tirado2010}, G. \cite{DellaValle2003}, H. \cite{Rol2003}}\\
{I. \cite{Thoene2007}, J. \cite{Wiersema2008}, K. \cite{Price2004}, L. \cite{Cenko2005}}\\
{M. \cite{Foley2005}, N. \cite{Prochaska2005}, O. \cite{Jakobsson2005}, P. \cite{Piranomonte2006}}\\
{Q. \cite{Fynbo2009}, R. \cite{Osip2006}, S. \cite{Bloom2006}, T. \cite{Cenko2007a}}\\
{U. \cite{Cenko2007b}, V. \cite{Jakobsson2007}, W. \cite{Vreeswijk2008a}, X. \cite{Thoene2008a}}\\
{Y. \cite{Thoene2008b}, Z. \cite{Vreeswijk2008b}, AA. \cite{Jakobsson2008a}, AB. \cite{Jakobsson2008b}}\\
{AC. \cite{Cucchiara2008a}, AD. \cite{Prochaska2008}, AE. \cite{Berger2008a}, AF. \cite{Cucchiara2008c}}\\
{AG. \cite{Berger2008b}, AH. \cite{Cucchiara2008d}, AI. \cite{Cucchiara2009a}, AJ. \cite{Chornock2009b}}\\
{AK. \cite{Tanvir2009}, AL. \cite{Chornock2009d}, AM. \cite{deUgartePostigo2009b}, AN. \cite{Thoene2009b}}\\
{AO. \cite{Cenko2009c}, AP. \cite{deUgartePostigo2009d}, AQ. \cite{Xu2009a}, AR. \cite{Chornock2009c}}\\
{AS. \cite{Cucchiara2009b}, AT. \cite{Wiersema2009a}, AU. \cite{Flores2010}, AV. \cite{Omeara2010}}\\
{AW. \cite{Cenko2011a}, AX. \cite{Milne2011}, AY. \cite{Malesani2011a}, AZ. \cite{deUgartePostigo2011a}}\\
{BA. \cite{Tanvir2011}, BB. \cite{Tello2012}, BC. \cite{Thoene2012}, BD. \cite{SanchezRamirez2012}}\\
{BE. \cite{Perley2012}, BF. \cite{Hjorth2013}, BG. \cite{deUgartepostigo2013}, BH. \cite{Flores2013}}\\
{BI. \cite{Tanvir2013}, BJ \cite{Smette2013}, BK. \cite{Xu2013b}, BL. \cite{Cucchiara2013}}\\
{BM. \cite{Xu2013d}, BN. \cite{Schulze2014}, BO. \cite{Tanvir2014a}, BP. \cite{Tanvir2014b}}\\
{BQ. \cite{Fynbo2014}, BR. \cite{deUgartePostigo2014b}, BS. \cite{CastroTirado2014}}\\
\hline
\hline
{}\\
{\bf References for the host extinction}\\
{}\\
\hline
{a) \cite{Japelj2014}, b) \cite{Christensen2004}, c) \cite{Schady2007a}, d) \cite{Kann2010}}\\
{e) \cite{Schady2010}, f) \cite{Schady2007b}, g) \cite{Greiner2011}}\\
\hline
\end{tabular}
\end{table*}


\begin{acknowledgements}
We gratefully acknowledge financial support from the OCEVU LabEx, France. We are also grateful to Yves Zolnierowski for insightfull discussion.
 \end{acknowledgements}


\bibliographystyle{aa} 
\bibliography{biblio-DT2014} 

\begin{thebibliography}{520}
\expandafter\ifx\csname natexlab\endcsname\relax\def\natexlab#1{#1}\fi

\bibitem[{{Ackermann} {et~al.}(2013){Ackermann}, {Ajello}, {Asano}, {Baldini},
  {Barbiellini}, {Baring}, {Bastieri}, {Bellazzini}, {Blandford}, {Bonamente},
  {Borgland}, {Bottacini}, {Bregeon}, {Brigida}, {Bruel}, {Buehler}, {Buson},
  {Caliandro}, {Cameron}, {Caraveo}, {Cecchi}, {Charles}, {Chaves},
  {Chekhtman}, {Chiang}, {Ciprini}, {Claus}, {Cohen-Tanugi}, {Conrad},
  {Cutini}, {D'Ammando}, {de Angelis}, {de Palma}, {Dermer}, {Silva}, {Drell},
  {Drlica-Wagner}, {Favuzzi}, {Fegan}, {Focke}, {Franckowiak}, {Fukazawa},
  {Fusco}, {Gargano}, {Gasparrini}, {Gehrels}, {Giglietto}, {Giordano},
  {Giroletti}, {Glanzman}, {Godfrey}, {Granot}, {Greiner}, {Grenier}, {Grove},
  {Guiriec}, {Hadasch}, {Hanabata}, {Hayashida}, {Hays}, {Hughes}, {Jackson},
  {Jogler}, {J{\'o}hannesson}, {Johnson}, {Kn{\"o}dlseder}, {Kocevski}, {Kuss},
  {Lande}, {Larsson}, {Latronico}, {Longo}, {Loparco}, {Lovellette}, {Lubrano},
  {Mazziotta}, {McEnery}, {Mehault}, {M{\'e}sz{\'a}ros}, {Michelson},
  {Mitthumsiri}, {Mizuno}, {Monte}, {Monzani}, {Moretti}, {Morselli},
  {Moskalenko}, {Murgia}, {Naumann-Godo}, {Norris}, {Nuss}, {Nymark}, {Ohno},
  {Ohsugi}, {Omodei}, {Orienti}, {Orlando}, {Paneque}, {Perkins},
  {Pesce-Rollins}, {Piron}, {Pivato}, {Racusin}, {Rain{\`o}}, {Rando},
  {Razzano}, {Razzaque}, {Reimer}, {Reimer}, {Romoli}, {Roth}, {Ryde},
  {Sanchez}, {Sgr{\`o}}, {Siskind}, {Sonbas}, {Spinelli}, {Stamatikos},
  {Takahashi}, {Tanaka}, {Thayer}, {Thayer}, {Tibaldo}, {Tinivella}, {Tosti},
  {Troja}, {Usher}, {Vandenbroucke}, {Vasileiou}, {Vianello}, {Vitale},
  {Waite}, {Winer}, {Wood}, {Yang}, {Gruber}, {Bhat}, {Bissaldi}, {Briggs},
  {Burgess}, {Connaughton}, {Foley}, {Kippen}, {Kouveliotou}, {McBreen},
  {McGlynn}, {Paciesas}, {Pelassa}, {Preece}, {Rau}, {van der Horst}, {von
  Kienlin}, {Kann}, {Filgas}, {Klose}, {Kr{\"u}hler}, {Fukui}, {Sako},
  {Tristram}, {Oates}, {Ukwatta}, \& {Littlejohns}}]{Ackermann2013}
{Ackermann}, M., {Ajello}, M., {Asano}, K., {et~al.} 2013, \apj, 763, 71

\bibitem[{{Akitaya} {et~al.}(2014){Akitaya}, {Moritani}, {Ui}, {Kanda}, \&
  {Yoshida}}]{Akitaya2014}
{Akitaya}, H., {Moritani}, Y., {Ui}, T., {Kanda}, Y., \& {Yoshida}, M. 2014,
  GRB Coordinates Network, 16163, 1

\bibitem[{{Amati} {et~al.}(2002){Amati}, {Frontera}, {Tavani}, {in't Zand},
  {Antonelli}, {Costa}, {Feroci}, {Guidorzi}, {Heise}, {Masetti}, {Montanari},
  {Nicastro}, {Palazzi}, {Pian}, {Piro}, \& {Soffitta}}]{Amati2002}
{Amati}, L., {Frontera}, F., {Tavani}, M., {et~al.} 2002, \aap, 390, 81

\bibitem[{{Amati} \& {Valle}(2013)}]{Amati2014}
{Amati}, L. \& {Valle}, M.~D. 2013, International Journal of Modern Physics D,
  22, 30028

\bibitem[{{Andreev} {et~al.}(2009{\natexlab{a}}){Andreev}, {Sergeev},
  {Parakhin}, {Karpov}, {Kuznietsova}, {Petkov}, \& {Pozanenko}}]{Andreev2009b}
{Andreev}, M., {Sergeev}, A., {Parakhin}, N., {et~al.} 2009{\natexlab{a}}, GRB
  Coordinates Network, 10273, 1

\bibitem[{{Andreev} {et~al.}(2009{\natexlab{b}}){Andreev}, {Sergeev}, \&
  {Pozanenko}}]{Andreev2009a}
{Andreev}, M., {Sergeev}, A., \& {Pozanenko}, A. 2009{\natexlab{b}}, GRB
  Coordinates Network, 10207, 1

\bibitem[{{Anumapa} {et~al.}(2009){Anumapa}, {Gurugubelli}, \&
  {Sahu}}]{Anumapa2009}
{Anumapa}, G.~C., {Gurugubelli}, U.~K., \& {Sahu}, D.~K. 2009, GRB Coordinates
  Network, 9576, 1

\bibitem[{{Atteia}(2003)}]{Atteia2003}
{Atteia}, J.-L. 2003, \aap, 407, L1

\bibitem[{{Band} {et~al.}(1993){Band}, {Matteson}, {Ford}, {Schaefer},
  {Palmer}, {Teegarden}, {Cline}, {Briggs}, {Paciesas}, {Pendleton}, {Fishman},
  {Kouveliotou}, {Meegan}, {Wilson}, \& {Lestrade}}]{Band1993}
{Band}, D., {Matteson}, J., {Ford}, L., {et~al.} 1993, \apj, 413, 281

\bibitem[{{Band} \& {Preece}(2005)}]{Band2005}
{Band}, D.~L. \& {Preece}, R.~D. 2005, \apj, 627, 319

\bibitem[{{Berger} {et~al.}(2008){Berger}, {Fox}, {Cucchiara}, \&
  {Cenko}}]{Berger2008a}
{Berger}, E., {Fox}, D.~B., {Cucchiara}, A., \& {Cenko}, S.~B. 2008, GRB
  Coordinates Network, 8335, 1

\bibitem[{{Berger} {et~al.}(2002){Berger}, {Kulkarni}, {Bloom}, {Price}, {Fox},
  {Frail}, {Axelrod}, {Chevalier}, {Colbert}, {Costa}, {Djorgovski},
  {Frontera}, {Galama}, {Halpern}, {Harrison}, {Holtzman}, {Hurley}, {Kimble},
  {McCarthy}, {Piro}, {Reichart}, {Ricker}, {Sari}, {Schmidt}, {Wheeler},
  {Vanderppek}, \& {Yost}}]{Berger2002}
{Berger}, E., {Kulkarni}, S.~R., {Bloom}, J.~S., {et~al.} 2002, \apj, 581, 981

\bibitem[{{Berger} \& {Rauch}(2008)}]{Berger2008b}
{Berger}, E. \& {Rauch}, M. 2008, GRB Coordinates Network, 8542, 1

\bibitem[{{Bersier}(2011)}]{Bersier2011}
{Bersier}, D. 2011, GRB Coordinates Network, 12216, 1

\bibitem[{{Bersier} {et~al.}(2003){Bersier}, {Stanek}, {Winn}, {Grav},
  {Holman}, {Matheson}, {Mochejska}, {Steeghs}, {Walker}, {Garnavich}, {Quinn},
  {Jha}, {Cook}, {Craig}, {Meintjes}, \& {Calitz}}]{Bersier2003}
{Bersier}, D., {Stanek}, K.~Z., {Winn}, J.~N., {et~al.} 2003, \apjl, 584, L43

\bibitem[{{Beskin} {et~al.}(2002){Beskin}, {Birjukov}, {Koposov},
  {Spiridonova}, \& {Pozanenko}}]{Beskin2002}
{Beskin}, G., {Birjukov}, A., {Koposov}, S., {Spiridonova}, O., \& {Pozanenko},
  A. 2002, GRB Coordinates Network, 1528, 1

\bibitem[{{Bikmaev} {et~al.}(2014){Bikmaev}, {Khamitov}, {Melnikov},
  {Sakhibullin}, {Burenin}, {Pavlinsky}, {Sunyaev}, {Kirbiyik}, {Kiziloglu}, \&
  {Gogus}}]{Bikmaev2014}
{Bikmaev}, I., {Khamitov}, I., {Melnikov}, S., {et~al.} 2014, GRB Coordinates
  Network, 16185, 1

\bibitem[{{Bikmaev} {et~al.}(2009){Bikmaev}, {Zhuchkov}, {Sakhibullin},
  {Khamitov}, {Eker}, {Kiziloglu}, {Gogus}, {Burenin}, {Pavlinsky}, \&
  {Sunyaev}}]{Bikmaev2009}
{Bikmaev}, I., {Zhuchkov}, R., {Sakhibullin}, N., {et~al.} 2009, GRB
  Coordinates Network, 9190, 1

\bibitem[{{Bloom} {et~al.}(2006){Bloom}, {Perley}, \& {Chen}}]{Bloom2006}
{Bloom}, J.~S., {Perley}, D.~A., \& {Chen}, H.~W. 2006, GRB Coordinates
  Network, 5826, 1

\bibitem[{{Bloom} {et~al.}(2009){Bloom}, {Perley}, {Li}, {Butler}, {Miller},
  {Kocevski}, {Kann}, {Foley}, {Chen}, {Filippenko}, {Starr}, {Macomber},
  {Prochaska}, {Chornock}, {Poznanski}, {Klose}, {Skrutskie}, {Lopez}, {Hall},
  {Glazebrook}, \& {Blake}}]{Bloom2009}
{Bloom}, J.~S., {Perley}, D.~A., {Li}, W., {et~al.} 2009, \apj, 691, 723

\bibitem[{{Brennan} {et~al.}(2008){Brennan}, {Reichart}, {Nysewander},
  {Lacluyze}, {Ivarsen}, {Crain}, {Schubel}, {Foster}, {Haislip}, {Styblova},
  \& {Trotter}}]{Brennan2008}
{Brennan}, T., {Reichart}, D., {Nysewander}, M., {et~al.} 2008, GRB Coordinates
  Network, 7629, 1

\bibitem[{{Broens} \& {Boyd}(2011)}]{Broens2011}
{Broens}, E. \& {Boyd}, D. 2011, GRB Coordinates Network, 12009, 1

\bibitem[{{Burenin} {et~al.}(2003){Burenin}, {Denissenko}, {Pavlinsky},
  {Sunyaev}, {Terekhov}, {Tkachenko}, {Aslan}, {Khamitov}, {Uluc}, {Kiziloglu},
  {Alpar}, {Baykal}, {Bikmaev}, {Sakhibullin}, \& {Suleymanov}}]{Burenin2003}
{Burenin}, R., {Denissenko}, D., {Pavlinsky}, M., {et~al.} 2003, GRB
  Coordinates Network, 1990, 1

\bibitem[{{Butler} {et~al.}(2012{\natexlab{a}}){Butler}, {Watson}, {Kutyrev},
  {Lee}, {Richer}, {Klein}, {Fox}, {Prochaska}, {Bloom}, {Cucchiara}, {Troja},
  {de Diego}, {Georgiev}, {Gonzalez}, {Roman-Zuniga}, {Gehrels}, \&
  {Moseley}}]{Butler2012a}
{Butler}, N., {Watson}, A.~M., {Kutyrev}, A., {et~al.} 2012{\natexlab{a}}, GRB
  Coordinates Network, 14077, 1

\bibitem[{{Butler} {et~al.}(2012{\natexlab{b}}){Butler}, {Watson}, {Kutyrev},
  {Lee}, {Richer}, {Klein}, {Fox}, {Prochaska}, {Bloom}, {Cucchiara}, {Troja},
  {de Diegoi}, {Georgiev}, {Gonzalez}, {Roman-Zuniga}, \&
  {Gehrels}}]{Butler2012b}
{Butler}, N., {Watson}, A.~M., {Kutyrev}, A., {et~al.} 2012{\natexlab{b}}, GRB
  Coordinates Network, 14080, 1

\bibitem[{{Butler} {et~al.}(2013{\natexlab{a}}){Butler}, {Watson}, {Kutyrev},
  {Lee}, {Richer}, {Klein}, {Fox}, {Prochaska}, {Bloom}, {Cucchiara}, {Troja},
  {Littlejohns}, {Ramirez-Ruiz}, {de Diego}, {Georgiev}, {Gonzalez},
  {Roman-Zuniga}, {Gehrels}, \& {Moseley}}]{Butler2013a}
{Butler}, N., {Watson}, A.~M., {Kutyrev}, A., {et~al.} 2013{\natexlab{a}}, GRB
  Coordinates Network, 14431, 1

\bibitem[{{Butler} {et~al.}(2013{\natexlab{b}}){Butler}, {Watson}, {Kutyrev},
  {Lee}, {Richer}, {Klein}, {Fox}, {Prochaska}, {Bloom}, {Cucchiara}, {Troja},
  {Littlejohns}, {Ramirez-Ruiz}, {de Diego}, {Georgiev}, {Gonzalez},
  {Roman-Zuniga}, {Gehrels}, \& {Moseley}}]{Butler2013b}
{Butler}, N., {Watson}, A.~M., {Kutyrev}, A., {et~al.} 2013{\natexlab{b}}, GRB
  Coordinates Network, 14441, 1

\bibitem[{{Butler} {et~al.}(2013{\natexlab{c}}){Butler}, {Watson}, {Kutyrev},
  {Lee}, {Richer}, {Klein}, {Fox}, {Prochaska}, {Bloom}, {Cucchiara}, {Troja},
  {Littlejohns}, {Ramirez-Ruiz}, {de Diego}, {Georgiev}, {Gonzalez},
  {Roman-Zuniga}, {Gehrels}, \& {Moseley}}]{Butler2013c}
{Butler}, N., {Watson}, A.~M., {Kutyrev}, A., {et~al.} 2013{\natexlab{c}}, GRB
  Coordinates Network, 15165, 1

\bibitem[{{Butler} {et~al.}(2014{\natexlab{a}}){Butler}, {Watson}, {Kutyrev},
  {Lee}, {Richer}, {Klein}, {Fox}, {Prochaska}, {Bloom}, {Cucchiara}, {Troja},
  {Littlejohns}, {Ramirez-Ruiz}, {de Diego}, {Georgiev}, {Gonzalez},
  {Roman-Zuniga}, {Gehrels}, \& {Moseley}}]{Butler2014a}
{Butler}, N., {Watson}, A.~M., {Kutyrev}, A., {et~al.} 2014{\natexlab{a}}, GRB
  Coordinates Network, 16121, 1

\bibitem[{{Butler} {et~al.}(2014{\natexlab{b}}){Butler}, {Watson}, {Kutyrev},
  {Lee}, {Richer}, {Klein}, {Fox}, {Prochaska}, {Bloom}, {Cucchiara}, {Troja},
  {Littlejohns}, {Ramirez-Ruiz}, {de Diego}, {Georgiev}, {Gonzalez},
  {Roman-Zuniga}, {Gehrels}, \& {Moseley}}]{Butler2014b}
{Butler}, N., {Watson}, A.~M., {Kutyrev}, A., {et~al.} 2014{\natexlab{b}}, GRB
  Coordinates Network, 16174, 1

\bibitem[{{Butler} {et~al.}(2014{\natexlab{c}}){Butler}, {Watson}, {Kutyrev},
  {Lee}, {Richer}, {Klein}, {Fox}, {Prochaska}, {Bloom}, {Cucchiara}, {Troja},
  {Littlejohns}, {Ramirez-Ruiz}, {de Diego}, {Georgiev}, {Gonzalez},
  {Roman-Zuniga}, {Gehrels}, \& {Moseley}}]{Butler2014c}
{Butler}, N., {Watson}, A.~M., {Kutyrev}, A., {et~al.} 2014{\natexlab{c}}, GRB
  Coordinates Network, 16513, 1

\bibitem[{{Butler} {et~al.}(2010){Butler}, {Bloom}, \&
  {Poznanski}}]{Butler2010}
{Butler}, N.~R., {Bloom}, J.~S., \& {Poznanski}, D. 2010, \apj, 711, 495

\bibitem[{{Butler} {et~al.}(2009){Butler}, {Kocevski}, \& {Bloom}}]{Butler2009}
{Butler}, N.~R., {Kocevski}, D., \& {Bloom}. 2009, \apj, 694, 76

\bibitem[{{Butler} {et~al.}(2007){Butler}, {Kocevski}, {Bloom}, \&
  {Curtis}}]{Butler2007}
{Butler}, N.~R., {Kocevski}, D., {Bloom}, J.~S., \& {Curtis}, J.~L. 2007, \apj,
  671, 656

\bibitem[{{Cabrera} {et~al.}(2007){Cabrera}, {Firmani}, {Avila-Reese},
  {Ghirlanda}, {Ghisellini}, \& {Nava}}]{Cabrera2007}
{Cabrera}, J.~I., {Firmani}, C., {Avila-Reese}, V., {et~al.} 2007, \mnras, 382,
  342

\bibitem[{{Cano} {et~al.}(2014{\natexlab{a}}){Cano}, {de Ugarte Postigo},
  {Pozanenko}, {Butler}, {Th{\"o}ne}, {Guidorzi}, {Kr{\"u}hler}, {Gorosabel},
  {Jakobsson}, {Leloudas}, {Malesani}, {Hjorth}, {Melandri}, {Mundell},
  {Wiersema}, {D'Avanzo}, {Schulze}, {Gomboc}, {Johansson}, {Zheng}, {Kann},
  {Knust}, {Varela}, {Akerlof}, {Bloom}, {Burkhonov}, {Cooke}, {de Diego},
  {Dhungana}, {Farina}, {Ferrante}, {Flewelling}, {Fox}, {Fynbo}, {Gehrels},
  {Georgiev}, {Gonz{\'a}lez}, {Greiner}, {G{\"u}ver}, {Hartoog}, {Hatch},
  {Jelinek}, {Kehoe}, {Klose}, {Klunko}, {Kopa{\v c}}, {Kutyrev}, {Krugly},
  {Lee}, {Levan}, {Linkov}, {Matkin}, {Minikulov}, {Molotov}, {Prochaska},
  {Richer}, {Rom{\'a}n-Z{\'u}{\~n}iga}, {Rumyantsev},
  {S{\'a}nchez-Ram{\'{\i}}rez}, {Steele}, {Tanvir}, {Volnova}, {Watson}, {Xu},
  \& {Yuan}}]{Cano2014a}
{Cano}, Z., {de Ugarte Postigo}, A., {Pozanenko}, A., {et~al.}
  2014{\natexlab{a}}, \aap, 568, A19

\bibitem[{{Cano} {et~al.}(2009{\natexlab{a}}){Cano}, {Guidorzi}, {Bersier},
  {Melandri}, {Steele}, {Smith}, \& {Mundell}}]{Cano2009a}
{Cano}, Z., {Guidorzi}, C., {Bersier}, D., {et~al.} 2009{\natexlab{a}}, GRB
  Coordinates Network, 9531, 1

\bibitem[{{Cano} {et~al.}(2014{\natexlab{b}}){Cano}, {Malesani}, {Geier},
  {Jensen}, {Taddia}, \& {Fremling}}]{Cano2014b}
{Cano}, Z., {Malesani}, D., {Geier}, S., {et~al.} 2014{\natexlab{b}}, GRB
  Coordinates Network, 16169, 1

\bibitem[{{Cano} {et~al.}(2009{\natexlab{b}}){Cano}, {Melandri}, {Mundell},
  {Gomboc}, {Bersier}, {Clay}, {Kobayashi}, {Mottram}, {Smith}, {Steele}, \&
  {Guidorzi}}]{Cano2009b}
{Cano}, Z., {Melandri}, A., {Mundell}, C.~G., {et~al.} 2009{\natexlab{b}}, GRB
  Coordinates Network, 10262, 1

\bibitem[{{Castro-Tirado} {et~al.}(2014){Castro-Tirado}, {Cunniffe},
  {Sanchez-Ramirez}, {Gorosabel}, {Jelinek}, {Oates}, {Jeong}, {Tello},
  {Pandey}, {Gomez-Velarde}, \& {Perez}}]{CastroTirado2014}
{Castro-Tirado}, A.~J., {Cunniffe}, R., {Sanchez-Ramirez}, R., {et~al.} 2014,
  GRB Coordinates Network, 16505, 1

\bibitem[{{Castro-Tirado} {et~al.}(2010){Castro-Tirado}, {M{\o}ller},
  {Garc{\'{\i}}a-Segura}, {Gorosabel}, {P{\'e}rez}, {de Ugarte Postigo},
  {Solano}, {Barrado}, {Klose}, {Kann}, {Castro Cer{\'o}n}, {Kouveliotou},
  {Fynbo}, {Hjorth}, {Pedersen}, {Pian}, {Rol}, {Palazzi}, {Masetti}, {Tanvir},
  {Vreeswijk}, {Andersen}, {Fruchter}, {Greiner}, {Wijers}, \& {van den
  Heuvel}}]{Castro-Tirado2010}
{Castro-Tirado}, A.~J., {M{\o}ller}, P., {Garc{\'{\i}}a-Segura}, G., {et~al.}
  2010, \aap, 517, A61

\bibitem[{{Cenko}(2009{\natexlab{a}})}]{Cenko2009a}
{Cenko}, S.~B. 2009{\natexlab{a}}, GRB Coordinates Network, 9201, 1

\bibitem[{{Cenko}(2009{\natexlab{b}})}]{Cenko2009b}
{Cenko}, S.~B. 2009{\natexlab{b}}, GRB Coordinates Network, 9513, 1

\bibitem[{{Cenko} {et~al.}(2007{\natexlab{a}}){Cenko}, {Cucchiara}, {Fox},
  {Berger}, \& {Price}}]{Cenko2007b}
{Cenko}, S.~B., {Cucchiara}, A., {Fox}, D.~B., {Berger}, E., \& {Price}, P.~A.
  2007{\natexlab{a}}, GRB Coordinates Network, 6888, 1

\bibitem[{{Cenko} {et~al.}(2007{\natexlab{b}}){Cenko}, {Fox}, {Cucchiara},
  {Schmidt}, {Berger}, {Price}, \& {Roth}}]{Cenko2007a}
{Cenko}, S.~B., {Fox}, D.~B., {Cucchiara}, A., {et~al.} 2007{\natexlab{b}}, GRB
  Coordinates Network, 6556, 1

\bibitem[{{Cenko} {et~al.}(2011){Cenko}, {Hora}, \& {Bloom}}]{Cenko2011a}
{Cenko}, S.~B., {Hora}, J.~L., \& {Bloom}, J.~S. 2011, GRB Coordinates Network,
  11638, 1

\bibitem[{{Cenko} {et~al.}(2006){Cenko}, {Kasliwal}, {Harrison}, {Pal'shin},
  {Frail}, {Cameron}, {Berger}, {Fox}, {Gal-Yam}, {Kulkarni}, {Moon}, {Nakar},
  {Ofek}, {Penprase}, {Price}, {Sari}, {Schmidt}, {Soderberg}, {Aptekar},
  {Frederiks}, {Golenetskii}, {Burrows}, {Chevalier}, {Gehrels}, {McCarthy},
  {Nousek}, \& {Piran}}]{Cenko2006a}
{Cenko}, S.~B., {Kasliwal}, M., {Harrison}, F.~A., {et~al.} 2006, \apj, 652,
  490

\bibitem[{{Cenko} {et~al.}(2005){Cenko}, {Kulkarni}, {Gal-Yam}, \&
  {Berger}}]{Cenko2005}
{Cenko}, S.~B., {Kulkarni}, S.~R., {Gal-Yam}, A., \& {Berger}, E. 2005, GRB
  Coordinates Network, 3542, 1

\bibitem[{{Cenko} \& {Perley}(2013{\natexlab{a}})}]{Cenko2013a}
{Cenko}, S.~B. \& {Perley}, D.~A. 2013{\natexlab{a}}, GRB Coordinates Network,
  14846, 1

\bibitem[{{Cenko} \& {Perley}(2013{\natexlab{b}})}]{Cenko2013b}
{Cenko}, S.~B. \& {Perley}, D.~A. 2013{\natexlab{b}}, GRB Coordinates Network,
  14960, 1

\bibitem[{{Cenko} \& {Perley}(2014{\natexlab{a}})}]{Cenko2014a}
{Cenko}, S.~B. \& {Perley}, D.~A. 2014{\natexlab{a}}, GRB Coordinates Network,
  16129, 1

\bibitem[{{Cenko} \& {Perley}(2014{\natexlab{b}})}]{Cenko2014b}
{Cenko}, S.~B. \& {Perley}, D.~A. 2014{\natexlab{b}}, GRB Coordinates Network,
  16153, 1

\bibitem[{{Cenko} {et~al.}(2009){Cenko}, {Perley}, {Junkkarinen}, {Burbidge},
  {Diego}, \& {Miller}}]{Cenko2009c}
{Cenko}, S.~B., {Perley}, D.~A., {Junkkarinen}, V., {et~al.} 2009, GRB
  Coordinates Network, 9518, 1

\bibitem[{{Cenko} \& {Rau}(2006)}]{Cenko2006b}
{Cenko}, S.~B. \& {Rau}, A. 2006, GRB Coordinates Network, 5512, 1

\bibitem[{{Chen} {et~al.}(2008){Chen}, {Huang}, {Chen}, {Huang}, {Urata}, \&
  {Marshall}}]{Chen2008}
{Chen}, T.-W., {Huang}, L.-C., {Chen}, Y.-T., {et~al.} 2008, GRB Coordinates
  Network, 7990, 1

\bibitem[{{Chen} {et~al.}(2014){Chen}, {King}, {Wen}, {Huang}, {Wang}, \&
  {Lehner}}]{Chen2014}
{Chen}, Y.~T., {King}, S.~K., {Wen}, C.~Y., {et~al.} 2014, GRB Coordinates
  Network, 15877, 1

\bibitem[{{Choi} {et~al.}(2014{\natexlab{a}}){Choi}, {Im}, {Sung}, \&
  {Urata}}]{Choi2014b}
{Choi}, C., {Im}, M., {Sung}, H.-I., \& {Urata}, Y. 2014{\natexlab{a}}, GRB
  Coordinates Network, 16149, 1

\bibitem[{{Choi} {et~al.}(2014{\natexlab{b}}){Choi}, {Im}, {Sung}, \&
  {Urata}}]{Choi2014a}
{Choi}, C., {Im}, M., {Sung}, H.-I., \& {Urata}, Y. 2014{\natexlab{b}}, GRB
  Coordinates Network, 15689, 1

\bibitem[{{Chornock} {et~al.}(2009{\natexlab{a}}){Chornock}, {Cenko},
  {Griffith}, {Kislak}, {Kleiser}, \& {Filippenko}}]{Chornock2009b}
{Chornock}, R., {Cenko}, S.~B., {Griffith}, C.~V., {et~al.} 2009{\natexlab{a}},
  GRB Coordinates Network, 9151, 1

\bibitem[{{Chornock} {et~al.}(2009{\natexlab{b}}){Chornock}, {Cenko}, {Li}, \&
  {Filippenko}}]{Chornock2009a}
{Chornock}, R., {Cenko}, S.~B., {Li}, W., \& {Filippenko}, A.~V.
  2009{\natexlab{b}}, GRB Coordinates Network, 9148, 1

\bibitem[{{Chornock} {et~al.}(2009{\natexlab{c}}){Chornock}, {Perley}, {Cenko},
  \& {Bloom}}]{Chornock2009d}
{Chornock}, R., {Perley}, D.~A., {Cenko}, S.~B., \& {Bloom}, J.~S.
  2009{\natexlab{c}}, GRB Coordinates Network, 9243, 1

\bibitem[{{Chornock} {et~al.}(2009{\natexlab{d}}){Chornock}, {Perley}, \&
  {Cobb}}]{Chornock2009c}
{Chornock}, R., {Perley}, D.~A., \& {Cobb}, B.~E. 2009{\natexlab{d}}, GRB
  Coordinates Network, 10100, 1

\bibitem[{{Christensen} {et~al.}(2004){Christensen}, {Hjorth}, {Gorosabel},
  {Vreeswijk}, {Fruchter}, {Sahu}, \& {Petro}}]{Christensen2004}
{Christensen}, L., {Hjorth}, J., {Gorosabel}, J., {et~al.} 2004, \aap, 413, 121

\bibitem[{{Christie} {et~al.}(2009{\natexlab{a}}){Christie}, {de Ugarte
  Postigo}, \& {Natusch}}]{Christie2009a}
{Christie}, G.~W., {de Ugarte Postigo}, A., \& {Natusch}, T.
  2009{\natexlab{a}}, GRB Coordinates Network, 9396, 1

\bibitem[{{Christie} {et~al.}(2009{\natexlab{b}}){Christie}, {Dong}, {de Ugarte
  Postigo}, \& {Natusch}}]{Christie2009b}
{Christie}, G.~W., {Dong}, S., {de Ugarte Postigo}, A., \& {Natusch}, T.
  2009{\natexlab{b}}, GRB Coordinates Network, 10137, 1

\bibitem[{{Ciabattari} {et~al.}(2014){Ciabattari}, {Donati}, {Mazzoni},
  {Petroni}, \& {Rossi}}]{Ciabattari2014}
{Ciabattari}, F., {Donati}, S., {Mazzoni}, E., {Petroni}, G., \& {Rossi}, M.
  2014, GRB Coordinates Network, 16511, 1

\bibitem[{{Cobb}(2008{\natexlab{a}})}]{Cobb2008a}
{Cobb}, B.~E. 2008{\natexlab{a}}, GRB Coordinates Network, 7609, 1

\bibitem[{{Cobb}(2008{\natexlab{b}})}]{Cobb2008b}
{Cobb}, B.~E. 2008{\natexlab{b}}, GRB Coordinates Network, 8356, 1

\bibitem[{{Cobb}(2008{\natexlab{c}})}]{Cobb2008c}
{Cobb}, B.~E. 2008{\natexlab{c}}, GRB Coordinates Network, 8547, 1

\bibitem[{{Cobb} \& {Bailyn}(2005)}]{Cobb2005}
{Cobb}, B.~E. \& {Bailyn}, C.~D. 2005, GRB Coordinates Network, 3506, 1

\bibitem[{{Cobb} {et~al.}(2010){Cobb}, {Bloom}, {Perley}, {Morgan}, {Cenko}, \&
  {Filippenko}}]{Cobb2010}
{Cobb}, B.~E., {Bloom}, J.~S., {Perley}, D.~A., {et~al.} 2010, \apjl, 718, L150

\bibitem[{{Collazzi} {et~al.}(2012){Collazzi}, {Schaefer}, {Goldstein}, \&
  {Preece}}]{Collazzi2012a}
{Collazzi}, A.~C., {Schaefer}, B.~E., {Goldstein}, A., \& {Preece}, R.~D. 2012,
  \apj, 747, 39

\bibitem[{{Covino} {et~al.}(2010){Covino}, {Campana}, {Conciatore}, {D'Elia},
  {Palazzi}, {Th{\"o}ne}, {Vergani}, {Wiersema}, {Brusasca}, {Cucchiara},
  {Cobb}, {Fern{\'a}ndez-Soto}, {Kann}, {Malesani}, {Tanvir}, {Antonelli},
  {Bremer}, {Castro-Tirado}, {de Ugarte Postigo}, {Molinari}, {Nicastro},
  {Stefanon}, {Testa}, {Tosti}, {Vitali}, {Amati}, {Chapman}, {Conconi},
  {Cutispoto}, {Fynbo}, {Goldoni}, {Henriksen}, {Horne}, {Malaspina}, {Meurs},
  {Pian}, {Stella}, {Tagliaferri}, {Ward}, \& {Zerbi}}]{Covino2010}
{Covino}, S., {Campana}, S., {Conciatore}, M.~L., {et~al.} 2010, \aap, 521, A53

\bibitem[{{Covino} {et~al.}(2013){Covino}, {Melandri}, {Salvaterra}, {Campana},
  {Vergani}, {Bernardini}, {D'Avanzo}, {D'Elia}, {Fugazza}, {Ghirlanda},
  {Ghisellini}, {Gomboc}, {Jin}, {Kr{\"u}hler}, {Malesani}, {Nava},
  {Sbarufatti}, \& {Tagliaferri}}]{Covino2013}
{Covino}, S., {Melandri}, A., {Salvaterra}, R., {et~al.} 2013, \mnras, 432,
  1231

\bibitem[{{Coward} {et~al.}(2014){Coward}, {Howell}, {Wan}, \&
  {Macpherson}}]{Coward2014}
{Coward}, D., {Howell}, E., {Wan}, L., \& {Macpherson}, D. 2014, ArXiv e-prints

\bibitem[{{Cucchiara} {et~al.}(2011{\natexlab{a}}){Cucchiara}, {Cenko},
  {Bloom}, {Melandri}, {Morgan}, {Kobayashi}, {Smith}, {Perley}, {Li}, {Hora},
  {da Silva}, {Prochaska}, {Milne}, {Butler}, {Cobb}, {Worseck}, {Mundell},
  {Steele}, {Filippenko}, {Fumagalli}, {Klein}, {Stephens}, {Bluck}, \&
  {Mason}}]{Cucchiara2011b}
{Cucchiara}, A., {Cenko}, S.~B., {Bloom}, J.~S., {et~al.} 2011{\natexlab{a}},
  \apj, 743, 154

\bibitem[{{Cucchiara} {et~al.}(2009){Cucchiara}, {Fox}, {Levan}, \&
  {Tanvir}}]{Cucchiara2009b}
{Cucchiara}, A., {Fox}, D., {Levan}, A., \& {Tanvir}, N. 2009, GRB Coordinates
  Network, 10202, 1

\bibitem[{{Cucchiara} \& {Fox}(2009)}]{Cucchiara2009a}
{Cucchiara}, A. \& {Fox}, D.~B. 2009, GRB Coordinates Network, 8774, 1

\bibitem[{{Cucchiara} {et~al.}(2008{\natexlab{a}}){Cucchiara}, {Fox}, {Cenko},
  \& {Berger}}]{Cucchiara2008b}
{Cucchiara}, A., {Fox}, D.~B., {Cenko}, S.~B., \& {Berger}, E.
  2008{\natexlab{a}}, GRB Coordinates Network, 8372, 1

\bibitem[{{Cucchiara} {et~al.}(2008{\natexlab{b}}){Cucchiara}, {Fox}, {Cenko},
  \& {Berger}}]{Cucchiara2008c}
{Cucchiara}, A., {Fox}, D.~B., {Cenko}, S.~B., \& {Berger}, E.
  2008{\natexlab{b}}, GRB Coordinates Network, 8346, 1

\bibitem[{{Cucchiara} {et~al.}(2008{\natexlab{c}}){Cucchiara}, {Fox}, {Cenko},
  \& {Berger}}]{Cucchiara2008d}
{Cucchiara}, A., {Fox}, D.~B., {Cenko}, S.~B., \& {Berger}, E.
  2008{\natexlab{c}}, GRB Coordinates Network, 8713, 1

\bibitem[{{Cucchiara} {et~al.}(2008{\natexlab{d}}){Cucchiara}, {Fox}, {Cenko},
  \& {Berger}}]{Cucchiara2008a}
{Cucchiara}, A., {Fox}, D.~B., {Cenko}, S.~B., \& {Berger}, E.
  2008{\natexlab{d}}, GRB Coordinates Network, 8065, 1

\bibitem[{{Cucchiara} {et~al.}(2011{\natexlab{b}}){Cucchiara}, {Levan}, {Fox},
  {Tanvir}, {Ukwatta}, {Berger}, {Kr{\"u}hler}, {K{\"u}pc{\"u} Yolda{\c s}},
  {Wu}, {Toma}, {Greiner}, {Olivares}, {Rowlinson}, {Amati}, {Sakamoto},
  {Roth}, {Stephens}, {Fritz}, {Fynbo}, {Hjorth}, {Malesani}, {Jakobsson},
  {Wiersema}, {O'Brien}, {Soderberg}, {Foley}, {Fruchter}, {Rhoads},
  {Rutledge}, {Schmidt}, {Dopita}, {Podsiadlowski}, {Willingale}, {Wolf},
  {Kulkarni}, \& {D'Avanzo}}]{Cucchiara2011a}
{Cucchiara}, A., {Levan}, A.~J., {Fox}, D.~B., {et~al.} 2011{\natexlab{b}},
  \apj, 736, 7

\bibitem[{{Cucchiara} \& {Perley}(2013)}]{Cucchiara2013}
{Cucchiara}, A. \& {Perley}, D. 2013, GRB Coordinates Network, 15144, 1

\bibitem[{{Cwiok} {et~al.}(2008){Cwiok}, {Dominik}, {Kasprowicz}, {Majcher},
  {Majczyna}, {Malek}, {Mankiewicz}, {Nawrocki}, {Piotrowski}, {Rybka},
  {Sokolowski}, {Uzycki}, {Wrochna}, {Zaremba}, \& {Zarnecki}}]{Cwiok2008}
{Cwiok}, M., {Dominik}, W., {Kasprowicz}, G., {et~al.} 2008, GRB Coordinates
  Network, 8707, 1

\bibitem[{{D'Avanzo} {et~al.}(2011){D'Avanzo}, {D'Elia}, {di Fabrizio}, \&
  {Gurtu}}]{Davanzo2011}
{D'Avanzo}, P., {D'Elia}, V., {di Fabrizio}, L., \& {Gurtu}, A. 2011, GRB
  Coordinates Network, 11997, 1

\bibitem[{{De Cia} {et~al.}(2010){De Cia}, {Malesani}, {Vreeswijk}, {Schulze},
  \& {Jakobsson}}]{Decia2010}
{De Cia}, A., {Malesani}, D., {Vreeswijk}, P.~M., {Schulze}, S., \&
  {Jakobsson}, P. 2010, GRB Coordinates Network, 11027, 1

\bibitem[{{de Pasquale} \& {Cannizzo}(2010)}]{dePasquale2010}
{de Pasquale}, M. \& {Cannizzo}, J. 2010, GRB Coordinates Network, 11338, 1

\bibitem[{{de Ugarte Postigo} {et~al.}(2005){de Ugarte Postigo},
  {Castro-Tirado}, {Gorosabel}, {J{\'o}hannesson}, {Bj{\"o}rnsson},
  {Gudmundsson}, {Bremer}, {Pak}, {Tanvir}, {Castro Cer{\'o}n}, {Guzyi},
  {Jel{\'{\i}}nek}, {Klose}, {P{\'e}rez-Ram{\'{\i}}rez}, {Aceituno}, {Campo
  Bagat{\'{\i}}n}, {Covino}, {Cardiel}, {Fathkullin}, {Henden}, {Huferath},
  {Kurata}, {Malesani}, {Mannucci}, {Ruiz-Lapuente}, {Sokolov}, {Thiele},
  {Wisotzki}, {Antonelli}, {Bartolini}, {Boattini}, {Guarnieri}, {Piccioni},
  {Pizzichini}, {del Principe}, {di Paola}, {Fugazza}, {Ghisellini}, {Hunt},
  {Konstantinova}, {Masetti}, {Palazzi}, {Pian}, {Stefanon}, {Testa}, \&
  {Tristram}}]{deUgartePostigo2005}
{de Ugarte Postigo}, A., {Castro-Tirado}, A.~J., {Gorosabel}, J., {et~al.}
  2005, \aap, 443, 841

\bibitem[{{de Ugarte Postigo} {et~al.}(2011){de Ugarte Postigo},
  {Castro-Tirado}, {Tello}, {Cabrera Lavers}, \&
  {Reverte}}]{deUgartePostigo2011a}
{de Ugarte Postigo}, A., {Castro-Tirado}, A.~J., {Tello}, J.~C., {Cabrera
  Lavers}, A., \& {Reverte}, D. 2011, GRB Coordinates Network, 11993, 1

\bibitem[{{de Ugarte Postigo} {et~al.}(2009{\natexlab{a}}){de Ugarte Postigo},
  {Gorosabel}, {D'Avanzo}, {Kubanek}, {Jelinek}, {Cunniffe}, {Guziy},
  {Castro-Tirado}, {Yock}, {Allen}, {Bond}, \&
  {Christie}}]{deUgartePostigo2009c}
{de Ugarte Postigo}, A., {Gorosabel}, J., {D'Avanzo}, P., {et~al.}
  2009{\natexlab{a}}, GRB Coordinates Network, 10104, 1

\bibitem[{{de Ugarte Postigo} {et~al.}(2009{\natexlab{b}}){de Ugarte Postigo},
  {Gorosabel}, {Fynbo}, {Wiersema}, \& {Tanvir}}]{deUgartePostigo2009d}
{de Ugarte Postigo}, A., {Gorosabel}, J., {Fynbo}, J.~P.~U., {Wiersema}, K., \&
  {Tanvir}, N. 2009{\natexlab{b}}, GRB Coordinates Network, 9771, 1

\bibitem[{{de Ugarte Postigo} {et~al.}(2009{\natexlab{c}}){de Ugarte Postigo},
  {Gorosabel}, {Malesani}, {Fynbo}, \& {Levan}}]{deUgartePostigo2009a}
{de Ugarte Postigo}, A., {Gorosabel}, J., {Malesani}, D., {Fynbo}, J.~P.~U., \&
  {Levan}, A.~J. 2009{\natexlab{c}}, GRB Coordinates Network, 9381, 1

\bibitem[{{de Ugarte Postigo} {et~al.}(2009{\natexlab{d}}){de Ugarte Postigo},
  {Gorosabel}, {Malesani}, {Fynbo}, \& {Levan}}]{deUgartePostigo2009b}
{de Ugarte Postigo}, A., {Gorosabel}, J., {Malesani}, D., {Fynbo}, J.~P.~U., \&
  {Levan}, A.~J. 2009{\natexlab{d}}, GRB Coordinates Network, 9383, 1

\bibitem[{{de Ugarte Postigo} {et~al.}(2014{\natexlab{a}}){de Ugarte Postigo},
  {Gorosabel}, {Xu}, {Kruehler}, {Djupvik}, {Gafton}, \&
  {Libbrecht}}]{deUgartePostigo2014a}
{de Ugarte Postigo}, A., {Gorosabel}, J., {Xu}, D., {et~al.}
  2014{\natexlab{a}}, GRB Coordinates Network, 16253, 1

\bibitem[{{de Ugarte Postigo} {et~al.}(2014{\natexlab{b}}){de Ugarte Postigo},
  {Gorosabel}, {Xu}, {Malesani}, {Leloudas}, {Jakobsson}, {Kruehler},
  {Djupvik}, {Gafton}, \& {Libbrecht}}]{deUgartePostigo2014b}
{de Ugarte Postigo}, A., {Gorosabel}, J., {Xu}, D., {et~al.}
  2014{\natexlab{b}}, GRB Coordinates Network, 16310, 1

\bibitem[{{de Ugarte Postigo} {et~al.}(2013){de Ugarte Postigo}, {Tanvir},
  {Sanchez-Ramirez}, {Thoene}, {Gorosabel}, \& {Fynbo}}]{deUgartepostigo2013}
{de Ugarte Postigo}, A., {Tanvir}, N., {Sanchez-Ramirez}, R., {et~al.} 2013,
  GRB Coordinates Network, 14437, 1

\bibitem[{{Della Valle} {et~al.}(2003){Della Valle}, {Benetti}, {Malesani},
  {Mason}, {Antonelli}, {Cocozza}, {Covino}, {Fugazza}, {Ghisellini}, {Israel},
  {Stella}, \& {Testa}}]{DellaValle2003}
{Della Valle}, M., {Benetti}, S., {Malesani}, D., {et~al.} 2003, GRB
  Coordinates Network, 1809, 1

\bibitem[{{Dereli} {et~al.}(2013){Dereli}, {Klotz}, {Macpherson}, {Coward},
  {Gendre}, {Boer}, {Siellez}, {Bardho}, {Williams}, \& {Martin}}]{Dereli2013}
{Dereli}, H., {Klotz}, A., {Macpherson}, D., {et~al.} 2013, GRB Coordinates
  Network, 14372, 1

\bibitem[{{Distefano} {et~al.}(2006){Distefano}, {Covino}, {Molinari},
  {Chincarini}, {Zerbi}, {Testa}, {Tosti}, {Vitali}, {Antonelli}, {Conconi},
  {Cutispoto}, {Malaspina}, {Nicastro}, {Palazzi}, {Meurs}, \&
  {Goldoni}}]{Distefano2006}
{Distefano}, E., {Covino}, S., {Molinari}, E., {et~al.} 2006, GRB Coordinates
  Network, 4526, 1

\bibitem[{{Durig} {et~al.}(2005){Durig}, Oksanen, {Pullen}, \&
  {Price}}]{Durig2005}
{Durig}, D.~T., Oksanen, A.~C., {Pullen}, C., \& {Price}, A. 2005, GRB
  Coordinates Network, 3478, 1

\bibitem[{{Eichler} \& {Levinson}(2004)}]{Eichler2004}
{Eichler}, D. \& {Levinson}, A. 2004, \apjl, 614, L13

\bibitem[{{Elenin} {et~al.}(2010){Elenin}, {Molotov}, \&
  {Pozanenko}}]{Elenin2010}
{Elenin}, L., {Molotov}, I., \& {Pozanenko}, A. 2010, GRB Coordinates Network,
  11012, 1

\bibitem[{{Elenin} {et~al.}(2013{\natexlab{a}}){Elenin}, {Savanevych},
  {Bryukhovetskiy}, {Molotov}, \& {Pozanenko}}]{Elenin2013b}
{Elenin}, L., {Savanevych}, V., {Bryukhovetskiy}, A., {Molotov}, I., \&
  {Pozanenko}, A. 2013{\natexlab{a}}, GRB Coordinates Network, 14860, 1

\bibitem[{{Elenin} {et~al.}(2012){Elenin}, {Volnova}, {Molotov}, \&
  {Pozanenko}}]{Elenin2012}
{Elenin}, L., {Volnova}, A., {Molotov}, I., \& {Pozanenko}, A. 2012, GRB
  Coordinates Network, 13679, 1

\bibitem[{{Elenin} {et~al.}(2014){Elenin}, {Volnova}, {Molotov}, \&
  {Pozanenko}}]{Elenin2014}
{Elenin}, L., {Volnova}, A., {Molotov}, I., \& {Pozanenko}, A. 2014, GRB
  Coordinates Network, 16148, 1

\bibitem[{{Elenin} {et~al.}(2013{\natexlab{b}}){Elenin}, {Volnova},
  {Savanevych}, {Bryukhovetskiy}, {Molotov}, \& {Pozanenko}}]{Elenin2013a}
{Elenin}, L., {Volnova}, A., {Savanevych}, V., {et~al.} 2013{\natexlab{b}}, GRB
  Coordinates Network, 14428, 1

\bibitem[{{Elliott} {et~al.}(2014){Elliott}, {Varela}, {Kann}, \&
  {Greiner}}]{Elliott2014}
{Elliott}, J., {Varela}, K., {Kann}, D.~A., \& {Greiner}, J. 2014, GRB
  Coordinates Network, 15829, 1

\bibitem[{{Elunko} \& {Pozanenko}(2011)}]{Elunko2011}
{Elunko}, E. \& {Pozanenko}, A. 2011, GRB Coordinates Network, 11958, 1

\bibitem[{{Fatkhullin} {et~al.}(2009){Fatkhullin}, {Moskvitin},
  {Castro-Tirado}, \& {de Ugarte Postigo}}]{Fatkhullin2009}
{Fatkhullin}, T., {Moskvitin}, A., {Castro-Tirado}, A.~J., \& {de Ugarte
  Postigo}, A. 2009, GRB Coordinates Network, 9542, 1

\bibitem[{{Fernandez-Soto} {et~al.}(2009){Fernandez-Soto}, {Peris}, \&
  {Alonso-Lorite}}]{Fernandez-soto2009}
{Fernandez-Soto}, A., {Peris}, V., \& {Alonso-Lorite}, J. 2009, GRB Coordinates
  Network, 9536, 1

\bibitem[{{Ferrante} {et~al.}(2014){Ferrante}, {Guver}, {Flewelling}, {Kehoe},
  \& {Dhungana}}]{Ferrante2014}
{Ferrante}, F.~V., {Guver}, T., {Flewelling}, H., {Kehoe}, R., \& {Dhungana},
  G. 2014, GRB Coordinates Network, 16145, 1

\bibitem[{{Filgas} {et~al.}(2009{\natexlab{a}}){Filgas}, {Afonso}, {Klose}, \&
  {Greiner}}]{Filgas2009b}
{Filgas}, R., {Afonso}, P., {Klose}, S., \& {Greiner}, J. 2009{\natexlab{a}},
  GRB Coordinates Network, 10286, 1

\bibitem[{{Filgas} {et~al.}(2011){Filgas}, {Greiner}, {Schady}, {Kr{\"u}hler},
  {Updike}, {Klose}, {Nardini}, {Kann}, {Rossi}, {Sudilovsky}, {Afonso},
  {Clemens}, {Elliott}, {Nicuesa Guelbenzu}, {Olivares E.}, \&
  {Rau}}]{Filgas2011}
{Filgas}, R., {Greiner}, J., {Schady}, P., {et~al.} 2011, \aap, 535, A57

\bibitem[{{Filgas} {et~al.}(2010){Filgas}, {Kruehler}, {Greiner}, \&
  {Yoldas}}]{Filgas2010}
{Filgas}, R., {Kruehler}, T., {Greiner}, J., \& {Yoldas}, A. 2010, GRB
  Coordinates Network, 10592, 1

\bibitem[{{Filgas} {et~al.}(2009{\natexlab{b}}){Filgas}, {Updike}, \&
  {Greiner}}]{Filgas2009a}
{Filgas}, R., {Updike}, A., \& {Greiner}, J. 2009{\natexlab{b}}, GRB
  Coordinates Network, 10098, 1

\bibitem[{{Fiore} {et~al.}(2002){Fiore}, {Savaglio}, {Antonelli}, {Fontana},
  {Marconi}, {Stella}, {Di Paola}, {Stratta}, {Israel}, {Covino}, {Chincarini},
  {Ghisellini}, {Saracco}, {Zerbi}, {Lazzati}, {Perna}, {Vietri}, {Frontera},
  {Mereghetti}, {Meurs}, \& {Kawai}}]{Fiore2002}
{Fiore}, F., {Savaglio}, S., {Antonelli}, L.~A., {et~al.} 2002, GRB Coordinates
  Network, 1524, 1

\bibitem[{{Firmani} {et~al.}(2009){Firmani}, {Cabrera}, {Avila-Reese},
  {Ghisellini}, {Ghirlanda}, {Nava}, \& {Bosnjak}}]{Firmani2009}
{Firmani}, C., {Cabrera}, J.~I., {Avila-Reese}, V., {et~al.} 2009, \mnras, 393,
  1209

\bibitem[{{Flores} {et~al.}(2013){Flores}, {Covino}, {Xu}, {Kruehler}, {Fynbo},
  {Milvang-Jensen}, {de Ugarte Postigo}, {Kaper}, \& {Wiersema}}]{Flores2013}
{Flores}, H., {Covino}, S., {Xu}, D., {et~al.} 2013, GRB Coordinates Network,
  14491, 1

\bibitem[{{Flores} {et~al.}(2010){Flores}, {Fynbo}, {de Ugarte Postigo},
  {Milvang-Jensen}, {Malesani}, {Goldoni}, {Thoene}, {Piranomonte}, \&
  {Vergani}}]{Flores2010}
{Flores}, H., {Fynbo}, J.~P.~U., {de Ugarte Postigo}, A., {et~al.} 2010, GRB
  Coordinates Network, 11317, 1

\bibitem[{{Foley} {et~al.}(2005){Foley}, {Chen}, {Bloom}, \&
  {Prochaska}}]{Foley2005}
{Foley}, R.~J., {Chen}, H.-W., {Bloom}, J., \& {Prochaska}, J.~X. 2005, GRB
  Coordinates Network, 3483, 1

\bibitem[{{Fox} {et~al.}(2003){Fox}, {Price}, {Soderberg}, {Berger},
  {Kulkarni}, {Sari}, {Frail}, {Harrison}, {Yost}, {Matthews}, {Peterson},
  {Tanaka}, {Christiansen}, \& {Moriarty-Schieven}}]{Fox2003}
{Fox}, D.~W., {Price}, P.~A., {Soderberg}, A.~M., {et~al.} 2003, \apjl, 586, L5

\bibitem[{{Fugazza} {et~al.}(2003){Fugazza}, {Antonelli}, {Fiore}, {Covino},
  {Ghisellini}, {Pian}, {Masetti}, \& {Buzzoni}}]{Fugazza2003}
{Fugazza}, D., {Antonelli}, L.~A., {Fiore}, F., {et~al.} 2003, GRB Coordinates
  Network, 1982, 1

\bibitem[{{Fujiwara} {et~al.}(2014){Fujiwara}, {Yoshii}, {Saito}, {Tachibana},
  {Ohuchi}, {Kurita}, {Ono}, {Yatsu}, \& {Kawai}}]{Fujiwara2014}
{Fujiwara}, T., {Yoshii}, T., {Saito}, Y., {et~al.} 2014, GRB Coordinates
  Network, 16173, 1

\bibitem[{{Fynbo} {et~al.}(2009){Fynbo}, {Jakobsson}, {Prochaska}, {Malesani},
  {Ledoux}, {de Ugarte Postigo}, {Nardini}, {Vreeswijk}, {Wiersema}, {Hjorth},
  {Sollerman}, {Chen}, {Th{\"o}ne}, {Bj{\"o}rnsson}, {Bloom}, {Castro-Tirado},
  {Christensen}, {De Cia}, {Fruchter}, {Gorosabel}, {Graham}, {Jaunsen},
  {Jensen}, {Kann}, {Kouveliotou}, {Levan}, {Maund}, {Masetti},
  {Milvang-Jensen}, {Palazzi}, {Perley}, {Pian}, {Rol}, {Schady}, {Starling},
  {Tanvir}, {Watson}, {Xu}, {Augusteijn}, {Grundahl}, {Telting}, \&
  {Quirion}}]{Fynbo2009}
{Fynbo}, J.~P.~U., {Jakobsson}, P., {Prochaska}, J.~X., {et~al.} 2009, \apjs,
  185, 526

\bibitem[{{Fynbo} {et~al.}(2014){Fynbo}, {Kr{\"u}hler}, {Leighly}, {Ledoux},
  {Vreeswijk}, {Schulze}, {Noterdaeme}, {Watson}, {Wijers}, {Bolmer}, {Cano},
  {Christensen}, {Covino}, {D'Elia}, {Flores}, {Friis}, {Goldoni}, {Greiner},
  {Hammer}, {Hjorth}, {Jakobsson}, {Japelj}, {Kaper}, {Klose}, {Knust},
  {Leloudas}, {Levan}, {Malesani}, {Milvang-Jensen}, {M{\o}ller}, {Nicuesa
  Guelbenzu}, {Oates}, {Pian}, {Schady}, {Sparre}, {Tagliaferri}, {Tanvir},
  {Th{\"o}ne}, {de Ugarte Postigo}, {Vergani}, {Wiersema}, {Xu}, \&
  {Zafar}}]{Fynbo2014}
{Fynbo}, J.~P.~U., {Kr{\"u}hler}, T., {Leighly}, K., {et~al.} 2014, \aap, 572,
  A12

\bibitem[{{Galama} {et~al.}(1999){Galama}, {Vreeswijk}, {Rol}, {Kaper},
  {Masetti}, {Pian}, {Palazzi}, {Frontera}, {van Paradijs}, \&
  {Kouveliotou}}]{Galama1999}
{Galama}, T.~J., {Vreeswijk}, P.~M., {Rol}, E., {et~al.} 1999, GRB Coordinates
  Network, 388, 1

\bibitem[{{Galeev} {et~al.}(2009){Galeev}, {Bikmaev}, {Sakhibullin}, {Burenin},
  {Pavlinsky}, {Sunyaev}, {Khamitov}, {Eker}, {Kiziloglu}, \&
  {Gogus}}]{Galeev2009}
{Galeev}, A., {Bikmaev}, I., {Sakhibullin}, N., {et~al.} 2009, GRB Coordinates
  Network, 9548, 1

\bibitem[{{Galeev} {et~al.}(2012{\natexlab{a}}){Galeev}, {Khamitov}, {Bikmaev},
  {Sakhibullin}, {Burenin}, {Pavlinsky}, {Sunyaev}, {Eker}, {Kiziloglu}, \&
  {Gogus}}]{Galeev2012a}
{Galeev}, A., {Khamitov}, I., {Bikmaev}, I., {et~al.} 2012{\natexlab{a}}, GRB
  Coordinates Network, 13626, 1

\bibitem[{{Galeev} {et~al.}(2012{\natexlab{b}}){Galeev}, {Khamitov}, {Bikmaev},
  {Sakhibullin}, {Burenin}, {Pavlinsky}, {Sunyaev}, {Eker}, {Kiziloglu}, \&
  {Gogus}}]{Galeev2012b}
{Galeev}, A., {Khamitov}, I., {Bikmaev}, I., {et~al.} 2012{\natexlab{b}}, GRB
  Coordinates Network, 13636, 1

\bibitem[{{Garnavich} \& {Quinn}(2002)}]{Garnavich2002}
{Garnavich}, P. \& {Quinn}, J. 2002, GRB Coordinates Network, 1661, 1

\bibitem[{{Gendre} {et~al.}(2012){Gendre}, {Atteia}, {Bo{\"e}r}, {Colas},
  {Klotz}, {Kugel}, {Laas-Bourez}, {Rinner}, {Strajnic}, {Stratta}, \&
  {Vachier}}]{Gendre2012}
{Gendre}, B., {Atteia}, J.~L., {Bo{\"e}r}, M., {et~al.} 2012, \apj, 748, 59

\bibitem[{{Gendre} {et~al.}(2010){Gendre}, {Klotz}, {Palazzi}, {Kr{\"u}hler},
  {Covino}, {Afonso}, {Antonelli}, {Atteia}, {D'Avanzo}, {Bo{\"e}r}, {Greiner},
  \& {Klose}}]{Gendre2010}
{Gendre}, B., {Klotz}, A., {Palazzi}, E., {et~al.} 2010, \mnras, 405, 2372

\bibitem[{{Ghirlanda} {et~al.}(2005){Ghirlanda}, {Ghisellini}, \&
  {Firmani}}]{Ghirlanda2005}
{Ghirlanda}, G., {Ghisellini}, G., \& {Firmani}, C. 2005, \mnras, 361, L10

\bibitem[{{Goldstein}(2012)}]{Goldstein2012b}
{Goldstein}, A. 2012, in Gamma-Ray Bursts 2012 Conference (GRB 2012)

\bibitem[{{Gomboc} {et~al.}(2008{\natexlab{a}}){Gomboc}, {Guidorzi},
  {Melandri}, {Steele}, {Mundell}, {Bersier}, {Bode}, {Burgdorf}, {Fraser},
  {Kobayashi}, {Mottram}, {Smith}, {O'Brien}, {Bannister}, \&
  {Tanvir}}]{Gomboc2008a}
{Gomboc}, A., {Guidorzi}, C., {Melandri}, A., {et~al.} 2008{\natexlab{a}}, GRB
  Coordinates Network, 7625, 1

\bibitem[{{Gomboc} {et~al.}(2008{\natexlab{b}}){Gomboc}, {Guidorzi},
  {Melandri}, {Steele}, {Mundell}, {Bersier}, {Bode}, {Burgdorf}, {Fraser},
  {Kobayashi}, {Mottram}, {Smith}, {O'Brien}, {Bannister}, \&
  {Tanvir}}]{Gomboc2008b}
{Gomboc}, A., {Guidorzi}, C., {Melandri}, A., {et~al.} 2008{\natexlab{b}}, GRB
  Coordinates Network, 7626, 1

\bibitem[{{Gomboc} {et~al.}(2008{\natexlab{c}}){Gomboc}, {Melandri}, {Smith},
  {Mundell}, {Steele}, {Bersier}, {Kobayashi}, {Carter}, {Burgdorf}, {Bode}, \&
  {Guidorzi}}]{Gomboc2008c}
{Gomboc}, A., {Melandri}, A., {Smith}, R.~J., {et~al.} 2008{\natexlab{c}}, GRB
  Coordinates Network, 7831, 1

\bibitem[{{Gorbovskoy} {et~al.}(2013){Gorbovskoy}, {Lipunov}, {Denisenko},
  {Kornilov}, {Kuznetsov}, {Kuvshinov}, {Tyurina}, {Shatskiy}, {Balanutsa},
  {Zimnukhov}, {Chazov}, {Tlatov}, {Parhomenko}, {Dormidontov}, {Sennik},
  {Ivanov}, {Yazev}, {Budnev}, {Gres}, {Chuvalaev}, {Poleshchuk}, {Yurkov},
  {Sergienko}, {Varda}, {Sinyakov}, {Krushinski}, {Zalozhnich}, {Popov},
  {Bourdanov}, {Levato}, {Saffe}, {Mallamaci}, {Lopez}, \&
  {Podest}}]{Gorbovskoy2013}
{Gorbovskoy}, E., {Lipunov}, V., {Denisenko}, D., {et~al.} 2013, GRB
  Coordinates Network, 15405, 1

\bibitem[{{Gorbovskoy} {et~al.}(2009){Gorbovskoy}, {Lipunov}, {Kornilov},
  {Belinski}, {Shatskiy}, {Tyurina}, {Kuvshinov}, {Balanutsa}, {Chazov},
  {Kortunov}, {Kuznetsov}, {Zemnukhov}, {Kornilov}, {Krushinski}, {Zalozhnih},
  {Kopytova}, {Tlatov}, {Parhomenko}, {Dormidontov}, {Ivanov}, {Yazev},
  {Budnev}, {Konstantinov}, {Lenok}, {Sergienko}, \& {Yurkov}}]{Gorbovskoy2009}
{Gorbovskoy}, E., {Lipunov}, V., {Kornilov}, V., {et~al.} 2009, GRB Coordinates
  Network, 10231, 1

\bibitem[{{Gorbovskoy} {et~al.}(2014){Gorbovskoy}, {Lipunov}, {Kornilov},
  {Kuvshinov}, {Belinski}, {Tyurina}, {Shatskiy}, {Balanutsa}, {Zimnukhov},
  {Kuznetsov}, {Chazov}, {Denisenko}, {Sankovich}, {Ivanov}, {Yazev}, {Budnev},
  {Gres}, {Chuvalaev}, {Poleshchuk}, {Yurkov}, {Varda}, {Sinyakov}, {Tlatov},
  {Parhomenko}, {Dormidontov}, {Sennik}, {Krushinsky}, {Zalozhnih}, {Popov},
  {Levato}, {Saffe}, {Mallamaci}, {Lopez}, \& {Podest}}]{Gorbovskoy2014}
{Gorbovskoy}, E., {Lipunov}, V., {Kornilov}, V., {et~al.} 2014, GRB Coordinates
  Network, 16250, 1

\bibitem[{{Gorosabel} {et~al.}(2006){Gorosabel}, {Castro-Tirado},
  {Ramirez-Ruiz}, {Granot}, {Caon}, {Cair{\'o}s}, {Rubio-Herrera}, {Guziy}, {de
  Ugarte Postigo}, \& {Jel{\'{\i}}nek}}]{Gorosabel2006}
{Gorosabel}, J., {Castro-Tirado}, A.~J., {Ramirez-Ruiz}, E., {et~al.} 2006,
  \apjl, 641, L13

\bibitem[{{Gorosabel} {et~al.}(2009{\natexlab{a}}){Gorosabel}, {de Ugarte
  Postigo}, {Montes}, {Klutsch}, \& {Castro-Tirado}}]{Gorosabel2009b}
{Gorosabel}, J., {de Ugarte Postigo}, A., {Montes}, D., {Klutsch}, A., \&
  {Castro-Tirado}, A.~J. 2009{\natexlab{a}}, GRB Coordinates Network, 9379, 1

\bibitem[{{Gorosabel} {et~al.}(2002){Gorosabel}, {Fynbo}, {Hjorth}, {Pedersen},
  {Jensen}, {Holland}, {Castro Cer{\'o}n}, {Andersen}, \&
  {Castro-Tirado}}]{Gorosabel2002}
{Gorosabel}, J., {Fynbo}, J.~P.~U., {Hjorth}, J., {et~al.} 2002, GRB
  Coordinates Network, 1651, 1

\bibitem[{{Gorosabel} {et~al.}(2009{\natexlab{b}}){Gorosabel}, {Kubanek},
  {Jelinek}, {de Ugarte Postigo}, \& {Aceituno}}]{Gorosabel2009a}
{Gorosabel}, J., {Kubanek}, P., {Jelinek}, M., {de Ugarte Postigo}, A., \&
  {Aceituno}, J. 2009{\natexlab{b}}, GRB Coordinates Network, 9236, 1

\bibitem[{{Graham} {et~al.}(2014){Graham}, {Varela}, {Delvaux}, \&
  {Greiner}}]{Graham2014}
{Graham}, J., {Varela}, K., {Delvaux}, C., \& {Greiner}, J. 2014, GRB
  Coordinates Network, 16257, 1

\bibitem[{{Granot} \& {Sari}(2002)}]{Granot2002}
{Granot}, J. \& {Sari}, R. 2002, \apj, 568, 820

\bibitem[{{Greiner} {et~al.}(2011){Greiner}, {Kr{\"u}hler}, {Klose}, {Afonso},
  {Clemens}, {Filgas}, {Hartmann}, {K{\"u}pc{\"u} Yolda{\c s}}, {Nardini},
  {Olivares E.}, {Rau}, {Rossi}, {Schady}, \& {Updike}}]{Greiner2011}
{Greiner}, J., {Kr{\"u}hler}, T., {Klose}, S., {et~al.} 2011, \aap, 526, A30

\bibitem[{{Gruber} {et~al.}(2014){Gruber}, {Goldstein}, {Weller von Ahlefeld},
  {Narayana Bhat}, {Bissaldi}, {Briggs}, {Byrne}, {Cleveland}, {Connaughton},
  {Diehl}, {Fishman}, {Fitzpatrick}, {Foley}, {Gibby}, {Giles}, {Greiner},
  {Guiriec}, {van der Horst}, {von Kienlin}, {Kouveliotou}, {Layden}, {Lin},
  {Meegan}, {McGlynn}, {Paciesas}, {Pelassa}, {Preece}, {Rau}, {Wilson-Hodge},
  {Xiong}, {Younes}, \& {Yu}}]{Gruber2014}
{Gruber}, D., {Goldstein}, A., {Weller von Ahlefeld}, V., {et~al.} 2014, \apjs,
  211, 12

\bibitem[{{Gruber et al.}(2012)}]{Gruber2012}
{Gruber et al.} 2012, ArXiv e-prints, arXiv:1207.4620

\bibitem[{{Guidorzi} \& {Melandri}(2013)}]{Guidorzi2013b}
{Guidorzi}, C. \& {Melandri}, A. 2013, GRB Coordinates Network, 15140, 1

\bibitem[{{Guidorzi} {et~al.}(2013){Guidorzi}, {Melandri}, {Kopac}, \&
  {Gomboc}}]{Guidorzi2013a}
{Guidorzi}, C., {Melandri}, A., {Kopac}, D., \& {Gomboc}, A. 2013, GRB
  Coordinates Network, 14405, 1

\bibitem[{{Guidorzi} {et~al.}(2010){Guidorzi}, {Smith}, {Tanvir}, {Steele},
  {Kobayashi}, {Mundell}, {Bersier}, {Cano}, {Clay}, {Melandri}, {Mottram}, \&
  {Gomboc}}]{Guidorzi2010}
{Guidorzi}, C., {Smith}, R.~J., {Tanvir}, N., {et~al.} 2010, GRB Coordinates
  Network, 10589, 1

\bibitem[{{Guidorzi} \& {Steele}(2008)}]{Guidorzi2008}
{Guidorzi}, C. \& {Steele}, I. 2008, GRB Coordinates Network, 8064, 1

\bibitem[{{Guidorzi} {et~al.}(2009){Guidorzi}, {Steele}, {Melandri}, {Bersier},
  {Mottram}, {Mundell}, {Smith}, {Gomboc}, {O'Brien}, {Bannister}, \&
  {Tanvir}}]{Guidorzi2009}
{Guidorzi}, C., {Steele}, I.~A., {Melandri}, A., {et~al.} 2009, GRB Coordinates
  Network, 9375, 1

\bibitem[{{Guver}(2013)}]{Guver2013}
{Guver}, T. 2013, GRB Coordinates Network, 14407, 1

\bibitem[{{Guver} {et~al.}(2014){Guver}, {Ferrante}, {Flewelling}, {Kehoe}, \&
  {Dhungana}}]{Guver2014}
{Guver}, T., {Ferrante}, F.~V., {Flewelling}, H., {Kehoe}, R., \& {Dhungana},
  G. 2014, GRB Coordinates Network, 16120, 1

\bibitem[{{Guziy} {et~al.}(2014){Guziy}, {Gorosabel}, {Castro-Tirado},
  {Cunniffe}, {Jelinek}, {Jeong}, {Lara-Gil}, {Sanchez-Ramirez}, {Tello},
  {Kubanek}, {Pandey}, {Fan}, {Zhao}, {Bai}, {Wang}, {Xin}, \&
  {Cui}}]{Guziy2014}
{Guziy}, S., {Gorosabel}, J., {Castro-Tirado}, A.~J., {et~al.} 2014, GRB
  Coordinates Network, 15685, 1

\bibitem[{{Haislip} {et~al.}(2005){Haislip}, {MacLeod}, {Nysewander}, {Foster},
  {Crain}, {Reichart}, {Bayliss}, {Kirschbrown}, \& {Mack}}]{Haislip2005}
{Haislip}, J., {MacLeod}, C., {Nysewander}, M., {et~al.} 2005, GRB Coordinates
  Network, 3568, 1

\bibitem[{{Haislip} {et~al.}(2009{\natexlab{a}}){Haislip}, {Reichart},
  {Ivarsen}, {Cominsky}, {McLin}, {Graves}, {Spear}, {Lacluyze}, {Foster},
  {Moore}, {Oza}, {Schubel}, {Styblova}, {Trotter}, {Crain}, \&
  {Nysewander}}]{Haislip2009a}
{Haislip}, J., {Reichart}, D., {Ivarsen}, K., {et~al.} 2009{\natexlab{a}}, GRB
  Coordinates Network, 9999, 1

\bibitem[{{Haislip} {et~al.}(2009{\natexlab{b}}){Haislip}, {Reichart},
  {Ivarsen}, {Lacluyze}, {Cominsky}, {McLin}, {Graves}, {Spear}, {Egger},
  {Foster}, {Moore}, {Oza}, {Schubel}, {Styblova}, {Trotter}, {Crain}, \&
  {Nysewander}}]{Haislip2009b}
{Haislip}, J., {Reichart}, D., {Ivarsen}, K., {et~al.} 2009{\natexlab{b}}, GRB
  Coordinates Network, 10219, 1

\bibitem[{{Haislip} {et~al.}(2009{\natexlab{c}}){Haislip}, {Reichart},
  {Ivarsen}, {Lacluyze}, {Egger}, {Foster}, {Moore}, {Oza}, {Schubel},
  {Styblova}, {Trotter}, {Crain}, \& {Nysewander}}]{Haislip2009c}
{Haislip}, J., {Reichart}, D., {Ivarsen}, K., {et~al.} 2009{\natexlab{c}}, GRB
  Coordinates Network, 10230, 1

\bibitem[{{Haislip} {et~al.}(2009{\natexlab{d}}){Haislip}, {Reichart},
  {Ivarsen}, {Lacluyze}, {Egger}, {Foster}, {Moore}, {Oza}, {Schubel},
  {Styblova}, {Trotter}, {Crain}, \& {Nysewander}}]{Haislip2009d}
{Haislip}, J., {Reichart}, D., {Ivarsen}, K., {et~al.} 2009{\natexlab{d}}, GRB
  Coordinates Network, 10294, 1

\bibitem[{{Halpern} \& {Armstrong}(2006{\natexlab{a}})}]{Halpern2006c}
{Halpern}, J. \& {Armstrong}, E. 2006{\natexlab{a}}, GRB Coordinates Network,
  5853, 1

\bibitem[{{Halpern} \& {Armstrong}(2006{\natexlab{b}})}]{Halpern2006b}
{Halpern}, J. \& {Armstrong}, E. 2006{\natexlab{b}}, GRB Coordinates Network,
  5851, 1

\bibitem[{{Halpern} {et~al.}(2006){Halpern}, {Mirabal}, \&
  {Armstrong}}]{Halpern2006a}
{Halpern}, J.~P., {Mirabal}, N., \& {Armstrong}, E. 2006, GRB Coordinates
  Network, 5847, 1

\bibitem[{{Harbeck} {et~al.}(2014{\natexlab{a}}){Harbeck}, {Kaur},
  {Delgado-Navarro}, {Orio}, \& {Hartmann}}]{Harbeck2014a}
{Harbeck}, D., {Kaur}, A., {Delgado-Navarro}, A., {Orio}, M., \& {Hartmann},
  D.~H. 2014{\natexlab{a}}, GRB Coordinates Network, 16165, 1

\bibitem[{{Harbeck} {et~al.}(2014{\natexlab{b}}){Harbeck}, {Kaur},
  {Delgado-Navarro}, {Orio}, \& {Hartmann}}]{Harbeck2014b}
{Harbeck}, D., {Kaur}, A., {Delgado-Navarro}, A., {Orio}, M., \& {Hartmann},
  D.~H. 2014{\natexlab{b}}, GRB Coordinates Network, 16175, 1

\bibitem[{{Harrison} {et~al.}(1999){Harrison}, {Bloom}, {Frail}, {Sari},
  {Kulkarni}, {Djorgovski}, {Axelrod}, {Mould}, {Schmidt}, {Wieringa}, {Wark},
  {Subrahmanyan}, {McConnell}, {McCarthy}, {Schaefer}, {McMahon}, {Markze},
  {Firth}, {Soffitta}, \& {Amati}}]{Harrison1999}
{Harrison}, F.~A., {Bloom}, J.~S., {Frail}, D.~A., {et~al.} 1999, \apjl, 523,
  L121

\bibitem[{{Henden} {et~al.}(2009{\natexlab{a}}){Henden}, {Gross}, {Denny},
  {Terrell}, \& {Cooney}}]{Henden2009b}
{Henden}, A., {Gross}, J., {Denny}, B., {Terrell}, D., \& {Cooney}, W.
  2009{\natexlab{a}}, GRB Coordinates Network, 9220, 1

\bibitem[{{Henden} {et~al.}(2009{\natexlab{b}}){Henden}, {Gross}, {Denny},
  {Terrell}, \& {Cooney}}]{Henden2009a}
{Henden}, A., {Gross}, J., {Denny}, B., {Terrell}, D., \& {Cooney}, W.
  2009{\natexlab{b}}, GRB Coordinates Network, 9211, 1

\bibitem[{{Hentunen}(2007)}]{Hentunen2007}
{Hentunen}, V.-P. 2007, GRB Coordinates Network, 6981, 1

\bibitem[{{Hentunen} \& {Nissinen}(2013)}]{Hentunen2013c}
{Hentunen}, V.-P. \& {Nissinen}, M. 2013, GRB Coordinates Network, 14891, 1

\bibitem[{{Hentunen} {et~al.}(2011{\natexlab{a}}){Hentunen}, {Nissinen}, \&
  {Salmi}}]{Hentunen2011a}
{Hentunen}, V.-P., {Nissinen}, M., \& {Salmi}, T. 2011{\natexlab{a}}, GRB
  Coordinates Network, 11709, 1

\bibitem[{{Hentunen} {et~al.}(2011{\natexlab{b}}){Hentunen}, {Nissinen}, \&
  {Salmi}}]{Hentunen2011c}
{Hentunen}, V.-P., {Nissinen}, M., \& {Salmi}, T. 2011{\natexlab{b}}, GRB
  Coordinates Network, 11966, 1

\bibitem[{{Hentunen} {et~al.}(2012){Hentunen}, {Nissinen}, \&
  {Salmi}}]{Hentunen2012}
{Hentunen}, V.-P., {Nissinen}, M., \& {Salmi}, T. 2012, GRB Coordinates
  Network, 13119, 1

\bibitem[{{Hentunen} {et~al.}(2013{\natexlab{a}}){Hentunen}, {Nissinen}, \&
  {Salmi}}]{Hentunen2013a}
{Hentunen}, V.-P., {Nissinen}, M., \& {Salmi}, T. 2013{\natexlab{a}}, GRB
  Coordinates Network, 14410, 1

\bibitem[{{Hentunen} {et~al.}(2013{\natexlab{b}}){Hentunen}, {Nissinen}, \&
  {Salmi}}]{Hentunen2013b}
{Hentunen}, V.-P., {Nissinen}, M., \& {Salmi}, T. 2013{\natexlab{b}}, GRB
  Coordinates Network, 14572, 1

\bibitem[{{Hentunen} {et~al.}(2013{\natexlab{c}}){Hentunen}, {Nissinen}, \&
  {Salmi}}]{Hentunen2013e}
{Hentunen}, V.-P., {Nissinen}, M., \& {Salmi}, T. 2013{\natexlab{c}}, GRB
  Coordinates Network, 15418, 1

\bibitem[{{Hentunen} {et~al.}(2014){Hentunen}, {Nissinen}, \&
  {Salmi}}]{Hentunen2014}
{Hentunen}, V.-P., {Nissinen}, M., \& {Salmi}, T. 2014, GRB Coordinates
  Network, 16126, 1

\bibitem[{{Hentunen} {et~al.}(2011{\natexlab{c}}){Hentunen}, {Nissinen},
  {Salmi}, \& {Vilokki}}]{Hentunen2011b}
{Hentunen}, V.-P., {Nissinen}, M., {Salmi}, T., \& {Vilokki}, H.
  2011{\natexlab{c}}, GRB Coordinates Network, 11717, 1

\bibitem[{{Hentunen} {et~al.}(2013{\natexlab{d}}){Hentunen}, {Nissinen},
  {Vilokki}, \& {Salmi}}]{Hentunen2013d}
{Hentunen}, V.-P., {Nissinen}, M., {Vilokki}, H., \& {Salmi}, T.
  2013{\natexlab{d}}, GRB Coordinates Network, 15156, 1

\bibitem[{{Heussaff} {et~al.}(2013){Heussaff}, {Atteia}, \&
  {Zolnierowski}}]{Heussaff2013}
{Heussaff}, V., {Atteia}, J.-L., \& {Zolnierowski}, Y. 2013, \aap, 557, A100

\bibitem[{{Hjorth} {et~al.}(2012){Hjorth}, {Malesani}, {Jakobsson}, {Jaunsen},
  {Fynbo}, {Gorosabel}, {Kr{\"u}hler}, {Levan}, {Micha{\l}owski},
  {Milvang-Jensen}, {M{\o}ller}, {Schulze}, {Tanvir}, \& {Watson}}]{Hjorth2012}
{Hjorth}, J., {Malesani}, D., {Jakobsson}, P., {et~al.} 2012, \apj, 756, 187

\bibitem[{{Hjorth} {et~al.}(2013){Hjorth}, {Melandri}, {Malesani}, {Kruehler},
  \& {Xu}}]{Hjorth2013}
{Hjorth}, J., {Melandri}, A., {Malesani}, D., {Kruehler}, T., \& {Xu}, D. 2013,
  GRB Coordinates Network, 14365, 1

\bibitem[{{Hjorth} {et~al.}(2003){Hjorth}, {M{\o}ller}, {Gorosabel}, {Fynbo},
  {Toft}, {Jaunsen}, {Kaas}, {Pursimo}, {Torii}, {Kato}, {Yamaoka}, {Yoshida},
  {Thomsen}, {Andersen}, {Burud}, {Castro Cer{\'o}n}, {Castro-Tirado},
  {Fruchter}, {Kaper}, {Kouveliotou}, {Masetti}, {Palazzi}, {Pedersen}, {Pian},
  {Rhoads}, {Rol}, {Tanvir}, {Vreeswijk}, {Wijers}, \& {van den
  Heuvel}}]{Hjorth2003}
{Hjorth}, J., {M{\o}ller}, P., {Gorosabel}, J., {et~al.} 2003, \apj, 597, 699

\bibitem[{{Holland} {et~al.}(2004){Holland}, {Bersier}, {Bloom}, {Garnavich},
  {Caldwell}, {Challis}, {Kirshner}, {Luhman}, {McLeod}, \&
  {Stanek}}]{Holland2004}
{Holland}, S.~T., {Bersier}, D., {Bloom}, J.~S., {et~al.} 2004, \aj, 128, 1955

\bibitem[{{Holland} {et~al.}(2003){Holland}, {Weidinger}, {Fynbo}, {Gorosabel},
  {Hjorth}, {Pedersen}, {M{\'e}ndez Alvarez}, {Augusteijn}, {Castro Cer{\'o}n},
  {Castro-Tirado}, {Dahle}, {Egholm}, {Jakobsson}, {Jensen}, {Levan},
  {M{\o}ller}, {Pedersen}, {Pursimo}, {Ruiz-Lapuente}, \&
  {Thomsen}}]{Holland2003}
{Holland}, S.~T., {Weidinger}, M., {Fynbo}, J.~P.~U., {et~al.} 2003, \aj, 125,
  2291

\bibitem[{{Huang} {et~al.}(2005){Huang}, {Urata}, {Filippenko}, {Hu}, {Ip},
  {Kuo}, {Li}, {Lin}, {Lin}, {Makishima}, {Onda}, {Qiu}, \&
  {Tamagawa}}]{Huang2005}
{Huang}, K.~Y., {Urata}, Y., {Filippenko}, A.~V., {et~al.} 2005, \apjl, 628,
  L93

\bibitem[{{Huang} {et~al.}(2008){Huang}, {Chen}, {Chen}, {Huang}, \&
  {Urata}}]{Huang2008}
{Huang}, L.-C., {Chen}, T.-W., {Chen}, Y.-T., {Huang}, K.~Y., \& {Urata}, Y.
  2008, GRB Coordinates Network, 7999, 1

\bibitem[{{Ibrahimov} {et~al.}(2003){Ibrahimov}, {Asfandiyarov}, {Kahharov},
  {Pozanenko}, {Rumyantsev}, \& {Beskin}}]{Ibrahimov2003}
{Ibrahimov}, M.~A., {Asfandiyarov}, I.~M., {Kahharov}, B.~B., {et~al.} 2003,
  GRB Coordinates Network, 2192, 1

\bibitem[{{Im} \& {Choi}(2013)}]{Im2013b}
{Im}, M. \& {Choi}, C. 2013, GRB Coordinates Network, 15432, 1

\bibitem[{{Im} {et~al.}(2007){Im}, {Lee}, \& {Urata}}]{Im2007}
{Im}, M., {Lee}, I., \& {Urata}, Y. 2007, GRB Coordinates Network, 6970, 1

\bibitem[{{Im} {et~al.}(2009{\natexlab{a}}){Im}, {Park}, {Jeon}, {Lee}, {Jeon},
  \& {Urata}}]{Im2009a}
{Im}, M., {Park}, W., {Jeon}, Y., {et~al.} 2009{\natexlab{a}}, GRB Coordinates
  Network, 9248, 1

\bibitem[{{Im} {et~al.}(2009{\natexlab{b}}){Im}, {Park}, \& {Urata}}]{Im2009b}
{Im}, M., {Park}, W.~K., \& {Urata}, Y. 2009{\natexlab{b}}, GRB Coordinates
  Network, 9522, 1

\bibitem[{{Im} {et~al.}(2011){Im}, {Sung}, \& {Urata}}]{Im2011}
{Im}, M., {Sung}, H.-I., \& {Urata}, Y. 2011, GRB Coordinates Network, 12004, 1

\bibitem[{{Im} {et~al.}(2013){Im}, {Sung}, \& {Urata}}]{Im2013a}
{Im}, M., {Sung}, H.-I., \& {Urata}, Y. 2013, GRB Coordinates Network, 14854, 1

\bibitem[{{Ishimura} {et~al.}(2007){Ishimura}, {Yatsu}, {Shimokawabe}, {Kawai},
  \& {Yoshida}}]{Ishimura2007}
{Ishimura}, T., {Yatsu}, Y., {Shimokawabe}, T., {Kawai}, N., \& {Yoshida}, M.
  2007, GRB Coordinates Network, 7026, 1

\bibitem[{{Ivanov} {et~al.}(2014){Ivanov}, {Yazev}, {Budnev}, {Gres},
  {Chuvalaev}, {Poleshchuk}, {Gorbovskoy}, {Lipunov}, {Pruzhinskaya},
  {Kornilov}, {Kuvshinov}, {Tyurina}, {Balanutsa}, {Kuznetsov}, {Chazov},
  {Denisenko}, {Yurkov}, {Sergienko}, {Varda}, {Sinyakov}, {Gabovich},
  {Krushinski}, {Zalozhnih}, {Popov}, {Tlatov}, {Parhomenko}, {Dormidontov},
  {Sennik}, {Levato}, {Saffe}, {Mallamaci}, {Lopez}, \& {Podest}}]{Ivanov2014}
{Ivanov}, K., {Yazev}, S., {Budnev}, N.~M., {et~al.} 2014, GRB Coordinates
  Network, 16552, 1

\bibitem[{{Ivanov} {et~al.}(2010){Ivanov}, {Yazev}, {Budnev}, {Konstantinov},
  {Lenok}, {Gres}, {Chuvalaev}, {Yurkov}, {Sergienko}, {Varda}, {Garusina},
  {Gorbovskoy}, {Lipunov}, {Kornilov}, {Belinski}, {Shatskiy}, {Tyurina},
  {Kuvshinov}, {Balanutsa}, {Chazov}, {Kortunov}, {Kuznetsov}, {Zemnukhov},
  {Kornilov}, {Tlatov}, {Parhomenko}, {Dormidontov}, {Krushinski}, {Zalozhnih},
  {Kopytova}, \& {Popov}}]{Ivanov2010}
{Ivanov}, K., {Yazev}, S., {Budnev}, N.~M., {et~al.} 2010, GRB Coordinates
  Network, 10582, 1

\bibitem[{{Izzo} \& {D'Avino}(2013)}]{Izzo2013}
{Izzo}, L. \& {D'Avino}, L. 2013, GRB Coordinates Network, 15153, 1

\bibitem[{{Jakobsson} {et~al.}(2005){Jakobsson}, {Fynbo}, {Paraficz},
  {Telting}, {Jensen}, {Hjorth}, \& {Castro Cer{\'o}n}}]{Jakobsson2005}
{Jakobsson}, P., {Fynbo}, J.~P.~U., {Paraficz}, D., {et~al.} 2005, GRB
  Coordinates Network, 4029, 1

\bibitem[{{Jakobsson} {et~al.}(2008{\natexlab{a}}){Jakobsson}, {Malesani},
  {Vreeswijk}, {Fynbo}, {Milvang-Jensen}, {de Ugarte Postigo}, {Nordstrom},
  {Stonkute}, \& {Sorensen}}]{Jakobsson2008b}
{Jakobsson}, P., {Malesani}, D., {Vreeswijk}, P.~M., {et~al.}
  2008{\natexlab{a}}, GRB Coordinates Network, 7998, 1

\bibitem[{{Jakobsson} {et~al.}(2007){Jakobsson}, {Vreeswijk}, {Hjorth},
  {Malesani}, {Fynbo}, \& {Thoene}}]{Jakobsson2007}
{Jakobsson}, P., {Vreeswijk}, P.~M., {Hjorth}, J., {et~al.} 2007, GRB
  Coordinates Network, 6952, 1

\bibitem[{{Jakobsson} {et~al.}(2008{\natexlab{b}}){Jakobsson}, {Vreeswijk},
  {Xu}, \& {Thoene}}]{Jakobsson2008a}
{Jakobsson}, P., {Vreeswijk}, P.~M., {Xu}, D., \& {Thoene}, C.~C.
  2008{\natexlab{b}}, GRB Coordinates Network, 7832, 1

\bibitem[{{Jang} {et~al.}(2012){Jang}, {Im}, \& {Urata}}]{Jang2012}
{Jang}, M., {Im}, M., \& {Urata}, Y. 2012, GRB Coordinates Network, 13139, 1

\bibitem[{{Japelj} {et~al.}(2012){Japelj}, {Kopac}, {Guidorzi}, {Mundell}, \&
  {Virgili}}]{Japelj2012}
{Japelj}, J., {Kopac}, D., {Guidorzi}, C., {Mundell}, C., \& {Virgili}, F.
  2012, GRB Coordinates Network, 14058, 1

\bibitem[{{Japelj} {et~al.}(2014){Japelj}, {Kopa{\v c}}, {Kobayashi},
  {Harrison}, {Guidorzi}, {Virgili}, {Mundell}, {Melandri}, \&
  {Gomboc}}]{Japelj2014}
{Japelj}, J., {Kopa{\v c}}, D., {Kobayashi}, S., {et~al.} 2014, \apj, 785, 84

\bibitem[{{Jelinek} \& {Kubanek}(2009)}]{Jelinek2009}
{Jelinek}, M. \& {Kubanek}, P. 2009, GRB Coordinates Network, 9404, 1

\bibitem[{{Jeon} {et~al.}(2011){Jeon}, {Im}, {Pak}, \& {Jeong}}]{Jeon2011}
{Jeon}, Y., {Im}, M., {Pak}, S., \& {Jeong}, H. 2011, GRB Coordinates Network,
  11967, 1

\bibitem[{{Jin} {et~al.}(2013){Jin}, {Covino}, {Della Valle}, {Ferrero},
  {Fugazza}, {Malesani}, {Melandri}, {Pian}, {Salvaterra}, {Bersier},
  {Campana}, {Cano}, {Castro-Tirado}, {D'Avanzo}, {Fynbo}, {Gomboc},
  {Gorosabel}, {Guidorzi}, {Haislip}, {Hjorth}, {Kobayashi}, {LaCluyze},
  {Marconi}, {Mazzali}, {Mundell}, {Piranomonte}, {Reichart},
  {S{\'a}nchez-Ram{\'{\i}}rez}, {Smith}, {Steele}, {Tagliaferri}, {Tanvir},
  {Valenti}, {Vergani}, {Vestrand}, {Walker}, \& {Wo{\'z}niak}}]{Jin2013}
{Jin}, Z.-P., {Covino}, S., {Della Valle}, M., {et~al.} 2013, \apj, 774, 114

\bibitem[{{Kaneko} {et~al.}(2007){Kaneko}, {Ramirez-Ruiz}, {Granot},
  {Kouveliotou}, {Woosley}, {Patel}, {Rol}, {in 't Zand}, {van der Horst},
  {Wijers}, \& {Strom}}]{Kaneko2007}
{Kaneko}, Y., {Ramirez-Ruiz}, E., {Granot}, J., {et~al.} 2007, \apj, 654, 385

\bibitem[{{Kann} {et~al.}(2007){Kann}, {Hoegner}, \& {Filgas}}]{Kann2007}
{Kann}, D.~A., {Hoegner}, C., \& {Filgas}, R. 2007, GRB Coordinates Network,
  6918, 1

\bibitem[{{Kann} {et~al.}(2010){Kann}, {Klose}, {Zhang}, {Malesani}, {Nakar},
  {Pozanenko}, {Wilson}, {Butler}, {Jakobsson}, {Schulze}, {Andreev},
  {Antonelli}, {Bikmaev}, {Biryukov}, {B{\"o}ttcher}, {Burenin}, {Castro
  Cer{\'o}n}, {Castro-Tirado}, {Chincarini}, {Cobb}, {Covino}, {D'Avanzo},
  {D'Elia}, {Della Valle}, {de Ugarte Postigo}, {Efimov}, {Ferrero}, {Fugazza},
  {Fynbo}, {G{\aa}lfalk}, {Grundahl}, {Gorosabel}, {Gupta}, {Guziy}, {Hafizov},
  {Hjorth}, {Holhjem}, {Ibrahimov}, {Im}, {Israel}, {Je{\'l}inek}, {Jensen},
  {Karimov}, {Khamitov}, {Kizilo{\v g}lu}, {Klunko}, {Kub{\'a}nek}, {Kutyrev},
  {Laursen}, {Levan}, {Mannucci}, {Martin}, {Mescheryakov}, {Mirabal},
  {Norris}, {Ovaldsen}, {Paraficz}, {Pavlenko}, {Piranomonte}, {Rossi},
  {Rumyantsev}, {Salinas}, {Sergeev}, {Sharapov}, {Sollerman}, {Stecklum},
  {Stella}, {Tagliaferri}, {Tanvir}, {Telting}, {Testa}, {Updike}, {Volnova},
  {Watson}, {Wiersema}, \& {Xu}}]{Kann2010}
{Kann}, D.~A., {Klose}, S., {Zhang}, B., {et~al.} 2010, \apj, 720, 1513

\bibitem[{{Kann} {et~al.}(2008){Kann}, {Laux}, \& {Ertel}}]{Kann2008}
{Kann}, D.~A., {Laux}, U., \& {Ertel}, S. 2008, GRB Coordinates Network, 7845,
  1

\bibitem[{{Kann} {et~al.}(2009{\natexlab{a}}){Kann}, {Laux}, {Roeder}, \&
  {Meusinger}}]{Kann2009a}
{Kann}, D.~A., {Laux}, U., {Roeder}, M., \& {Meusinger}, H. 2009{\natexlab{a}},
  GRB Coordinates Network, 10076, 1

\bibitem[{{Kann} {et~al.}(2009{\natexlab{b}}){Kann}, {Laux}, {Roeder}, \&
  {Meusinger}}]{Kann2009b}
{Kann}, D.~A., {Laux}, U., {Roeder}, M., \& {Meusinger}, H. 2009{\natexlab{b}},
  GRB Coordinates Network, 10090, 1

\bibitem[{{Kann} \& {Manohar}(2006)}]{Kann2006}
{Kann}, D.~A. \& {Manohar}, S. 2006, GRB Coordinates Network, 5278, 1

\bibitem[{{Kann} {et~al.}(2014){Kann}, {Nicuesa Guelbenzu}, \&
  {Greiner}}]{Kann2014}
{Kann}, D.~A., {Nicuesa Guelbenzu}, A., \& {Greiner}, J. 2014, GRB Coordinates
  Network, 16926, 1

\bibitem[{{Kann} {et~al.}(2011){Kann}, {Schmidl}, {Hoegner}, {Stecklum},
  {Schumann}, \& {Hartmann}}]{Kann2011}
{Kann}, D.~A., {Schmidl}, S., {Hoegner}, C., {et~al.} 2011, GRB Coordinates
  Network, 11996, 1

\bibitem[{{Kann} {et~al.}(2013){Kann}, {Stecklum}, \& {Ludwig}}]{Kann2013}
{Kann}, D.~A., {Stecklum}, B., \& {Ludwig}, F. 2013, GRB Coordinates Network,
  14593, 1

\bibitem[{{Khamitov} {et~al.}(2009){Khamitov}, {Parmaksizoglu}, {Ak}, {Eker},
  {Aslan}, {Kultur}, {Gogus}, {Burenin}, {Pavlinsky}, {Sunyaev}, {Bikmaev}, \&
  {Sakhibullin}}]{Khamitov2009}
{Khamitov}, I., {Parmaksizoglu}, M., {Ak}, T., {et~al.} 2009, GRB Coordinates
  Network, 9597, 1

\bibitem[{{Khorunzhev} {et~al.}(2013){Khorunzhev}, {Burenin}, {Pavlinsky},
  {Sunyaev}, {Bikmaev}, {Sakhibullin}, {Khamitov}, \&
  {Kirbiyik}}]{Khorunzhev2013}
{Khorunzhev}, G., {Burenin}, R., {Pavlinsky}, M., {et~al.} 2013, GRB
  Coordinates Network, 15244, 1

\bibitem[{{Kinugasa} {et~al.}(2009{\natexlab{a}}){Kinugasa}, {Honda},
  {Hashimoto}, {Takahashi}, \& {Taguchi}}]{Kinugasa2009a}
{Kinugasa}, K., {Honda}, S., {Hashimoto}, O., {Takahashi}, H., \& {Taguchi}, H.
  2009{\natexlab{a}}, GRB Coordinates Network, 10248, 1

\bibitem[{{Kinugasa} {et~al.}(2009{\natexlab{b}}){Kinugasa}, {Honda},
  {Takahashi}, {Taguchi}, \& {Hashimoto}}]{Kinugasa2009b}
{Kinugasa}, K., {Honda}, S., {Takahashi}, H., {Taguchi}, H., \& {Hashimoto}, O.
  2009{\natexlab{b}}, GRB Coordinates Network, 10275, 1

\bibitem[{{Kiziloglu} {et~al.}(2002){Kiziloglu}, {Alpar}, {Baykal}, {Kaan},
  {Ankay}, {Huseyin}, {Terekov}, {Sunyaev}, {Tkachenko}, {Denissenko},
  {Bikmaev}, {Sakhibullin}, {Suleymanov}, \& {Khamitov}}]{Kiziloglu2002}
{Kiziloglu}, U., {Alpar}, A., {Baykal}, A., {et~al.} 2002, GRB Coordinates
  Network, 1488, 1

\bibitem[{{Klotz} {et~al.}(2006){Klotz}, {Boer}, \& {Atteia}}]{Klotz2006}
{Klotz}, A., {Boer}, M., \& {Atteia}, J.~L. 2006, GRB Coordinates Network,
  5506, 1

\bibitem[{{Klotz} {et~al.}(2008){Klotz}, {Boer}, \& {Atteia}}]{Klotz2008}
{Klotz}, A., {Boer}, M., \& {Atteia}, J.~L. 2008, GRB Coordinates Network,
  7595, 1

\bibitem[{{Klotz} {et~al.}(2005){Klotz}, {Bo{\"e}r}, {Atteia}, {Stratta},
  {Behrend}, {Malacrino}, \& {Damerdji}}]{Klotz2005}
{Klotz}, A., {Bo{\"e}r}, M., {Atteia}, J.~L., {et~al.} 2005, \aap, 439, L35

\bibitem[{{Klotz} {et~al.}(2009{\natexlab{a}}){Klotz}, {Gendre}, {Boer}, \&
  {Atteia}}]{Klotz2009a}
{Klotz}, A., {Gendre}, B., {Boer}, M., \& {Atteia}, J.~L. 2009{\natexlab{a}},
  GRB Coordinates Network, 9401, 1

\bibitem[{{Klotz} {et~al.}(2009{\natexlab{b}}){Klotz}, {Gendre}, {Boer}, \&
  {Atteia}}]{Klotz2009c}
{Klotz}, A., {Gendre}, B., {Boer}, M., \& {Atteia}, J.~L. 2009{\natexlab{b}},
  GRB Coordinates Network, 10200, 1

\bibitem[{{Klotz} {et~al.}(2011){Klotz}, {Gendre}, {Boer}, \&
  {Atteia}}]{Klotz2011}
{Klotz}, A., {Gendre}, B., {Boer}, M., \& {Atteia}, J.~L. 2011, GRB Coordinates
  Network, 12011, 1

\bibitem[{{Klotz} {et~al.}(2012){Klotz}, {Gendre}, {Boer}, \&
  {Atteia}}]{Klotz2012}
{Klotz}, A., {Gendre}, B., {Boer}, M., \& {Atteia}, J.~L. 2012, GRB Coordinates
  Network, 13108, 1

\bibitem[{{Klotz} {et~al.}(2013){Klotz}, {Gendre}, {Boer}, {Siellez}, {Dereli},
  {Bardho}, \& {Atteia}}]{Klotz2013}
{Klotz}, A., {Gendre}, B., {Boer}, M., {et~al.} 2013, GRB Coordinates Network,
  14908, 1

\bibitem[{{Klotz} \& {Kugel}(2009)}]{Klotz2009b}
{Klotz}, A. \& {Kugel}, F. 2009, GRB Coordinates Network, 9402, 1

\bibitem[{{Klotz} {et~al.}(2014{\natexlab{a}}){Klotz}, {Turpin}, {Boer},
  {Gendre}, {Siellez}, {Dereli}, {Bardho}, \& {Atteia}}]{Klotz2014a}
{Klotz}, A., {Turpin}, D., {Boer}, M., {et~al.} 2014{\natexlab{a}}, GRB
  Coordinates Network, 16469, 1

\bibitem[{{Klotz} {et~al.}(2014{\natexlab{b}}){Klotz}, {Turpin}, {MacPherson},
  {Coward}, {Boer}, {Gendre}, {Siellez}, {Dereli}, {Bardho}, {Martin}, \&
  {Williams}}]{Klotz2014b}
{Klotz}, A., {Turpin}, D., {MacPherson}, D., {et~al.} 2014{\natexlab{b}}, GRB
  Coordinates Network, 16920, 1

\bibitem[{{Klunko} {et~al.}(2014){Klunko}, {Volnova}, {Eselevich}, {Korobtsev},
  \& {Pozanenko}}]{Klunko2014}
{Klunko}, E., {Volnova}, A., {Eselevich}, M., {Korobtsev}, I., \& {Pozanenko},
  A. 2014, GRB Coordinates Network, 16251, 1

\bibitem[{{Klunko} {et~al.}(2009){Klunko}, {Volnova}, \&
  {Pozanenko}}]{Klunko2009}
{Klunko}, E., {Volnova}, A., \& {Pozanenko}, A. 2009, GRB Coordinates Network,
  9613, 1

\bibitem[{{Kobayashi}(2000)}]{Kobayashi2000}
{Kobayashi}, S. 2000, \apj, 545, 807

\bibitem[{{Kocevski}(2012)}]{Kocevski2012}
{Kocevski}, D. 2012, \apj, 747, 146

\bibitem[{{Kocevski} {et~al.}(2007){Kocevski}, {Perley}, {Bloom}, {Modjaz}, \&
  {Poznanski}}]{Kocevski2007}
{Kocevski}, D., {Perley}, D.~A., {Bloom}, J.~S., {Modjaz}, M., \& {Poznanski},
  D. 2007, GRB Coordinates Network, 6919, 1

\bibitem[{{Krimm} {et~al.}(2009){Krimm}, {Yamaoka}, {Sugita}, {Ohno},
  {Sakamoto}, {Barthelmy}, {Gehrels}, {Hara}, {Norris}, {Ohmori}, {Onda},
  {Sato}, {Tanaka}, {Tashiro}, \& {Yamauchi}}]{Krimm2009}
{Krimm}, H.~A., {Yamaoka}, K., {Sugita}, S., {et~al.} 2009, \apj, 704, 1405

\bibitem[{{Kruehler} {et~al.}(2008{\natexlab{a}}){Kruehler}, {Kupcu Yoldas},
  {Greiner}, {Clemens}, {Yoldas}, \& {Szokoly}}]{Kruehler2008a}
{Kruehler}, T., {Kupcu Yoldas}, A., {Greiner}, J., {et~al.} 2008{\natexlab{a}},
  GRB Coordinates Network, 7586, 1

\bibitem[{{Kruehler} {et~al.}(2008{\natexlab{b}}){Kruehler}, {Schrey},
  {Greiner}, {Yoldas}, {Clemens}, {McBreen}, {Kupcu Yoldas}, \&
  {Szokoly}}]{Kruehler2008b}
{Kruehler}, T., {Schrey}, F., {Greiner}, J., {et~al.} 2008{\natexlab{b}}, GRB
  Coordinates Network, 8075, 1

\bibitem[{{Krugly} {et~al.}(2013){Krugly}, {Slyusarev}, {Molotov}, \&
  {Pozanenko}}]{Krugly2013}
{Krugly}, Y., {Slyusarev}, I., {Molotov}, I., \& {Pozanenko}, A. 2013, GRB
  Coordinates Network, 14585, 1

\bibitem[{{Kulkarni} {et~al.}(1999){Kulkarni}, {Djorgovski}, {Odewahn},
  {Bloom}, {Gal}, {Koresko}, {Harrison}, {Lubin}, {Armus}, {Sari},
  {Illingworth}, {Kelson}, {Magee}, {van Dokkum}, {Frail}, {Mulchaey},
  {Malkan}, {McClean}, {Teplitz}, {Koerner}, {Kirkpatrick}, {Kobayashi},
  {Yadigaroglu}, {Halpern}, {Piran}, {Goodrich}, {Chaffee}, {Feroci}, \&
  {Costa}}]{Kulkarni1999}
{Kulkarni}, S.~R., {Djorgovski}, S.~G., {Odewahn}, S.~C., {et~al.} 1999, \nat,
  398, 389

\bibitem[{{Kumar} \& {Zhang}(2014)}]{Kumar2014}
{Kumar}, P. \& {Zhang}, B. 2014, ArXiv e-prints

\bibitem[{{Kuroda} {et~al.}(2011{\natexlab{a}}){Kuroda}, {Hanayama}, {Miyaji},
  {Watanabe}, {Yanagisawa}, {Nagayama}, {Yoshida}, {Ohta}, \&
  {Kawai}}]{Kuroda2011b}
{Kuroda}, D., {Hanayama}, H., {Miyaji}, T., {et~al.} 2011{\natexlab{a}}, GRB
  Coordinates Network, 11972, 1

\bibitem[{{Kuroda} {et~al.}(2011{\natexlab{b}}){Kuroda}, {Hanayama}, {Miyaji},
  {Watanabe}, {Yanagisawa}, {Nagayama}, {Yoshida}, {Ohta}, \&
  {Kawai}}]{Kuroda2011c}
{Kuroda}, D., {Hanayama}, H., {Miyaji}, T., {et~al.} 2011{\natexlab{b}}, GRB
  Coordinates Network, 12226, 1

\bibitem[{{Kuroda} {et~al.}(2014{\natexlab{a}}){Kuroda}, {Hanayama}, {Miyaji},
  {Watanabe}, {Yanagisawa}, {Nagayama}, {Yoshida}, {Ohta}, \&
  {Kawai}}]{Kuroda2014b}
{Kuroda}, D., {Hanayama}, H., {Miyaji}, T., {et~al.} 2014{\natexlab{a}}, GRB
  Coordinates Network, 16132, 1

\bibitem[{{Kuroda} {et~al.}(2011{\natexlab{c}}){Kuroda}, {Yanagisawa},
  {Shimizu}, {Nagayama}, {Yoshida}, {Ohta}, \& {Kawai}}]{Kuroda2011a}
{Kuroda}, D., {Yanagisawa}, K., {Shimizu}, H., {et~al.} 2011{\natexlab{c}}, GRB
  Coordinates Network, 11719, 1

\bibitem[{{Kuroda} {et~al.}(2012){Kuroda}, {Yanagisawa}, {Shimizu}, {Toda},
  {Nagayama}, {Yoshida}, {Ohta}, \& {Kawai}}]{Kuroda2012}
{Kuroda}, D., {Yanagisawa}, K., {Shimizu}, Y., {et~al.} 2012, GRB Coordinates
  Network, 14062, 1

\bibitem[{{Kuroda} {et~al.}(2013){Kuroda}, {Yanagisawa}, {Shimizu}, {Toda},
  {Nagayama}, {Yoshida}, {Ohta}, \& {Kawai}}]{Kuroda2013}
{Kuroda}, D., {Yanagisawa}, K., {Shimizu}, Y., {et~al.} 2013, GRB Coordinates
  Network, 14568, 1

\bibitem[{{Kuroda} {et~al.}(2014{\natexlab{b}}){Kuroda}, {Yanagisawa},
  {Shimizu}, {Toda}, {Nagayama}, {Yoshida}, {Ohta}, \& {Kawai}}]{Kuroda2014a}
{Kuroda}, D., {Yanagisawa}, K., {Shimizu}, Y., {et~al.} 2014{\natexlab{b}}, GRB
  Coordinates Network, 16131, 1

\bibitem[{{Kuroda} {et~al.}(2014{\natexlab{c}}){Kuroda}, {Yanagisawa},
  {Shimizu}, {Toda}, {Nagayama}, {Yoshida}, {Ohta}, \& {Kawai}}]{Kuroda2014c}
{Kuroda}, D., {Yanagisawa}, K., {Shimizu}, Y., {et~al.} 2014{\natexlab{c}}, GRB
  Coordinates Network, 16160, 1

\bibitem[{{Kuroda} {et~al.}(2008){Kuroda}, {Yoshida}, {Yanagisawa}, {Shimizu},
  {Nagayama}, {Toda}, \& {Kawai}}]{Kuroda2008}
{Kuroda}, D., {Yoshida}, M., {Yanagisawa}, K., {et~al.} 2008, GRB Coordinates
  Network, 8724, 1

\bibitem[{{Kuvshinov} {et~al.}(2008){Kuvshinov}, {Lipunov}, {Kornilov},
  {Gorbovskoy}, {Belinski}, {Shatskiy}, {Tyurina}, {Tlatov}, {Golubov},
  {Krushinski}, {Zalognikh}, {Yazev}, \& {Ivanov}}]{Kuvshinov2008}
{Kuvshinov}, D., {Lipunov}, V., {Kornilov}, V., {et~al.} 2008, GRB Coordinates
  Network, 7836, 1

\bibitem[{{Laas-Bourez} {et~al.}(2010){Laas-Bourez}, {Coward}, {Blair},
  {Gendre}, {Boer}, {Klotz}, \& {Thierry}}]{Laas-Bourez2010}
{Laas-Bourez}, M., {Coward}, D., {Blair}, D., {et~al.} 2010, GRB Coordinates
  Network, 11336, 1

\bibitem[{{Lamb} {et~al.}(2005){Lamb}, {Donaghy}, \& {Graziani}}]{Lamb2005}
{Lamb}, D.~Q., {Donaghy}, T.~Q., \& {Graziani}, C. 2005, \apj, 620, 355

\bibitem[{{Laursen} \& {Stanek}(2003)}]{Laursen2003}
{Laursen}, L.~T. \& {Stanek}, K.~Z. 2003, \apjl, 597, L107

\bibitem[{{Lee} {et~al.}(2010){Lee}, {Im}, \& {Urata}}]{Lee2010}
{Lee}, I., {Im}, M., \& {Urata}, Y. 2010, Journal of Korean Astronomical
  Society, 43, 95

\bibitem[{{Lee} {et~al.}(2013){Lee}, {Butler}, {Watson}, {Kutyrev}, {Richer},
  {Klein}, {Fox}, {Prochaska}, {Bloom}, {Cucchiara}, {Troja}, {Littlejohns},
  {Ramirez-Ruiz}, {de Diego}, {Georgiev}, {Gonzalez}, {Roman-Zuniga},
  {Gehrels}, \& {Moseley}}]{Lee2013}
{Lee}, W.~H., {Butler}, N., {Watson}, A.~M., {et~al.} 2013, GRB Coordinates
  Network, 15321, 1

\bibitem[{{Leloudas} {et~al.}(2013){Leloudas}, {Tanvir}, {Xu}, {Malesani},
  {Jakobsson}, {Smirnova}, \& {Pedersen}}]{Leloudas2013}
{Leloudas}, G., {Tanvir}, N.~R., {Xu}, D., {et~al.} 2013, GRB Coordinates
  Network, 14954, 1

\bibitem[{{Leloudas} {et~al.}(2011){Leloudas}, {Xu}, {de Ugarte Postigo},
  {Jakobsson}, {Guaita}, \& {Sandberg}}]{Leloudas2011}
{Leloudas}, G., {Xu}, D., {de Ugarte Postigo}, A., {et~al.} 2011, GRB
  Coordinates Network, 11994, 1

\bibitem[{{Leonini} {et~al.}(2013){Leonini}, {Guerrini}, {Rosi}, \& {Tinjaca
  Ramirez}}]{Leonini2013}
{Leonini}, S., {Guerrini}, G., {Rosi}, P., \& {Tinjaca Ramirez}, L.~M. 2013,
  GRB Coordinates Network, 15150, 1

\bibitem[{{Li} {et~al.}(2003){Li}, {Filippenko}, {Chornock}, \& {Jha}}]{Li2003}
{Li}, W., {Filippenko}, A.~V., {Chornock}, R., \& {Jha}, S. 2003, \apjl, 586,
  L9

\bibitem[{{Li} {et~al.}(2009){Li}, {Perley}, \& {Filippenko}}]{Li2009}
{Li}, W., {Perley}, D.~A., \& {Filippenko}, A.~V. 2009, GRB Coordinates
  Network, 9517, 1

\bibitem[{{Li} {et~al.}(2002{\natexlab{a}}){Li}, {Filippenko}, \&
  {Chornock}}]{Li2002a}
{Li}, W.~D., {Filippenko}, A.~V., \& {Chornock}, R. 2002{\natexlab{a}}, GRB
  Coordinates Network, 1473, 1

\bibitem[{{Li} {et~al.}(2002{\natexlab{b}}){Li}, {Filippenko}, \&
  {Chornock}}]{Li2002b}
{Li}, W.~D., {Filippenko}, A.~V., \& {Chornock}, R. 2002{\natexlab{b}}, GRB
  Coordinates Network, 1491, 1

\bibitem[{{Lipkin} {et~al.}(2004){Lipkin}, {Ofek}, {Gal-Yam}, {Leibowitz},
  {Poznanski}, {Kaspi}, {Polishook}, {Kulkarni}, {Fox}, {Berger}, {Mirabal},
  {Halpern}, {Bureau}, {Fathi}, {Price}, {Peterson}, {Frebel}, {Schmidt},
  {Orosz}, {Fitzgerald}, {Bloom}, {van Dokkum}, {Bailyn}, {Buxton}, \&
  {Barsony}}]{Lipkin2004}
{Lipkin}, Y.~M., {Ofek}, E.~O., {Gal-Yam}, A., {et~al.} 2004, \apj, 606, 381

\bibitem[{{Littlejohns} {et~al.}(2013{\natexlab{a}}){Littlejohns}, {Butler},
  {Watson}, {Kutyrev}, {Lee}, {Richer}, {Klein}, {Fox}, {Prochaska}, {Bloom},
  {Cucchiara}, {Troja}, {Ramirez-Ruiz}, {de Diego}, {Georgiev}, {Gonzalez},
  {Roman-Zuniga}, {Gehrels}, \& {Moseley}}]{Littlejohns2013a}
{Littlejohns}, O., {Butler}, N., {Watson}, A.~M., {et~al.} 2013{\natexlab{a}},
  GRB Coordinates Network, 15420, 1

\bibitem[{{Littlejohns} {et~al.}(2013{\natexlab{b}}){Littlejohns}, {Butler},
  {Watson}, {Kutyrev}, {Lee}, {Richer}, {Klein}, {Fox}, {Prochaska}, {Bloom},
  {Cucchiara}, {Troja}, {Ramirez-Ruiz}, {de Diego}, {Georgiev}, {Gonzalez},
  {Roman-Zuniga}, {Gehrels}, \& {Moseley}}]{Littlejohns2013b}
{Littlejohns}, O., {Butler}, N., {Watson}, A.~M., {et~al.} 2013{\natexlab{b}},
  GRB Coordinates Network, 15436, 1

\bibitem[{{Littlejohns} {et~al.}(2013{\natexlab{c}}){Littlejohns}, {Butler},
  {Watson}, {Kutyrev}, {Lee}, {Richer}, {Klein}, {Fox}, {Prochaska}, {Bloom},
  {Cucchiara}, {Troja}, {Ramirez-Ruiz}, {de Diego}, {Georgiev}, {Gonzalez},
  {Roman-Zuniga}, {Gehrels}, \& {Moseley}}]{Littlejohns2013c}
{Littlejohns}, O., {Butler}, N., {Watson}, A.~M., {et~al.} 2013{\natexlab{c}},
  GRB Coordinates Network, 15444, 1

\bibitem[{{Littlejohns} {et~al.}(2013{\natexlab{d}}){Littlejohns}, {Butler},
  {Watson}, {Kutyrev}, {Lee}, {Richer}, {Klein}, {Fox}, {Prochaska}, {Bloom},
  {Cucchiara}, {Troja}, {Ramirez-Ruiz}, {de Diego}, {Georgiev}, {Gonzalez},
  {Roman-Zuniga}, {Gehrels}, \& {Moseley}}]{Littlejohns2013d}
{Littlejohns}, O., {Butler}, N., {Watson}, A.~M., {et~al.} 2013{\natexlab{d}},
  GRB Coordinates Network, 15462, 1

\bibitem[{{Littlejohns} {et~al.}(2013{\natexlab{e}}){Littlejohns}, {Butler},
  {Watson}, {Kutyrev}, {Lee}, {Richer}, {Klein}, {Fox}, {Prochaska}, {Bloom},
  {Cucchiara}, {Troja}, {Ramirez-Ruiz}, {de Diego}, {Georgiev}, {Gonzalez},
  {Roman-Zuniga}, {Gehrels}, \& {Moseley}}]{Littlejohns2013e}
{Littlejohns}, O., {Butler}, N., {Watson}, A.~M., {et~al.} 2013{\natexlab{e}},
  GRB Coordinates Network, 15463, 1

\bibitem[{{Littlejohns} {et~al.}(2014{\natexlab{a}}){Littlejohns}, {Butler},
  {Watson}, {Kutyrev}, {Lee}, {Richer}, {Klein}, {Fox}, {Prochaska}, {Bloom},
  {Cucchiara}, {Troja}, {Ramirez-Ruiz}, {de Diego}, {Georgiev}, {Gonzalez},
  {Roman-Zuniga}, {Gehrels}, \& {Moseley}}]{Littlejohns2014a}
{Littlejohns}, O., {Butler}, N., {Watson}, A.~M., {et~al.} 2014{\natexlab{a}},
  GRB Coordinates Network, 16136, 1

\bibitem[{{Littlejohns} {et~al.}(2014{\natexlab{b}}){Littlejohns}, {Butler},
  {Watson}, {Kutyrev}, {Lee}, {Richer}, {Klein}, {Fox}, {Prochaska}, {Bloom},
  {Cucchiara}, {Troja}, {Ramirez-Ruiz}, {de Diego}, {Georgiev}, {Gonzalez},
  {Roman-Zuniga}, {Gehrels}, \& {Moseley}}]{Littlejohns2014b}
{Littlejohns}, O., {Butler}, N., {Watson}, A.~M., {et~al.} 2014{\natexlab{b}},
  GRB Coordinates Network, 16139, 1

\bibitem[{{Littlejohns} {et~al.}(2014{\natexlab{c}}){Littlejohns}, {Butler},
  {Watson}, {Kutyrev}, {Lee}, {Richer}, {Klein}, {Fox}, {Prochaska}, {Bloom},
  {Cucchiara}, {Troja}, {Ramirez-Ruiz}, {de Diego}, {Georgiev}, {Gonzalez},
  {Roman-Zuniga}, {Gehrels}, \& {Moseley}}]{Littlejohns2014c}
{Littlejohns}, O., {Butler}, N., {Watson}, A.~M., {et~al.} 2014{\natexlab{c}},
  GRB Coordinates Network, 16170, 1

\bibitem[{{Litvinenko} {et~al.}(2012){Litvinenko}, {Volnova}, {Molotov}, \&
  {Pozanenko}}]{Litvinenko2012}
{Litvinenko}, E., {Volnova}, A., {Molotov}, I., \& {Pozanenko}, A. 2012, GRB
  Coordinates Network, 13693, 1

\bibitem[{{Loew} {et~al.}(2008){Loew}, {Kruehler}, \& {Greiner}}]{Loew2008}
{Loew}, S., {Kruehler}, T., \& {Greiner}, J. 2008, GRB Coordinates Network,
  8540, 1

\bibitem[{{Maiorano} {et~al.}(2006){Maiorano}, {Masetti}, {Palazzi},
  {Savaglio}, {Rol}, {Vreeswijk}, {Pian}, {Price}, {Peterson},
  {Jel{\'{\i}}nek}, {Amati}, {Andersen}, {Castro-Tirado}, {Castro Cer{\'o}n},
  {de Ugarte Postigo}, {Frontera}, {Fruchter}, {Fynbo}, {Gorosabel}, {Henden},
  {Hjorth}, {Jensen}, {Klose}, {Kouveliotou}, {Masi}, {M{\o}ller}, {Nicastro},
  {Ofek}, {Pandey}, {Rhoads}, {Tanvir}, {Wijers}, \& {van den
  Heuvel}}]{Maiorano2006}
{Maiorano}, E., {Masetti}, N., {Palazzi}, E., {et~al.} 2006, \aap, 455, 423

\bibitem[{{Malesani} {et~al.}(2011{\natexlab{a}}){Malesani}, {Fugazza},
  {D'Avanzo}, {D'Elia}, {Melandri}, {Piranomonte}, {Fynbo}, {Cecconi}, \&
  {Mainella}}]{Malesani2011a}
{Malesani}, D., {Fugazza}, D., {D'Avanzo}, P., {et~al.} 2011{\natexlab{a}}, GRB
  Coordinates Network, 11977, 1

\bibitem[{{Malesani} {et~al.}(2007){Malesani}, {Hjorth}, {Fynbo}, {Sollerman},
  {Olofsson}, {Paraficz}, \& {Durant}}]{Malesani2007}
{Malesani}, D., {Hjorth}, J., {Fynbo}, J.~P.~U., {et~al.} 2007, GRB Coordinates
  Network, 6555, 1

\bibitem[{{Malesani} {et~al.}(2011{\natexlab{b}}){Malesani}, {Leloudas}, {Xu},
  {de Ugarte Postigo}, {Hjorth}, {Jakobsson}, \& {Nielsen}}]{Malesani2011b}
{Malesani}, D., {Leloudas}, G., {Xu}, D., {et~al.} 2011{\natexlab{b}}, GRB
  Coordinates Network, 12220, 1

\bibitem[{{Malesani} {et~al.}(2005){Malesani}, {Piranomonte}, {Fiore},
  {Tagliaferri}, {Fugazza}, \& {Cosentino}}]{Malesani2005}
{Malesani}, D., {Piranomonte}, S., {Fiore}, F., {et~al.} 2005, GRB Coordinates
  Network, 3469, 1

\bibitem[{{Mao} {et~al.}(2009){Mao}, {Cha}, \& {Bai}}]{Mao2009}
{Mao}, J., {Cha}, G., \& {Bai}, J. 2009, GRB Coordinates Network, 9305, 1

\bibitem[{{Margutti} {et~al.}(2013){Margutti}, {Zaninoni}, {Bernardini},
  {Chincarini}, {Pasotti}, {Guidorzi}, {Angelini}, {Burrows}, {Capalbi},
  {Evans}, {Gehrels}, {Kennea}, {Mangano}, {Moretti}, {Nousek}, {Osborne},
  {Page}, {Perri}, {Racusin}, {Romano}, {Sbarufatti}, {Stafford}, \&
  {Stamatikos}}]{Margutti2013}
{Margutti}, R., {Zaninoni}, E., {Bernardini}, M.~G., {et~al.} 2013, \mnras,
  428, 729

\bibitem[{{Martini} {et~al.}(2003){Martini}, {Garnavich}, \&
  {Stanek}}]{Martini2003}
{Martini}, P., {Garnavich}, P., \& {Stanek}, K.~Z. 2003, GRB Coordinates
  Network, 1979, 1

\bibitem[{{Maselli} {et~al.}(2014){Maselli}, {Melandri}, {Nava}, {Mundell},
  {Kawai}, {Campana}, {Covino}, {Cummings}, {Cusumano}, {Evans}, {Ghirlanda},
  {Ghisellini}, {Guidorzi}, {Kobayashi}, {Kuin}, {La Parola}, {Mangano},
  {Oates}, {Sakamoto}, {Serino}, {Virgili}, {Zhang}, {Barthelmy}, {Beardmore},
  {Bernardini}, {Bersier}, {Burrows}, {Calderone}, {Capalbi}, {Chiang},
  {D'Avanzo}, {D'Elia}, {De Pasquale}, {Fugazza}, {Gehrels}, {Gomboc},
  {Harrison}, {Hanayama}, {Japelj}, {Kennea}, {Kopac}, {Kouveliotou}, {Kuroda},
  {Levan}, {Malesani}, {Marshall}, {Nousek}, {O'Brien}, {Osborne}, {Pagani},
  {Page}, {Page}, {Perri}, {Pritchard}, {Romano}, {Saito}, {Sbarufatti},
  {Salvaterra}, {Steele}, {Tanvir}, {Vianello}, {Weigand}, {Wiersema}, {Yatsu},
  {Yoshii}, \& {Tagliaferri}}]{Maselli2014}
{Maselli}, A., {Melandri}, A., {Nava}, L., {et~al.} 2014, Science, 343, 48

\bibitem[{{Masi} \& {Nocentini}(2013)}]{Masi2013}
{Masi}, G. \& {Nocentini}, F. 2013, GRB Coordinates Network, 15152, 1

\bibitem[{{McMahon} {et~al.}(2006){McMahon}, {Kumar}, \& {Piran}}]{McMahon2006}
{McMahon}, E., {Kumar}, P., \& {Piran}, T. 2006, \mnras, 366, 575

\bibitem[{{Melandri} {et~al.}(2011){Melandri}, {D'Avanzo}, {Fugazza}, \&
  {Palazzi}}]{Melandri2011}
{Melandri}, A., {D'Avanzo}, P., {Fugazza}, D., \& {Palazzi}, E. 2011, GRB
  Coordinates Network, 11963, 1

\bibitem[{{Melandri} {et~al.}(2009{\natexlab{a}}){Melandri}, {Guidorzi},
  {Bersier}, {Cano}, {Steele}, {Mundell}, {O'Brien}, \&
  {Tanvir}}]{Melandri2009b}
{Melandri}, A., {Guidorzi}, C., {Bersier}, D., {et~al.} 2009{\natexlab{a}}, GRB
  Coordinates Network, 9520, 1

\bibitem[{{Melandri} {et~al.}(2008){Melandri}, {Guidorzi}, \&
  {Carter}}]{Melandri2008}
{Melandri}, A., {Guidorzi}, C., \& {Carter}, D. 2008, GRB Coordinates Network,
  8699, 1

\bibitem[{{Melandri} {et~al.}(2009{\natexlab{b}}){Melandri}, {Guidorzi},
  {Mundell}, {Bersier}, {O'Brien}, \& {Tanvir}}]{Melandri2009a}
{Melandri}, A., {Guidorzi}, C., {Mundell}, C.~G., {et~al.} 2009{\natexlab{b}},
  GRB Coordinates Network, 9200, 1

\bibitem[{{Melandri} {et~al.}(2006){Melandri}, {Guidorzi}, {Mundell}, {Steele},
  {Smith}, {Monfardini}, {Carter}, {Kobayashi}, {Bersier}, {Bode}, {Gomboc},
  {O'Brien}, {Rol}, \& {Bannister}}]{Melandri2006}
{Melandri}, A., {Guidorzi}, C., {Mundell}, C.~G., {et~al.} 2006, GRB
  Coordinates Network, 5827, 1

\bibitem[{{Melandri} {et~al.}(2013{\natexlab{a}}){Melandri}, {Japelj},
  {Virgili}, \& {Mundell}}]{Melandri2013a}
{Melandri}, A., {Japelj}, J., {Virgili}, F.~J., \& {Mundell}, C.~G.
  2013{\natexlab{a}}, GRB Coordinates Network, 14362, 1

\bibitem[{{Melandri} {et~al.}(2014){Melandri}, {Virgili}, {Guidorzi},
  {Bernardini}, {Kobayashi}, {Mundell}, {Gomboc}, {Dintinjana}, {Hentunen},
  {Japelj}, {Kopa{\v c}}, {Kuroda}, {Morgan}, {Steele}, {Quadri}, {Arici},
  {Arnold}, {Girelli}, {Hanayama}, {Kawai}, {Miku{\v z}}, {Nissinen}, {Salmi},
  {Smith}, {Strabla}, {Tonincelli}, \& {Quadri}}]{Melandri2014}
{Melandri}, A., {Virgili}, F.~J., {Guidorzi}, C., {et~al.} 2014, \aap, 572, A55

\bibitem[{{Melandri} {et~al.}(2013{\natexlab{b}}){Melandri}, {Virgili},
  {Mundell}, \& {Gomboc}}]{Melandri2013b}
{Melandri}, A., {Virgili}, F.~J., {Mundell}, C.~G., \& {Gomboc}, A.
  2013{\natexlab{b}}, GRB Coordinates Network, 14843, 1

\bibitem[{{Meszaros} \& {Rees}(1993)}]{Meszaros1993}
{Meszaros}, P. \& {Rees}, M.~J. 1993, \apjl, 418, L59

\bibitem[{{Milne} \& {Cenko}(2011)}]{Milne2011}
{Milne}, P.~A. \& {Cenko}, S.~B. 2011, GRB Coordinates Network, 11708, 1

\bibitem[{{Mirabal} {et~al.}(2005){Mirabal}, {Bonfield}, \&
  {Schawinski}}]{Mirabal2005}
{Mirabal}, N., {Bonfield}, D., \& {Schawinski}, K. 2005, GRB Coordinates
  Network, 3488, 1

\bibitem[{{Mirabal} {et~al.}(2007){Mirabal}, {McGreer}, {Halpern}, {Dietrich},
  \& {Peterson}}]{Mirabal2007}
{Mirabal}, N., {McGreer}, I.~D., {Halpern}, J.~P., {Dietrich}, M., \&
  {Peterson}, B.~M. 2007, GRB Coordinates Network, 6526, 1

\bibitem[{{Misra1} {et~al.}(2005){Misra1}, {Resmi}, {Pandey}, {Bhattacharya},
  \& {Sagar}}]{Misra2005}
{Misra1}, K., {Resmi}, L., {Pandey}, S.~B., {Bhattacharya}, D., \& {Sagar}, R.
  2005, Bulletin of the Astronomical Society of India, 33, 487

\bibitem[{{Monfardini} {et~al.}(2006){Monfardini}, {Kobayashi}, {Guidorzi},
  {Carter}, {Mundell}, {Bersier}, {Gomboc}, {Melandri}, {Mottram}, {Smith}, \&
  {Steele}}]{Monfardini2006}
{Monfardini}, A., {Kobayashi}, S., {Guidorzi}, C., {et~al.} 2006, \apj, 648,
  1125

\bibitem[{{Moskvitin}(2011)}]{Moskvitin2011a}
{Moskvitin}, A.~S. 2011, GRB Coordinates Network, 11962, 1

\bibitem[{{Moskvitin}(2013)}]{Moskvitin2013}
{Moskvitin}, A.~S. 2013, GRB Coordinates Network, 15412, 1

\bibitem[{{Moskvitin} {et~al.}(2011){Moskvitin}, {Sokolov}, {Spiridonova},
  {Andreev}, {Sergeev}, \& {Parakhin}}]{Moskvitin2011b}
{Moskvitin}, A.~S., {Sokolov}, V.~V., {Spiridonova}, O.~I., {et~al.} 2011, GRB
  Coordinates Network, 12333, 1

\bibitem[{{Mundell} {et~al.}(2007){Mundell}, {Melandri}, {Guidorzi},
  {Kobayashi}, {Steele}, {Malesani}, {Amati}, {D'Avanzo}, {Bersier}, {Gomboc},
  {Rol}, {Bode}, {Carter}, {Mottram}, {Monfardini}, {Smith}, {Malhotra},
  {Wang}, {Bannister}, {O'Brien}, \& {Tanvir}}]{Mundell2007}
{Mundell}, C.~G., {Melandri}, A., {Guidorzi}, C., {et~al.} 2007, \apj, 660, 489

\bibitem[{{Nakajima} {et~al.}(2009){Nakajima}, {Yatsu}, {Mori}, {Endo},
  {Shimokawabe}, \& {Kawai}}]{Nakajima2009}
{Nakajima}, H., {Yatsu}, Y., {Mori}, Y.~A., {et~al.} 2009, GRB Coordinates
  Network, 10260, 1

\bibitem[{{Nakar} \& {Piran}(2004)}]{Nakar2004}
{Nakar}, E. \& {Piran}, T. 2004, \mnras, 353, 647

\bibitem[{{Nakar} \& {Piran}(2005)}]{Nakar2005}
{Nakar}, E. \& {Piran}, T. 2005, \mnras, 360, L73

\bibitem[{{Narayan} {et~al.}(2011){Narayan}, {Kumar}, \&
  {Tchekhovskoy}}]{Narayan2011}
{Narayan}, R., {Kumar}, P., \& {Tchekhovskoy}, A. 2011, \mnras, 416, 2193

\bibitem[{{Nardini} {et~al.}(2014){Nardini}, {Elliott}, {Filgas}, {Schady},
  {Greiner}, {Kr{\"u}hler}, {Klose}, {Afonso}, {Kann}, {Nicuesa Guelbenzu},
  {Olivares E.}, {Rau}, {Rossi}, {Sudilovsky}, \& {Schmidl}}]{Nardini2014}
{Nardini}, M., {Elliott}, J., {Filgas}, R., {et~al.} 2014, \aap, 562, A29

\bibitem[{{Nardini} {et~al.}(2010){Nardini}, {Kruehler}, {Klose}, {Rossi}, \&
  {Greiner}}]{Nardini2010}
{Nardini}, M., {Kruehler}, T., {Klose}, S., {Rossi}, A., \& {Greiner}, J. 2010,
  GRB Coordinates Network, 11337, 1

\bibitem[{{Nava} {et~al.}(2012){Nava}, {Salvaterra}, {Ghirlanda}, {Ghisellini},
  {Campana}, {Covino}, {Cusumano}, {D'Avanzo}, {D'Elia}, {Fugazza}, {Melandri},
  {Sbarufatti}, {Vergani}, \& {Tagliaferri}}]{Nava2012}
{Nava}, L., {Salvaterra}, R., {Ghirlanda}, G., {et~al.} 2012, \mnras, 421, 1256

\bibitem[{{Nevski} {et~al.}(2012){Nevski}, {Volnova}, {Molotov}, \&
  {Pozanenko}}]{Nevski2012}
{Nevski}, V., {Volnova}, A., {Molotov}, I., \& {Pozanenko}, A. 2012, GRB
  Coordinates Network, 13761, 1

\bibitem[{{Nysewander} {et~al.}(2009){Nysewander}, {Fruchter}, \&
  {Pe'er}}]{Nysewander2009}
{Nysewander}, M., {Fruchter}, A.~S., \& {Pe'er}, A. 2009, \apj, 701, 824

\bibitem[{{Nysewander} {et~al.}(2006{\natexlab{a}}){Nysewander}, {Lacluyze},
  {Reichart}, {Crain}, {Foster}, \& {Ivarson}}]{Nysewander2006b}
{Nysewander}, M., {Lacluyze}, A., {Reichart}, D., {et~al.} 2006{\natexlab{a}},
  GRB Coordinates Network, 4548, 1

\bibitem[{{Nysewander} {et~al.}(2006{\natexlab{b}}){Nysewander}, {Lacluyze},
  {Reichart}, {Crain}, {Foster}, {Ivarson}, {Haislip}, {MacLeod}, \&
  {Kirschbrown}}]{Nysewander2006a}
{Nysewander}, M., {Lacluyze}, A., {Reichart}, D., {et~al.} 2006{\natexlab{b}},
  GRB Coordinates Network, 4530, 1

\bibitem[{{O Meara} {et~al.}(2010){O Meara}, {Chen}, \&
  {Prochaska}}]{Omeara2010}
{O Meara}, J., {Chen}, H.-W., \& {Prochaska}, J.~X. 2010, GRB Coordinates
  Network, 11089, 1

\bibitem[{{Oksanen}(2009)}]{Oksanen2009}
{Oksanen}, A. 2009, GRB Coordinates Network, 9239, 1

\bibitem[{{Oksanen} {et~al.}(2008){Oksanen}, {Templeton}, {Henden}, \&
  {Kann}}]{Oksanen2008}
{Oksanen}, A., {Templeton}, M., {Henden}, A.~A., \& {Kann}, D.~A. 2008, Journal
  of the American Association of Variable Star Observers (JAAVSO), 36, 53

\bibitem[{{Olivares} {et~al.}(2010){Olivares}, {Filgas}, \&
  {Greiner}}]{Olivares2010}
{Olivares}, E.~F., {Filgas}, R., \& {Greiner}, J. 2010, GRB Coordinates
  Network, 11013, 1

\bibitem[{{Olivares} {et~al.}(2009{\natexlab{a}}){Olivares}, {Kruehler},
  {Greiner}, \& {Filgas}}]{Olivares2009a}
{Olivares}, F., {Kruehler}, T., {Greiner}, J., \& {Filgas}, R.
  2009{\natexlab{a}}, GRB Coordinates Network, 9215, 1

\bibitem[{{Olivares} {et~al.}(2009{\natexlab{b}}){Olivares}, {Kupcu Yoldas},
  {Greiner}, \& {Yoldas}}]{Olivares2009b}
{Olivares}, F., {Kupcu Yoldas}, A., {Greiner}, J., \& {Yoldas}, A.
  2009{\natexlab{b}}, GRB Coordinates Network, 9245, 1

\bibitem[{{Osip} {et~al.}(2006){Osip}, {Chen}, \& {Prochaska}}]{Osip2006}
{Osip}, D., {Chen}, H.-W., \& {Prochaska}, J.~X. 2006, GRB Coordinates Network,
  5715, 1

\bibitem[{{Paczynski} \& {Xu}(1994)}]{Paczynski1994}
{Paczynski}, B. \& {Xu}, G. 1994, \apj, 427, 708

\bibitem[{{Page} {et~al.}(2009){Page}, {Willingale}, {Bissaldi}, {Postigo},
  {Holland}, {McBreen}, {O'Brien}, {Osborne}, {Prochaska}, {Rol}, {Rykoff},
  {Starling}, {Tanvir}, {van der Horst}, {Wiersema}, {Zhang}, {Aceituno},
  {Akerlof}, {Beardmore}, {Briggs}, {Burrows}, {Castro-Tirado}, {Connaughton},
  {Evans}, {Fynbo}, {Gehrels}, {Guidorzi}, {Howard}, {Kennea}, {Kouveliotou},
  {Pagani}, {Preece}, {Perley}, {Steele}, \& {Yuan}}]{Page2009}
{Page}, K.~L., {Willingale}, R., {Bissaldi}, E., {et~al.} 2009, \mnras, 400,
  134

\bibitem[{{Pandey} {et~al.}(2003{\natexlab{a}}){Pandey}, {Anupama}, {Sagar},
  {Bhattacharya}, {Castro-Tirado}, {Sahu}, {Parihar}, \&
  {Prabhu}}]{Pandey2003b}
{Pandey}, S.~B., {Anupama}, G.~C., {Sagar}, R., {et~al.} 2003{\natexlab{a}},
  \aap, 408, L21

\bibitem[{{Pandey} \& {Kumar}(2014)}]{Pandey2014b}
{Pandey}, S.~B. \& {Kumar}, B. 2014, GRB Coordinates Network, 16133, 1

\bibitem[{{Pandey} {et~al.}(2013){Pandey}, {Kumar}, \& {Joshi}}]{Pandey2013}
{Pandey}, S.~B., {Kumar}, B., \& {Joshi}, Y.~C. 2013, GRB Coordinates Network,
  15501, 1

\bibitem[{{Pandey} {et~al.}(2014){Pandey}, {Kumar}, \& {Kumar}}]{Pandey2014a}
{Pandey}, S.~B., {Kumar}, B., \& {Kumar}, P. 2014, GRB Coordinates Network,
  15677, 1

\bibitem[{{Pandey} {et~al.}(2003{\natexlab{b}}){Pandey}, {Sahu}, {Resmi},
  {Sagar}, {Anupama}, {Bhattacharya}, {Mohan}, {Prabhu}, {Bhatt}, {Pandey},
  {Parihar}, \& {Castro-Tirado}}]{Pandey2003a}
{Pandey}, S.~B., {Sahu}, D.~K., {Resmi}, L., {et~al.} 2003{\natexlab{b}},
  Bulletin of the Astronomical Society of India, 31, 19

\bibitem[{{Pavlenko} {et~al.}(2009){Pavlenko}, {Rumyantsev}, \&
  {Pozanenko}}]{Pavlenko2009}
{Pavlenko}, E., {Rumyantsev}, V., \& {Pozanenko}, A. 2009, GRB Coordinates
  Network, 9179, 1

\bibitem[{{Pavlenko} {et~al.}(2011){Pavlenko}, {Volnova}, {Baklanov}, \&
  {Pozanenko}}]{Pavlenko2011}
{Pavlenko}, E., {Volnova}, A., {Baklanov}, A., \& {Pozanenko}, A. 2011, GRB
  Coordinates Network, 12005, 1

\bibitem[{{P{\'e}langeon} {et~al.}(2008){P{\'e}langeon}, {Atteia}, {Nakagawa},
  {Hurley}, {Yoshida}, {Vanderspek}, {Suzuki}, {Kawai}, {Pizzichini},
  {Bo{\"e}r}, {Braga}, {Crew}, {Donaghy}, {Dezalay}, {Doty}, {Fenimore},
  {Galassi}, {Graziani}, {Jernigan}, {Lamb}, {Levine}, {Manchanda}, {Martel},
  {Matsuoka}, {Olive}, {Prigozhin}, {Ricker}, {Sakamoto}, {Shirasaki},
  {Sugita}, {Takagishi}, {Tamagawa}, {Villasenor}, {Woosley}, \&
  {Yamauchi}}]{Pelangeon2008}
{P{\'e}langeon}, A., {Atteia}, J.-L., {Nakagawa}, Y.~E., {et~al.} 2008, \aap,
  491, 157

\bibitem[{{Perley}(2009{\natexlab{a}})}]{Perley2009b}
{Perley}, D.~A. 2009{\natexlab{a}}, GRB Coordinates Network, 10060, 1

\bibitem[{{Perley}(2009{\natexlab{b}})}]{Perley2009a}
{Perley}, D.~A. 2009{\natexlab{b}}, GRB Coordinates Network, 10058, 1

\bibitem[{{Perley}(2014)}]{Perley2014c}
{Perley}, D.~A. 2014, GRB Coordinates Network, 16884, 1

\bibitem[{{Perley} \& {Cenko}(2013)}]{Perley2013}
{Perley}, D.~A. \& {Cenko}, S.~B. 2013, GRB Coordinates Network, 15423, 1

\bibitem[{{Perley} \& {Cenko}(2014{\natexlab{a}})}]{Perley2014a}
{Perley}, D.~A. \& {Cenko}, S.~B. 2014{\natexlab{a}}, GRB Coordinates Network,
  15674, 1

\bibitem[{{Perley} \& {Cenko}(2014{\natexlab{b}})}]{Perley2014b}
{Perley}, D.~A. \& {Cenko}, S.~B. 2014{\natexlab{b}}, GRB Coordinates Network,
  16514, 1

\bibitem[{{Perley} {et~al.}(2010{\natexlab{a}}){Perley}, {Klein}, \&
  {Morgan}}]{Perley2010b}
{Perley}, D.~A., {Klein}, C.~R., \& {Morgan}, A.~N. 2010{\natexlab{a}}, GRB
  Coordinates Network, 11025, 1

\bibitem[{{Perley} {et~al.}(2010{\natexlab{b}}){Perley}, {Li}, \&
  {Filippenko}}]{Perley2010a}
{Perley}, D.~A., {Li}, W., \& {Filippenko}, A.~V. 2010{\natexlab{b}}, GRB
  Coordinates Network, 11024, 1

\bibitem[{{Perley} {et~al.}(2012){Perley}, {Prochaska}, \&
  {Morgan}}]{Perley2012}
{Perley}, D.~A., {Prochaska}, J.~X., \& {Morgan}, A.~N. 2012, GRB Coordinates
  Network, 14059, 1

\bibitem[{{Perri} {et~al.}(2009){Perri}, {Barthelmy}, {Burrows}, {Gronwall},
  {Holland}, {Kennea}, {Marshall}, {O'Brien}, {Pagani}, {Palmer}, {Rowlinson},
  {Siegel}, {Stamatikos}, \& {Ukwatta}}]{Perri2009}
{Perri}, M., {Barthelmy}, S.~D., {Burrows}, D.~N., {et~al.} 2009, GRB
  Coordinates Network, 9400, 1

\bibitem[{{Piran}(1999)}]{Piran1999}
{Piran}, T. 1999, \physrep, 314, 575

\bibitem[{{Piranomonte} {et~al.}(2006){Piranomonte}, {D'Elia}, {Fiore},
  {Covino}, {Fugazza}, {Chincarini}, {Malesani}, {D'Avanzo}, {Antonelli},
  {Ledoux}, {Lopez}, \& {Naef}}]{Piranomonte2006}
{Piranomonte}, S., {D'Elia}, V., {Fiore}, F., {et~al.} 2006, GRB Coordinates
  Network, 4520, 1

\bibitem[{{Pozanenko}(2005)}]{Pozanenko2005}
{Pozanenko}, A. 2005, GRB Coordinates Network, 4087, 1

\bibitem[{{Price} {et~al.}(2002){Price}, {Bloom}, {Goodrich}, {Barth}, {Cohen},
  \& {Fox}}]{Price2002b}
{Price}, P.~A., {Bloom}, J.~S., {Goodrich}, R.~W., {et~al.} 2002, GRB
  Coordinates Network, 1475, 1

\bibitem[{{Price} \& {Fox}(2002)}]{Price2002a}
{Price}, P.~A. \& {Fox}, D.~W. 2002, GRB Coordinates Network, 1732, 1

\bibitem[{{Price} {et~al.}(2004){Price}, {Roth}, {Rich}, {Schmidt}, {Peterson},
  {Cowie}, {Smith}, \& {Rest}}]{Price2004}
{Price}, P.~A., {Roth}, K., {Rich}, J., {et~al.} 2004, GRB Coordinates Network,
  2791, 1

\bibitem[{{Prochaska} {et~al.}(2005){Prochaska}, {Bloom}, {Wright}, {Paul
  Butler}, {Chen}, {Vogt}, \& {Marcy}}]{Prochaska2005}
{Prochaska}, J.~X., {Bloom}, J.~S., {Wright}, J.~T., {et~al.} 2005, GRB
  Coordinates Network, 3833, 1

\bibitem[{{Prochaska} {et~al.}(2008){Prochaska}, {Perley}, {Howard}, {Chen},
  {Marcy}, {Fischer}, \& {Wilburn}}]{Prochaska2008}
{Prochaska}, J.~X., {Perley}, D., {Howard}, A., {et~al.} 2008, GRB Coordinates
  Network, 8083, 1

\bibitem[{{Quadri} {et~al.}(2012){Quadri}, {Strabla}, {Girelli}, \&
  {Quadri}}]{Quadri2012}
{Quadri}, U., {Strabla}, L., {Girelli}, R., \& {Quadri}, A. 2012, GRB
  Coordinates Network, 13178, 1

\bibitem[{{Rees} \& {Meszaros}(1994)}]{Rees1994}
{Rees}, M.~J. \& {Meszaros}, P. 1994, \apjl, 430, L93

\bibitem[{{Rees} \& {M{\'e}sz{\'a}ros}(2005)}]{Rees2005}
{Rees}, M.~J. \& {M{\'e}sz{\'a}ros}, P. 2005, \apj, 628, 847

\bibitem[{{Resmi} {et~al.}(2012){Resmi}, {Misra}, {J{\'o}hannesson}, {Castro
  Tirado}, {Gorosabel}, {Jel{\'{\i}}nek}, {Bhattacharya}, {Kub{\'a}nek},
  {Anupama}, {Sota}, {Sahu}, {de Ugarte Postigo}, {Pandey}, {S{\'a}nchez
  Ram{\'{\i}}rez}, {Bremer}, \& {Sagar}}]{Resmi2012}
{Resmi}, L., {Misra}, K., {J{\'o}hannesson}, G., {et~al.} 2012, \mnras, 427,
  288

\bibitem[{{Rol} {et~al.}(2003){Rol}, {Vreeswijk}, \& {Jaunsen}}]{Rol2003}
{Rol}, E., {Vreeswijk}, P., \& {Jaunsen}, A. 2003, GRB Coordinates Network,
  1981, 1

\bibitem[{{Rossi} {et~al.}(2009{\natexlab{a}}){Rossi}, {Afonso}, \&
  {Greiner}}]{Rossi2009b}
{Rossi}, A., {Afonso}, P., \& {Greiner}, J. 2009{\natexlab{a}}, GRB Coordinates
  Network, 9382, 1

\bibitem[{{Rossi} {et~al.}(2009{\natexlab{b}}){Rossi}, {Kruehler}, {Greiner},
  \& {Yoldas}}]{Rossi2009a}
{Rossi}, A., {Kruehler}, T., {Greiner}, J., \& {Yoldas}, A. 2009{\natexlab{b}},
  GRB Coordinates Network, 9408, 1

\bibitem[{{Roy} {et~al.}(2008){Roy}, {Kumar}, {Pandey}, \& {Kumar}}]{Roy2008}
{Roy}, R., {Kumar}, B., {Pandey}, S.~B., \& {Kumar}, B. 2008, GRB Coordinates
  Network, 8717, 1

\bibitem[{{Roy} {et~al.}(2009){Roy}, {Kumar}, {Pandey}, \& {Kumar}}]{Roy2009}
{Roy}, R., {Kumar}, B., {Pandey}, S.~B., \& {Kumar}, B. 2009, GRB Coordinates
  Network, 9278, 1

\bibitem[{{Rujopakarn} {et~al.}(2009){Rujopakarn}, {Guver}, {Pandey}, \&
  {Yuan}}]{Rujopakarn2009}
{Rujopakarn}, W., {Guver}, T., {Pandey}, S.~B., \& {Yuan}, F. 2009, GRB
  Coordinates Network, 9515, 1

\bibitem[{{Rujopakarn} \& {Rykoff}(2008)}]{Rujopakarn2008}
{Rujopakarn}, W. \& {Rykoff}, E.~S. 2008, GRB Coordinates Network, 8056, 1

\bibitem[{{Rujopakarn} {et~al.}(2011){Rujopakarn}, {Schaefer}, \&
  {Rykoff}}]{Rujopakarn2011}
{Rujopakarn}, W., {Schaefer}, B.~E., \& {Rykoff}, E.~S. 2011, GRB Coordinates
  Network, 11707, 1

\bibitem[{{Rumyantsev} {et~al.}(2009){Rumyantsev}, {Antoniuk}, \&
  {Pozanenko}}]{Rumyantsev2009a}
{Rumyantsev}, V., {Antoniuk}, K., \& {Pozanenko}, A. 2009, GRB Coordinates
  Network, 9320, 1

\bibitem[{{Rumyantsev} {et~al.}(2011{\natexlab{a}}){Rumyantsev}, {Antoniuk}, \&
  {Pozanenko}}]{Rumyantsev2011a}
{Rumyantsev}, V., {Antoniuk}, K., \& {Pozanenko}, A. 2011{\natexlab{a}}, GRB
  Coordinates Network, 11973, 1

\bibitem[{{Rumyantsev} {et~al.}(2003){Rumyantsev}, {Biryukov}, \&
  {Pozanenko}}]{Rumyantsev2003}
{Rumyantsev}, V., {Biryukov}, V., \& {Pozanenko}, A. 2003, GRB Coordinates
  Network, 1991, 1

\bibitem[{{Rumyantsev} {et~al.}(2005){Rumyantsev}, {Biryukov}, \&
  {Pozanenko}}]{Rumyantsev2005}
{Rumyantsev}, V., {Biryukov}, V., \& {Pozanenko}, A. 2005, GRB Coordinates
  Network, 4081, 1

\bibitem[{{Rumyantsev} \& {Pozanenko}(2008)}]{Rumyantsev2008}
{Rumyantsev}, V. \& {Pozanenko}, A. 2008, GRB Coordinates Network, 7857, 1

\bibitem[{{Rumyantsev} \& {Pozanenko}(2009)}]{Rumyantsev2009b}
{Rumyantsev}, V. \& {Pozanenko}, A. 2009, GRB Coordinates Network, 9539, 1

\bibitem[{{Rumyantsev} \& {Pozanenko}(2013)}]{Rumyantsev2013}
{Rumyantsev}, V. \& {Pozanenko}, A. 2013, GRB Coordinates Network, 14907, 1

\bibitem[{{Rumyantsev} {et~al.}(2011{\natexlab{b}}){Rumyantsev}, {Pozanenko},
  \& {Klunko}}]{Rumyantsev2011b}
{Rumyantsev}, V., {Pozanenko}, A., \& {Klunko}, E. 2011{\natexlab{b}}, GRB
  Coordinates Network, 11986, 1

\bibitem[{{Rykoff} \& {Rujopakarn}(2008)}]{Rykoff2008}
{Rykoff}, E.~S. \& {Rujopakarn}, W. 2008, GRB Coordinates Network, 7593, 1

\bibitem[{{Rykoff} {et~al.}(2005){Rykoff}, {Yost}, \& {Swan}}]{Rykoff2005}
{Rykoff}, E.~S., {Yost}, S.~A., \& {Swan}, H. 2005, GRB Coordinates Network,
  3465, 1

\bibitem[{{Rykoff} \& {BenDaniel}(2005)}]{Torii2005}
{Rykoff}, K. \& {BenDaniel}, M. 2005, GRB Coordinates Network, 3470, 1

\bibitem[{{Sahu}(2014)}]{Sahu2014}
{Sahu}, D.~K. 2014, GRB Coordinates Network, 16272, 1

\bibitem[{{Sahu} {et~al.}(2012){Sahu}, {Anupama}, \& {Pandey}}]{Sahu2012}
{Sahu}, D.~K., {Anupama}, G.~C., \& {Pandey}, S.~B. 2012, GRB Coordinates
  Network, 13185, 1

\bibitem[{{Sahu} {et~al.}(2000){Sahu}, {Vreeswijk}, {Bakos}, {Menzies},
  {Bragaglia}, {Frontera}, {Piro}, {Albrow}, {Bond}, {Bower}, {Caldwell},
  {Castro-Tirado}, {Courbin}, {Dominik}, {Fynbo}, {Galama}, {Glazebrook},
  {Greenhill}, {Gorosabel}, {Hearnshaw}, {Hill}, {Hjorth}, {Kane}, {Kilmartin},
  {Kouveliotou}, {Martin}, {Masetti}, {Maxted}, {Minniti}, {M{\o}ller},
  {Muraki}, {Nakamura}, {Noda}, {Ohnishi}, {Palazzi}, {van Paradijs}, {Pian},
  {Pollard}, {Rattenbury}, {Reid}, {Rol}, {Saito}, {Sackett}, {Saizar},
  {Tinney}, {Vermaak}, {Watson}, {Williams}, {Yock}, \& {Dar}}]{Sahu2000}
{Sahu}, K.~C., {Vreeswijk}, P., {Bakos}, G., {et~al.} 2000, \apj, 540, 74

\bibitem[{{Sakamoto} {et~al.}(2006){Sakamoto}, {Barbier}, {Barthelmy},
  {Cummings}, {Fenimore}, {Gehrels}, {Hullinger}, {Krimm}, {Markwardt},
  {Palmer}, {Parsons}, {Sato}, \& {Tueller}}]{Sakamoto2006}
{Sakamoto}, T., {Barbier}, L., {Barthelmy}, S.~D., {et~al.} 2006, \apjl, 636,
  L73

\bibitem[{{Sanchez-Ramirez} {et~al.}(2012){Sanchez-Ramirez}, {Gorosabel}, {de
  Ugarte Postigo}, \& {Gonzalez Perez}}]{SanchezRamirez2012}
{Sanchez-Ramirez}, R., {Gorosabel}, J., {de Ugarte Postigo}, A., \& {Gonzalez
  Perez}, J.~M. 2012, GRB Coordinates Network, 13723, 1

\bibitem[{{Sari} {et~al.}(1998){Sari}, {Piran}, \& {Narayan}}]{Sari1998}
{Sari}, R., {Piran}, T., \& {Narayan}, R. 1998, \apjl, 497, L17

\bibitem[{{Schady} {et~al.}(2007{\natexlab{a}}){Schady}, {de Pasquale}, {Page},
  {Vetere}, {Pandey}, {Wang}, {Cummings}, {Zhang}, {Zane}, {Breeveld},
  {Burrows}, {Gehrels}, {Gronwall}, {Hunsberger}, {Markwardt}, {Mason},
  {M{\'e}sz{\'a}ros}, {Norris}, {Oates}, {Pagani}, {Poole}, {Roming}, {Smith},
  \& {vanden Berk}}]{Schady2007b}
{Schady}, P., {de Pasquale}, M., {Page}, M.~J., {et~al.} 2007{\natexlab{a}},
  \mnras, 380, 1041

\bibitem[{{Schady} {et~al.}(2007{\natexlab{b}}){Schady}, {Mason}, {Page}, {de
  Pasquale}, {Morris}, {Romano}, {Roming}, {Immler}, \& {vanden
  Berk}}]{Schady2007a}
{Schady}, P., {Mason}, K.~O., {Page}, M.~J., {et~al.} 2007{\natexlab{b}},
  \mnras, 377, 273

\bibitem[{{Schady} {et~al.}(2010){Schady}, {Page}, {Oates}, {Still}, {de
  Pasquale}, {Dwelly}, {Kuin}, {Holland}, {Marshall}, \& {Roming}}]{Schady2010}
{Schady}, P., {Page}, M.~J., {Oates}, S.~R., {et~al.} 2010, \mnras, 401, 2773

\bibitem[{{Schaefer} \& {Collazzi}(2007)}]{Schaefer2007a}
{Schaefer}, B.~E. \& {Collazzi}, A.~C. 2007, \apjl, 656, L53

\bibitem[{{Schaefer} {et~al.}(2007){Schaefer}, {McKay}, \&
  {Yuan}}]{Schaefer2007b}
{Schaefer}, B.~E., {McKay}, T.~A., \& {Yuan}, F. 2007, GRB Coordinates Network,
  6948, 1

\bibitem[{{Schlegel} {et~al.}(1998){Schlegel}, {Finkbeiner}, \&
  {Davis}}]{Schlegel1998}
{Schlegel}, D.~J., {Finkbeiner}, D.~P., \& {Davis}, M. 1998, \apj, 500, 525

\bibitem[{{Schmidl} {et~al.}(2014){Schmidl}, {Graham}, \&
  {Greiner}}]{Schmidl2014}
{Schmidl}, S., {Graham}, J.~F., \& {Greiner}, J. 2014, GRB Coordinates Network,
  16898, 1

\bibitem[{{Schulze} {et~al.}(2014){Schulze}, {Wiersema}, {Xu}, \&
  {Fynbo}}]{Schulze2014}
{Schulze}, S., {Wiersema}, K., {Xu}, D., \& {Fynbo}, J.~P.~U. 2014, GRB
  Coordinates Network, 15831, 1

\bibitem[{{Shahmoradi}(2013)}]{Shahmoradi2013}
{Shahmoradi}, A. 2013, ArXiv e-prints

\bibitem[{{Shahmoradi} \& {Nemiroff}(2011)}]{Shahmoradi2011}
{Shahmoradi}, A. \& {Nemiroff}, R.~J. 2011, \mnras, 411, 1843

\bibitem[{{Singer} {et~al.}(2015){Singer}, {Kasliwal}, {Cenko}, {Perley},
  {Anderson}, {Anupama}, {Arcavi}, {Bhalerao}, {Bue}, {Cao}, {Connaughton},
  {Corsi}, {Cucchiara}, {Fender}, {Fox}, {Gehrels}, {Goldstein}, {Gorosabel},
  {Horesh}, {Hurley}, {Johansson}, {Kann}, {Kouveliotou}, {Huang}, {Kulkarni},
  {Masci}, {Nugent}, {Rau}, {Rebbapragada}, {Staley}, {Svinkin}, {Th{\"o}ne},
  {de Ugarte Postigo}, {Urata}, \& {Weinstein}}]{Singer2015}
{Singer}, L.~P., {Kasliwal}, M.~M., {Cenko}, S.~B., {et~al.} 2015, ArXiv
  e-prints

\bibitem[{{Smette} {et~al.}(2013){Smette}, {Ledoux}, {Vreeswijk}, {De Cia},
  {Petitjean}, {Fynbo}, {Malesani}, \& {Fox}}]{Smette2013}
{Smette}, A., {Ledoux}, C., {Vreeswijk}, P., {et~al.} 2013, GRB Coordinates
  Network, 14848, 1

\bibitem[{{Smith} {et~al.}(2009{\natexlab{a}}){Smith}, {Gomboc}, {Guidorzi},
  {Mundell}, {Steele}, {Melandri}, {Kobayashi}, {Mottram}, {Bersier}, \&
  {Cano}}]{Smith2009a}
{Smith}, R.~J., {Gomboc}, A., {Guidorzi}, C., {et~al.} 2009{\natexlab{a}}, GRB
  Coordinates Network, 9770, 1

\bibitem[{{Smith} {et~al.}(2009{\natexlab{b}}){Smith}, {Steele}, {Cano},
  {Guidorzi}, {Mundell}, {Melandri}, {Kobayashi}, {Mottram}, {Bersier}, \&
  {Gomboc}}]{Smith2009b}
{Smith}, R.~J., {Steele}, I.~A., {Cano}, Z., {et~al.} 2009{\natexlab{b}}, GRB
  Coordinates Network, 9784, 1

\bibitem[{{Soderberg} {et~al.}(2006){Soderberg}, {Kulkarni}, {Price}, {Fox},
  {Berger}, {Moon}, {Cenko}, {Gal-Yam}, {Frail}, {Chevalier}, {Cowie}, {Da
  Costa}, {MacFadyen}, {McCarthy}, {Noel}, {Park}, {Peterson}, {Phillips},
  {Rauch}, {Rest}, {Rich}, {Roth}, {Roth}, {Schmidt}, {Smith}, \&
  {Wood}}]{Soderberg2006}
{Soderberg}, A.~M., {Kulkarni}, S.~R., {Price}, P.~A., {et~al.} 2006, \apj,
  636, 391

\bibitem[{{Sonbas} {et~al.}(2013){Sonbas}, {Temiz}, {Guver}, {Eker}, {Kaynar},
  {Gogus}, \& {Kirbiyik}}]{Sonbas2013}
{Sonbas}, E., {Temiz}, U., {Guver}, T., {et~al.} 2013, GRB Coordinates Network,
  15161, 1

\bibitem[{{Sonoda} {et~al.}(2008){Sonoda}, {Tanaka}, {Hara}, {Ohmori}, {Kono},
  {Hayasi}, {Daikyuji}, {Noda}, {Nisioka}, \& {Yamauchi}}]{Sonoda2008}
{Sonoda}, E., {Tanaka}, H., {Hara}, R., {et~al.} 2008, GRB Coordinates Network,
  8697, 1

\bibitem[{{Soulier}(2012)}]{Soulier2012}
{Soulier}, J.-F. 2012, GRB Coordinates Network, 13126, 1

\bibitem[{{Stanek} {et~al.}(2005){Stanek}, {Garnavich}, {Nutzman}, {Hartman},
  {Garg}, {Adelberger}, {Berlind}, {Bonanos}, {Calkins}, {Challis}, {Gaudi},
  {Holman}, {Kirshner}, {McLeod}, {Osip}, {Pimenova}, {Reiprich}, {Romanishin},
  {Spahr}, {Tegler}, \& {Zhao}}]{Stanek2005}
{Stanek}, K.~Z., {Garnavich}, P.~M., {Nutzman}, P.~A., {et~al.} 2005, \apjl,
  626, L5

\bibitem[{{Starling} {et~al.}(2009){Starling}, {Rol}, {van der Horst}, {Yoon},
  {Pal'Shin}, {Ledoux}, {Page}, {Fynbo}, {Wiersema}, {Tanvir}, {Jakobsson},
  {Guidorzi}, {Curran}, {Levan}, {O'Brien}, {Osborne}, {Svinkin}, {de Ugarte
  Postigo}, {Oosting}, \& {Howarth}}]{Starling2009}
{Starling}, R.~L.~C., {Rol}, E., {van der Horst}, A.~J., {et~al.} 2009, \mnras,
  400, 90

\bibitem[{{Sudilovsky} {et~al.}(2013){Sudilovsky}, {Nicuesa Guelbenzu}, \&
  {Greiner}}]{Sudilovsky2013}
{Sudilovsky}, V., {Nicuesa Guelbenzu}, A., \& {Greiner}, J. 2013, GRB
  Coordinates Network, 14364, 1

\bibitem[{{Tanga} {et~al.}(2014{\natexlab{a}}){Tanga}, {Graham}, {Kann}, \&
  {Greiner}}]{Tanga2014b}
{Tanga}, M., {Graham}, J.~F., {Kann}, D.~A., \& {Greiner}, J.
  2014{\natexlab{a}}, GRB Coordinates Network, 16458, 1

\bibitem[{{Tanga} {et~al.}(2014{\natexlab{b}}){Tanga}, {Klose}, \&
  {Greiner}}]{Tanga2014a}
{Tanga}, M., {Klose}, S., \& {Greiner}, J. 2014{\natexlab{b}}, GRB Coordinates
  Network, 15665, 1

\bibitem[{{Tanigawa} {et~al.}(2013){Tanigawa}, {Yoshii}, {Ito}, {Saito},
  {Yano}, {Usui}, {Tachibana}, {Kurita}, {Yatsu}, \& {Kawai}}]{Tanigawa2013}
{Tanigawa}, T., {Yoshii}, T., {Ito}, K., {et~al.} 2013, GRB Coordinates
  Network, 15481, 1

\bibitem[{{Tanvir} {et~al.}(2009){Tanvir}, {Fox}, {Levan}, {Berger},
  {Wiersema}, {Fynbo}, {Cucchiara}, {Kr{\"u}hler}, {Gehrels}, {Bloom},
  {Greiner}, {Evans}, {Rol}, {Olivares}, {Hjorth}, {Jakobsson}, {Farihi},
  {Willingale}, {Starling}, {Cenko}, {Perley}, {Maund}, {Duke}, {Wijers},
  {Adamson}, {Allan}, {Bremer}, {Burrows}, {Castro-Tirado}, {Cavanagh}, {de
  Ugarte Postigo}, {Dopita}, {Fatkhullin}, {Fruchter}, {Foley}, {Gorosabel},
  {Kennea}, {Kerr}, {Klose}, {Krimm}, {Komarova}, {Kulkarni}, {Moskvitin},
  {Mundell}, {Naylor}, {Page}, {Penprase}, {Perri}, {Podsiadlowski}, {Roth},
  {Rutledge}, {Sakamoto}, {Schady}, {Schmidt}, {Soderberg}, {Sollerman},
  {Stephens}, {Stratta}, {Ukwatta}, {Watson}, {Westra}, {Wold}, \&
  {Wolf}}]{Tanvir2009}
{Tanvir}, N.~R., {Fox}, D.~B., {Levan}, A.~J., {et~al.} 2009, \nat, 461, 1254

\bibitem[{{Tanvir} {et~al.}(2014{\natexlab{a}}){Tanvir}, {Levan}, {Cucchiarra},
  {Perley}, \& {Cenko}}]{Tanvir2014a}
{Tanvir}, N.~R., {Levan}, A.~J., {Cucchiarra}, A., {Perley}, D., \& {Cenko},
  S.~B. 2014{\natexlab{a}}, GRB Coordinates Network, 16125, 1

\bibitem[{{Tanvir} {et~al.}(2013){Tanvir}, {Levan}, {Matulonis}, \&
  {Smith}}]{Tanvir2013}
{Tanvir}, N.~R., {Levan}, A.~J., {Matulonis}, T., \& {Smith}, A.~B. 2013, GRB
  Coordinates Network, 14567, 1

\bibitem[{{Tanvir} {et~al.}(2014{\natexlab{b}}){Tanvir}, {Levan}, {Wiersema},
  {Petric}, {Chiboucas}, \& {Miller}}]{Tanvir2014b}
{Tanvir}, N.~R., {Levan}, A.~J., {Wiersema}, K., {et~al.} 2014{\natexlab{b}},
  GRB Coordinates Network, 16150, 1

\bibitem[{{Tanvir} {et~al.}(2011){Tanvir}, {Wiersema}, {Levan}, {Cenko}, \&
  {Geballe}}]{Tanvir2011}
{Tanvir}, N.~R., {Wiersema}, K., {Levan}, A.~J., {Cenko}, S.~B., \& {Geballe},
  T. 2011, GRB Coordinates Network, 12225, 1

\bibitem[{{Tasselli}(2011)}]{Tasselli2011}
{Tasselli}, A. 2011, GRB Coordinates Network, 12014, 1

\bibitem[{{Tello} {et~al.}(2010){Tello}, {Jelinek}, {Castro-Tirado},
  {Gorosabel}, {de Ugarte Postigo}, {Allen}, {Yock}, \& {Lin}}]{Tello2010}
{Tello}, J.~C., {Jelinek}, M., {Castro-Tirado}, A.~J., {et~al.} 2010, GRB
  Coordinates Network, 11343, 1

\bibitem[{{Tello} {et~al.}(2012){Tello}, {Sanchez-Ramirez}, {Gorosabel},
  {Castro-Tirado}, {Rivero}, {Gomez-Velarde}, \& {Klotz}}]{Tello2012}
{Tello}, J.~C., {Sanchez-Ramirez}, R., {Gorosabel}, J., {et~al.} 2012, GRB
  Coordinates Network, 13118, 1

\bibitem[{{Terron} {et~al.}(2013){Terron}, {Fernandez}, \&
  {Gorosabel}}]{Terron2013}
{Terron}, V., {Fernandez}, M., \& {Gorosabel}, J. 2013, GRB Coordinates
  Network, 15411, 1

\bibitem[{{Testa} {et~al.}(2003){Testa}, {Fugazza}, {Della Valle}, {Malesani},
  {Mason}, {Pian}, {Antonelli}, {Benetti}, {Cocozza}, {Covino}, {Ghisellini},
  {Israel}, {Masetti}, {Palazzi}, \& {Stella}}]{Testa2003}
{Testa}, V., {Fugazza}, D., {Della Valle}, M., {et~al.} 2003, GRB Coordinates
  Network, 1821, 1

\bibitem[{{Thoene} {et~al.}(2008{\natexlab{a}}){Thoene}, {De Cia}, {Malesani},
  \& {Vreeswijk}}]{Thoene2008a}
{Thoene}, C.~C., {De Cia}, A., {Malesani}, D., \& {Vreeswijk}, P.~M.
  2008{\natexlab{a}}, GRB Coordinates Network, 7587, 1

\bibitem[{{Thoene} {et~al.}(2012){Thoene}, {de Ugarte Postigo}, {Gorosabel},
  {Sanchez-Ramirez}, {Fynbo}, \& {Gomez Velarde}}]{Thoene2012}
{Thoene}, C.~C., {de Ugarte Postigo}, A., {Gorosabel}, J., {et~al.} 2012, GRB
  Coordinates Network, 13628, 1

\bibitem[{{Thoene} {et~al.}(2009{\natexlab{a}}){Thoene}, {Goldoni}, {Covino},
  {Antonelli}, {Malesani}, {Fynbo}, {Levan}, {Jakobsson}, {Flores},
  {Milvang-Jensen}, {Hjorth}, {Watson}, {Wiersema}, {Tanvir}, \& {de Ugarte
  Postigo}}]{Thoene2009c}
{Thoene}, C.~C., {Goldoni}, P., {Covino}, S., {et~al.} 2009{\natexlab{a}}, GRB
  Coordinates Network, 10233, 1

\bibitem[{{Thoene} {et~al.}(2009{\natexlab{b}}){Thoene}, {Jakobsson}, {De Cia},
  {Levan}, {Fynbo}, {Hjorth}, {Malesani}, {Tanvir}, {Fugazza}, \&
  {D'Avanzo}}]{Thoene2009b}
{Thoene}, C.~C., {Jakobsson}, P., {De Cia}, A., {et~al.} 2009{\natexlab{b}},
  GRB Coordinates Network, 9409, 1

\bibitem[{{Thoene} {et~al.}(2009{\natexlab{c}}){Thoene}, {Malesani}, {Levan},
  {Jakobsson}, {Hjorth}, \& {Tanvir}}]{Thoene2009a}
{Thoene}, C.~C., {Malesani}, D., {Levan}, A.~J., {et~al.} 2009{\natexlab{c}},
  GRB Coordinates Network, 9403, 1

\bibitem[{{Thoene} {et~al.}(2008{\natexlab{b}}){Thoene}, {Malesani},
  {Vreeswijk}, {Fynbo}, {Jakobsson}, {Ledoux}, \& {Smette}}]{Thoene2008b}
{Thoene}, C.~C., {Malesani}, D., {Vreeswijk}, P.~M., {et~al.}
  2008{\natexlab{b}}, GRB Coordinates Network, 7602, 1

\bibitem[{{Th{\"o}ne} {et~al.}(2007){Th{\"o}ne}, {Greiner}, {Savaglio}, \&
  {Jehin}}]{Thoene2007}
{Th{\"o}ne}, C.~C., {Greiner}, J., {Savaglio}, S., \& {Jehin}, E. 2007, \apj,
  671, 628

\bibitem[{{Toma} {et~al.}(2005){Toma}, {Yamazaki}, \& {Nakamura}}]{Toma2005}
{Toma}, K., {Yamazaki}, R., \& {Nakamura}, T. 2005, \apj, 635, 481

\bibitem[{{Troja} {et~al.}(2012){Troja}, {Sakamoto}, {Guidorzi}, {Norris},
  {Panaitescu}, {Kobayashi}, {Omodei}, {Brown}, {Burrows}, {Evans}, {Gehrels},
  {Marshall}, {Mawson}, {Melandri}, {Mundell}, {Oates}, {Pal'shin}, {Preece},
  {Racusin}, {Steele}, {Tanvir}, {Vasileiou}, {Wilson-Hodge}, \&
  {Yamaoka}}]{Troja2012}
{Troja}, E., {Sakamoto}, T., {Guidorzi}, C., {et~al.} 2012, \apj, 761, 50

\bibitem[{{Trotter} {et~al.}(2013{\natexlab{a}}){Trotter}, {Frank}, {Lacluyze},
  {Reichart}, {Haislip}, {Ivarsen}, {Moore}, {Cromartie}, {Egger}, {Foster},
  {Nysewander}, {Oza}, {Speckhard}, \& {Crain}}]{Trotter2013a}
{Trotter}, A., {Frank}, N., {Lacluyze}, A., {et~al.} 2013{\natexlab{a}}, GRB
  Coordinates Network, 14375, 1

\bibitem[{{Trotter} {et~al.}(2013{\natexlab{b}}){Trotter}, {Frank}, {Lacluyze},
  {Reichart}, {McLin}, {Cominsky}, {Berger}, {Cromartie}, {Egger}, {Foster},
  {Haislip}, {Ivarsen}, {Maples}, {Moore}, {Nysewander}, {Speckhard}, \&
  {Crain}}]{Trotter2013b}
{Trotter}, A., {Frank}, N., {Lacluyze}, A., {et~al.} 2013{\natexlab{b}}, GRB
  Coordinates Network, 14445, 1

\bibitem[{{Trotter} {et~al.}(2014){Trotter}, {Haislip}, {Reichart}, {Lacluyze},
  {Verveer}, {Spuck}, {Foster}, {Frank}, {Ivarsen}, {Moore}, {Nysewander},
  {Beauchemin}, {Berger}, {Carroll}, {Cromartie}, {Egger}, {Hinckle},
  {Ireland}, {Maples}, {Scott}, \& {Crain}}]{Trotter2014}
{Trotter}, A., {Haislip}, J., {Reichart}, D., {et~al.} 2014, GRB Coordinates
  Network, 15859, 1

\bibitem[{{Uemura} {et~al.}(2006){Uemura}, {Arai}, \& {Uehara}}]{Uemura2006}
{Uemura}, M., {Arai}, A., \& {Uehara}, T. 2006, GRB Coordinates Network, 5828,
  1

\bibitem[{{Ukwatta} {et~al.}(2011){Ukwatta}, {Sonbas}, {Gehrels}, {Linnemann},
  {Tollefson}, \& {Abeysekara}}]{Ukwatta2011}
{Ukwatta}, T.~N., {Sonbas}, E., {Gehrels}, N., {et~al.} 2011, GRB Coordinates
  Network, 11715, 1

\bibitem[{{Updike} {et~al.}(2009{\natexlab{a}}){Updike}, {Brittain},
  {Hartmann}, {Colson}, {Cumbee}, {Hackett}, {Lewis}, \&
  {Kronberg}}]{Updike2009a}
{Updike}, A., {Brittain}, S., {Hartmann}, D., {et~al.} 2009{\natexlab{a}}, GRB
  Coordinates Network, 9529, 1

\bibitem[{{Updike} {et~al.}(2009{\natexlab{b}}){Updike}, {Rau}, {Kruehler},
  {Olivares}, \& {Greiner}}]{Updike2009b}
{Updike}, A., {Rau}, A., {Kruehler}, T., {Olivares}, F., \& {Greiner}, J.
  2009{\natexlab{b}}, GRB Coordinates Network, 9773, 1

\bibitem[{{Updike} {et~al.}(2009{\natexlab{c}}){Updike}, {Rossi}, \&
  {Greiner}}]{Updike2009c}
{Updike}, A., {Rossi}, A., \& {Greiner}, J. 2009{\natexlab{c}}, GRB Coordinates
  Network, 10271, 1

\bibitem[{{Updike} {et~al.}(2007{\natexlab{a}}){Updike}, {Hartmann}, {Henson},
  {Mesler}, {Bunker}, \& {Carson}}]{Updike2007a}
{Updike}, A.~C., {Hartmann}, D.~H., {Henson}, G., {et~al.} 2007{\natexlab{a}},
  GRB Coordinates Network, 6515, 1

\bibitem[{{Updike} {et~al.}(2011){Updike}, {Kann}, {Hartmann}, \&
  {Rumstay}}]{Updike2011}
{Updike}, A.~C., {Kann}, D.~A., {Hartmann}, D.~H., \& {Rumstay}, K. 2011, GRB
  Coordinates Network, 12001, 1

\bibitem[{{Updike} {et~al.}(2007{\natexlab{b}}){Updike}, {Milne}, {Williams},
  \& {Hartmann}}]{Updike2007c}
{Updike}, A.~C., {Milne}, P.~A., {Williams}, G.~G., \& {Hartmann}, D.~H.
  2007{\natexlab{b}}, GRB Coordinates Network, 6535, 1

\bibitem[{{Updike} {et~al.}(2007{\natexlab{c}}){Updike}, {Puls}, {Hartmann},
  {Wood}, {Cardenzana}, \& {Pederson}}]{Updike2007b}
{Updike}, A.~C., {Puls}, J., {Hartmann}, D.~H., {et~al.} 2007{\natexlab{c}},
  GRB Coordinates Network, 6530, 1

\bibitem[{{Urata}(2006{\natexlab{a}})}]{Urata2006b}
{Urata}, J. 2006{\natexlab{a}}, GRB Coordinates Network, 5547, 1

\bibitem[{{Urata}(2006{\natexlab{b}})}]{Urata2006a}
{Urata}, J. 2006{\natexlab{b}}, GRB Coordinates Network, 4430, 1

\bibitem[{{Urata} {et~al.}(2007){Urata}, {Huang}, {Qiu}, {Hu}, {Kuo},
  {Tamagawa}, {Ip}, {Kinoshita}, {Fukushi}, {Isogai}, {Miyata}, {Nakada},
  {Aoki}, {Soyano}, {Tarusawa}, {Mito}, {Onda}, {Ibrahimov}, {Pozanenko}, \&
  {Makishima}}]{Urata2007}
{Urata}, Y., {Huang}, K.~Y., {Qiu}, Y.~L., {et~al.} 2007, \apjl, 655, L81

\bibitem[{{Urata} {et~al.}(2003){Urata}, {Nishiura}, {Miyata}, {Mito},
  {Kawabata}, {Nakada}, {Aoki}, {Soyano}, {Tarusawa}, {Yoshida}, {Tamagawa}, \&
  {Makishima}}]{Urata2003}
{Urata}, Y., {Nishiura}, S., {Miyata}, T., {et~al.} 2003, \apjl, 595, L21

\bibitem[{{Urata} {et~al.}(2009){Urata}, {Zhang}, {Wen}, {Huang}, \&
  {Wang}}]{Urata2009}
{Urata}, Y., {Zhang}, Z.~W., {Wen}, C.~Y., {Huang}, K.~Y., \& {Wang}, S.~Y.
  2009, GRB Coordinates Network, 9240, 1

\bibitem[{{Vaalsta} {et~al.}(2009{\natexlab{a}}){Vaalsta}, {Coward}, {Ward},
  {Moore}, {Imerito}, {Blair}, {Burman}, {Gordon}, {Fletcher}, {Ahmet},
  {Burrell}, {Smith}, {Todd}, {Zadnik}, {Boer}, {Laas-Bourez}, {Klotz}, \&
  {Thierry}}]{Vaalsta2009b}
{Vaalsta}, T.~P., {Coward}, D.~M., {Ward}, I., {et~al.} 2009{\natexlab{a}}, GRB
  Coordinates Network, 10238, 1

\bibitem[{{Vaalsta} {et~al.}(2009{\natexlab{b}}){Vaalsta}, {Yan}, {Zadko},
  {Miao}, {Moore}, {Frost}, {Coward}, {Imerito}, {Blair}, {Burman}, {Luckas},
  {Gordon}, {Fletcher}, {Ahmet}, {Todd}, {Zadnik}, {Boer}, \&
  {Klotz}}]{Vaalsta2009a}
{Vaalsta}, T.~P., {Yan}, L., {Zadko}, J., {et~al.} 2009{\natexlab{b}}, GRB
  Coordinates Network, 9380, 1

\bibitem[{{Vestrand} {et~al.}(2014){Vestrand}, {Wren}, {Panaitescu}, {Wozniak},
  {Davis}, {Palmer}, {Vianello}, {Omodei}, {Xiong}, {Briggs}, {Elphick},
  {Paciesas}, \& {Rosing}}]{Vestrand2014}
{Vestrand}, W.~T., {Wren}, J.~A., {Panaitescu}, A., {et~al.} 2014, Science,
  343, 38

\bibitem[{{Virgili} {et~al.}(2013){Virgili}, {Mundell}, {Japelj}, {Gomboc},
  {Smith}, \& {Melandri}}]{Virgili2013}
{Virgili}, F.~J., {Mundell}, C.~G., {Japelj}, J., {et~al.} 2013, GRB
  Coordinates Network, 15406, 1

\bibitem[{{Volnova} {et~al.}(2014{\natexlab{a}}){Volnova}, {Inasaridze},
  {Inasaridze}, {Zhuzhunadze}, {Krugly}, {Molotov}, \&
  {Pozanenko}}]{Volnova2014g}
{Volnova}, A., {Inasaridze}, R., {Inasaridze}, G., {et~al.} 2014{\natexlab{a}},
  GRB Coordinates Network, 15730, 1

\bibitem[{{Volnova} {et~al.}(2014{\natexlab{b}}){Volnova}, {Inasaridze},
  {Kvaratskhelia}, {Ayvazian}, {Krugly}, {Molotov}, \&
  {Pozanenko}}]{Volnova2014c}
{Volnova}, A., {Inasaridze}, R., {Kvaratskhelia}, O., {et~al.}
  2014{\natexlab{b}}, GRB Coordinates Network, 16264, 1

\bibitem[{{Volnova} {et~al.}(2014{\natexlab{c}}){Volnova}, {Inasaridze},
  {Kvaratskhelia}, {Ayvazian}, {Krugly}, {Molotov}, \&
  {Pozanenko}}]{Volnova2014e}
{Volnova}, A., {Inasaridze}, R., {Kvaratskhelia}, O., {et~al.}
  2014{\natexlab{c}}, GRB Coordinates Network, 16536, 1

\bibitem[{{Volnova} {et~al.}(2014{\natexlab{d}}){Volnova}, {Klunko},
  {Eselevich}, {Korobtsev}, \& {Pozanenko}}]{Volnova2014a}
{Volnova}, A., {Klunko}, E., {Eselevich}, M., {Korobtsev}, I., \& {Pozanenko},
  A. 2014{\natexlab{d}}, GRB Coordinates Network, 16168, 1

\bibitem[{{Volnova} {et~al.}(2014{\natexlab{e}}){Volnova}, {Klunko},
  {Eselevich}, {Korobtsev}, \& {Pozanenko}}]{Volnova2014d}
{Volnova}, A., {Klunko}, E., {Eselevich}, M., {Korobtsev}, I., \& {Pozanenko},
  A. 2014{\natexlab{e}}, GRB Coordinates Network, 16247, 1

\bibitem[{{Volnova} {et~al.}(2014{\natexlab{f}}){Volnova}, {Kusakin},
  {Khruslov}, \& {Pozanenko}}]{Volnova2014b}
{Volnova}, A., {Kusakin}, A.~V., {Khruslov}, A.~V., \& {Pozanenko}, A.
  2014{\natexlab{f}}, GRB Coordinates Network, 16318, 1

\bibitem[{{Volnova} {et~al.}(2013{\natexlab{a}}){Volnova}, {Linkov},
  {Polyakov}, {Ivanov}, {Molotov}, \& {Pozanenko}}]{Volnova2013c}
{Volnova}, A., {Linkov}, V., {Polyakov}, K., {et~al.} 2013{\natexlab{a}}, GRB
  Coordinates Network, 15188, 1

\bibitem[{{Volnova} {et~al.}(2013{\natexlab{b}}){Volnova}, {Matkin}, {Stepura},
  {Molotov}, \& {Pozanenko}}]{Volnova2013a}
{Volnova}, A., {Matkin}, A., {Stepura}, A., {Molotov}, I., \& {Pozanenko}, A.
  2013{\natexlab{b}}, GRB Coordinates Network, 15185, 1

\bibitem[{{Volnova} {et~al.}(2013{\natexlab{c}}){Volnova}, {Minikulov},
  {Gulyamov}, {Molotov}, \& {Pozanenko}}]{Volnova2013b}
{Volnova}, A., {Minikulov}, N., {Gulyamov}, M., {Molotov}, I., \& {Pozanenko},
  A. 2013{\natexlab{c}}, GRB Coordinates Network, 15186, 1

\bibitem[{{Volnova} {et~al.}(2014{\natexlab{g}}){Volnova}, {Mundrzyjewski},
  {Kusakin}, \& {Pozanenko}}]{Volnova2014j}
{Volnova}, A., {Mundrzyjewski}, W., {Kusakin}, A., \& {Pozanenko}, A.
  2014{\natexlab{g}}, GRB Coordinates Network, 16651, 1

\bibitem[{{Volnova} {et~al.}(2009){Volnova}, {Pavlenko}, {Antoniuk},
  {Rumyantsev}, \& {Pozanenko}}]{Volnova2009}
{Volnova}, A., {Pavlenko}, E., {Antoniuk}, O., {Rumyantsev}, V., \&
  {Pozanenko}, A. 2009, GRB Coordinates Network, 9811, 1

\bibitem[{{Volnova} {et~al.}(2011){Volnova}, {Pozanenko}, {Korobtsev}, \&
  {Elunko}}]{Volnova2011}
{Volnova}, A., {Pozanenko}, A., {Korobtsev}, I., \& {Elunko}, E. 2011, GRB
  Coordinates Network, 11742, 1

\bibitem[{{Volnova} {et~al.}(2014{\natexlab{h}}){Volnova}, {Schmalz},
  {Tungalag}, {Mas}, {Molotov}, \& {Pozanenko}}]{Volnova2014i}
{Volnova}, A., {Schmalz}, S., {Tungalag}, N., {et~al.} 2014{\natexlab{h}}, GRB
  Coordinates Network, 16558, 1

\bibitem[{{Volnova} {et~al.}(2014{\natexlab{i}}){Volnova}, {Pozanenko},
  {Gorosabel}, {Perley}, {Frederiks}, {Kann}, {Rumyantsev}, {Biryukov},
  {Burkhonov}, {Castro-Tirado}, {Ferrero}, {Golenetskii}, {Klose}, {Loznikov},
  {Minaev}, {Stecklum}, {Svinkin}, {Tsvetkova}, {de Ugarte Postigo}, \&
  {Ulanov}}]{Volnova2014f}
{Volnova}, A.~A., {Pozanenko}, A.~S., {Gorosabel}, J., {et~al.}
  2014{\natexlab{i}}, \mnras, 442, 2586

\bibitem[{{Vreeswijk} {et~al.}(1999){Vreeswijk}, {Galama}, {Rol}, {Stappers},
  {Palazzi}, {Pian}, {Masetti}, {Frontera}, {van Paradijs}, {Kouveliotou}, \&
  {Boehnhardt}}]{Vreeswijk1999}
{Vreeswijk}, P.~M., {Galama}, T.~J., {Rol}, E., {et~al.} 1999, GRB Coordinates
  Network, 324, 1

\bibitem[{{Vreeswijk} {et~al.}(2008{\natexlab{a}}){Vreeswijk},
  {Milvang-Jensen}, {Smette}, {Malesani}, {Fynbo}, {Jakobsson}, {Jaunsen}, \&
  {Ledoux}}]{Vreeswijk2008a}
{Vreeswijk}, P.~M., {Milvang-Jensen}, B., {Smette}, A., {et~al.}
  2008{\natexlab{a}}, GRB Coordinates Network, 7451, 1

\bibitem[{{Vreeswijk} {et~al.}(2008{\natexlab{b}}){Vreeswijk}, {Thoene},
  {Malesani}, {Fynbo}, {Hjorth}, {Jakobsson}, {Tanvir}, \&
  {Levan}}]{Vreeswijk2008b}
{Vreeswijk}, P.~M., {Thoene}, C.~C., {Malesani}, D., {et~al.}
  2008{\natexlab{b}}, GRB Coordinates Network, 7601, 1

\bibitem[{{Wang} {et~al.}(2008){Wang}, {Schwamb}, {Huang}, {Wen}, {Zhang},
  {Wang}, {Chen}, {Bianco}, {Dave}, {Lehner}, {Marshall}, {Porrata}, {Alcock},
  {Byun}, {Cook}, {King}, {Lee}, \& {Urata}}]{Wang2008}
{Wang}, J.~H., {Schwamb}, M.~E., {Huang}, K.~Y., {et~al.} 2008, \apjl, 679, L5

\bibitem[{{Watson} {et~al.}(2013{\natexlab{a}}){Watson}, {Butler}, {Kutyrev},
  {Lee}, {Richer}, {Klein}, {Fox}, {Prochaska}, {Bloom}, {Cucchiara}, {Troja},
  {Littlejohns}, {Ramirez-Ruiz}, {de Diego}, {Georgiev}, {Gonzalez},
  {Roman-Zuniga}, {Gehrels}, \& {Moseley}}]{Watson2013a}
{Watson}, A.~M., {Butler}, N., {Kutyrev}, A., {et~al.} 2013{\natexlab{a}}, GRB
  Coordinates Network, 14439, 1

\bibitem[{{Watson} {et~al.}(2013{\natexlab{b}}){Watson}, {Butler}, {Kutyrev},
  {Lee}, {Richer}, {Klein}, {Fox}, {Prochaska}, {Bloom}, {Cucchiara}, {Troja},
  {Littlejohns}, {Ramirez-Ruiz}, {de Diego}, {Georgiev}, {Gonzalez},
  {Roman-Zuniga}, {Gehrels}, \& {Moseley}}]{Watson2013b}
{Watson}, A.~M., {Butler}, N., {Kutyrev}, A., {et~al.} 2013{\natexlab{b}}, GRB
  Coordinates Network, 14595, 1

\bibitem[{{Watson} {et~al.}(2013{\natexlab{c}}){Watson}, {Littlejohns},
  {Butler}, {Kutyrev}, {Lee}, {Richer}, {Klein}, {Fox}, {Prochaska}, {Bloom},
  {Cucchiara}, {Troja}, {Ramirez-Ruiz}, {de Diego}, {Georgiev}, {Gonzalez},
  {Roman-Zuniga}, {Gehrels}, \& {Moseley}}]{Watson2013c}
{Watson}, A.~M., {Littlejohns}, O., {Butler}, N., {et~al.} 2013{\natexlab{c}},
  GRB Coordinates Network, 15179, 1

\bibitem[{{Wiersema} {et~al.}(2009){Wiersema}, {Tanvir}, {Cucchiara}, {Levan},
  \& {Fox}}]{Wiersema2009a}
{Wiersema}, K., {Tanvir}, N.~R., {Cucchiara}, A., {Levan}, A.~J., \& {Fox}, D.
  2009, GRB Coordinates Network, 10263, 1

\bibitem[{{Wiersema} {et~al.}(2008){Wiersema}, {van der Horst}, {Kann}, {Rol},
  {Starling}, {Curran}, {Gorosabel}, {Levan}, {Fynbo}, {de Ugarte Postigo},
  {Wijers}, {Castro-Tirado}, {Guziy}, {Hornstrup}, {Hjorth}, {Jel{\'{\i}}nek},
  {Jensen}, {Kidger}, {Mart{\'{\i}}n-Luis}, {Tanvir}, {Tristram}, \&
  {Vreeswijk}}]{Wiersema2008}
{Wiersema}, K., {van der Horst}, A.~J., {Kann}, D.~A., {et~al.} 2008, \aap,
  481, 319

\bibitem[{{Williams} {et~al.}(2002){Williams}, {Blake}, {Hartmann}, \& {S-LOTIS
  collaboration}}]{Williams2002}
{Williams}, G.~G., {Blake}, C., {Hartmann}, D., \& {S-LOTIS collaboration}.
  2002, GRB Coordinates Network, 1492, 1

\bibitem[{{Wo{\'z}niak} {et~al.}(2009){Wo{\'z}niak}, {Vestrand}, {Panaitescu},
  {Wren}, {Davis}, \& {White}}]{Wozniak2009}
{Wo{\'z}niak}, P.~R., {Vestrand}, W.~T., {Panaitescu}, A.~D., {et~al.} 2009,
  \apj, 691, 495

\bibitem[{{Wo{\'z}niak} {et~al.}(2006){Wo{\'z}niak}, {Vestrand}, {Wren},
  {White}, {Evans}, \& {Casperson}}]{Wozniak2006}
{Wo{\'z}niak}, P.~R., {Vestrand}, W.~T., {Wren}, J.~A., {et~al.} 2006, \apjl,
  642, L99

\bibitem[{{Wren} {et~al.}(2011){Wren}, {Vestrand}, {Wozniak}, \&
  {Davis}}]{Wren2011}
{Wren}, J., {Vestrand}, W.~T., {Wozniak}, P.~R., \& {Davis}, H. 2011, GRB
  Coordinates Network, 11730, 1

\bibitem[{{Wren} {et~al.}(2009){Wren}, {Vestrand}, {Wozniak}, {Davis}, \&
  {Norman}}]{Wren2009}
{Wren}, J., {Vestrand}, W.~T., {Wozniak}, P.~R., {Davis}, H., \& {Norman}, B.
  2009, GRB Coordinates Network, 9778, 1

\bibitem[{{Xiao} \& {Schaefer}(2009)}]{Xiao2009}
{Xiao}, L. \& {Schaefer}, B.~E. 2009, \apj, 707, 387

\bibitem[{{Xin} {et~al.}(2010){Xin}, {Liu}, {Qiu}, {Wei}, {Wang}, {Deng}, {Wu},
  {Hu}, \& {Zheng}}]{Xin2010}
{Xin}, L., {Liu}, H., {Qiu}, Y., {et~al.} 2010, GRB Coordinates Network, 10583,
  1

\bibitem[{{Xin} {et~al.}(2013{\natexlab{a}}){Xin}, {Han}, {Qiu}, {Wei}, {Wang},
  {Deng}, \& {Wu}}]{Xin2013b}
{Xin}, L.~P., {Han}, X.~H., {Qiu}, Y.~L., {et~al.} 2013{\natexlab{a}}, GRB
  Coordinates Network, 15146, 1

\bibitem[{{Xin} {et~al.}(2009{\natexlab{a}}){Xin}, {Qian}, {Qiu}, {Wang},
  {Wei}, {Zheng}, {Deng}, \& {Hu}}]{Xin2009b}
{Xin}, L.~P., {Qian}, S.~B., {Qiu}, Y.~L., {et~al.} 2009{\natexlab{a}}, GRB
  Coordinates Network, 10279, 1

\bibitem[{{Xin} {et~al.}(2009{\natexlab{b}}){Xin}, {Wang}, {Wang}, {Qiu},
  {Wei}, {Zheng}, {Deng}, {Wu}, \& {Hu}}]{Xin2009a}
{Xin}, L.~P., {Wang}, X.~F., {Wang}, J., {et~al.} 2009{\natexlab{b}}, GRB
  Coordinates Network, 10080, 1

\bibitem[{{Xin} {et~al.}(2013{\natexlab{b}}){Xin}, {Wei}, {Qiu}, {Deng},
  {Wang}, {Han}, \& {Wu}}]{Xin2013c}
{Xin}, L.~P., {Wei}, J.~Y., {Qiu}, Y.~L., {et~al.} 2013{\natexlab{b}}, GRB
  Coordinates Network, 15425, 1

\bibitem[{{Xin} {et~al.}(2012{\natexlab{a}}){Xin}, {Wei}, {Qiu}, {Wang},
  {Deng}, {Wu}, \& {Han}}]{Xin2012b}
{Xin}, L.~P., {Wei}, J.~Y., {Qiu}, Y.~L., {et~al.} 2012{\natexlab{a}}, GRB
  Coordinates Network, 13149, 1

\bibitem[{{Xin} {et~al.}(2012{\natexlab{b}}){Xin}, {Wei}, {Qiu}, {Wang},
  {Deng}, {Wu}, \& {Han}}]{Xin2012a}
{Xin}, L.~P., {Wei}, J.~Y., {Qiu}, Y.~L., {et~al.} 2012{\natexlab{b}}, GRB
  Coordinates Network, 13131, 1

\bibitem[{{Xin} {et~al.}(2013{\natexlab{c}}){Xin}, {Wei}, {Qiu}, {Wang},
  {Deng}, {Wu}, \& {Han}}]{Xin2013a}
{Xin}, L.~P., {Wei}, J.~Y., {Qiu}, Y.~L., {et~al.} 2013{\natexlab{c}}, GRB
  Coordinates Network, 14571, 1

\bibitem[{{Xin} {et~al.}(2014){Xin}, {Yan}, {Wei}, {Qiu}, {Deng}, {Wang},
  {Han}, \& {Wu}}]{Xin2014}
{Xin}, L.~P., {Yan}, J.~Z., {Wei}, J.~Y., {et~al.} 2014, GRB Coordinates
  Network, 16586, 1

\bibitem[{{Xu}(2014)}]{Xu2014b}
{Xu}, D. 2014, GRB Coordinates Network, 16140, 1

\bibitem[{{Xu} {et~al.}(2013{\natexlab{a}}){Xu}, {Cao}, {Hu}, \&
  {Ai}}]{Xu2013a}
{Xu}, D., {Cao}, C., {Hu}, S.-M., \& {Ai}, J.-M. 2013{\natexlab{a}}, GRB
  Coordinates Network, 14570, 1

\bibitem[{{Xu} {et~al.}(2013{\natexlab{b}}){Xu}, {de Ugarte Postigo},
  {Malesani}, {Vreeswijk}, {D'Elia}, {Fynbo}, {Milvang-Jensen}, {Flores},
  {Hartoog}, {Goldoni}, {Kaper}, \& {Vergani}}]{Xu2013b}
{Xu}, D., {de Ugarte Postigo}, A., {Malesani}, D., {et~al.} 2013{\natexlab{b}},
  GRB Coordinates Network, 14956, 1

\bibitem[{{Xu} {et~al.}(2013{\natexlab{c}}){Xu}, {Fynbo}, {Jakobsson}, {Cano},
  {Milvang-Jensen}, {Malesani}, {de Ugarte Postigo}, \& {Hayes}}]{Xu2013d}
{Xu}, D., {Fynbo}, J.~P.~U., {Jakobsson}, P., {et~al.} 2013{\natexlab{c}}, GRB
  Coordinates Network, 15407, 1

\bibitem[{{Xu} {et~al.}(2011{\natexlab{a}}){Xu}, {Fynbo}, {Nielsen}, \&
  {Jakobsson}}]{Xu2011a}
{Xu}, D., {Fynbo}, J.~P.~U., {Nielsen}, M., \& {Jakobsson}, P.
  2011{\natexlab{a}}, GRB Coordinates Network, 11970, 1

\bibitem[{{Xu} {et~al.}(2009{\natexlab{a}}){Xu}, {Fynbo}, {Tanvir}, {Hjorth},
  {Leloudas}, {Malesani}, {Jakobsson}, {Wilson}, \& {Andersen}}]{Xu2009a}
{Xu}, D., {Fynbo}, J.~P.~U., {Tanvir}, N.~R., {et~al.} 2009{\natexlab{a}}, GRB
  Coordinates Network, 10053, 1

\bibitem[{{Xu} {et~al.}(2011{\natexlab{b}}){Xu}, {Kankare}, {Kangas}, \&
  {Jakobsson}}]{Xu2011b}
{Xu}, D., {Kankare}, E., {Kangas}, T., \& {Jakobsson}, P. 2011{\natexlab{b}},
  GRB Coordinates Network, 11974, 1

\bibitem[{{Xu} {et~al.}(2009{\natexlab{b}}){Xu}, {Leloudas}, {Malesani},
  {Jakobsson}, {Lindberg}, \& {Andersen}}]{Xu2009c}
{Xu}, D., {Leloudas}, G., {Malesani}, D., {et~al.} 2009{\natexlab{b}}, GRB
  Coordinates Network, 10269, 1

\bibitem[{{Xu} {et~al.}(2009{\natexlab{c}}){Xu}, {Malesani}, {Hjorth},
  {Djupvik}, {Datson}, {Jakobsson}, {Carmona}, \&
  {Baldovin-Saavedra}}]{Xu2009b}
{Xu}, D., {Malesani}, D., {Hjorth}, J., {et~al.} 2009{\natexlab{c}}, GRB
  Coordinates Network, 10196, 1

\bibitem[{{Xu} {et~al.}(2013{\natexlab{d}}){Xu}, {Zhang}, {Cao}, \&
  {Hu}}]{Xu2013c}
{Xu}, D., {Zhang}, C.~M., {Cao}, C., \& {Hu}, S.~M. 2013{\natexlab{d}}, GRB
  Coordinates Network, 15142, 1

\bibitem[{{Xu} {et~al.}(2014){Xu}, {Zhang}, {Bai}, {Niu}, {Esamdin}, \&
  {Ma}}]{Xu2014a}
{Xu}, D., {Zhang}, X., {Bai}, C.-H., {et~al.} 2014, GRB Coordinates Network,
  15655, 1

\bibitem[{{Yanagisawa} {et~al.}(2005){Yanagisawa}, {Toda}, \&
  {Kawai}}]{Yanagisawa2005}
{Yanagisawa}, K., {Toda}, H., \& {Kawai}, N. 2005, GRB Coordinates Network,
  3489, 1

\bibitem[{{Yanagisawa} {et~al.}(2006){Yanagisawa}, {Toda}, \&
  {Kawai}}]{Yanagisawa2006}
{Yanagisawa}, K., {Toda}, H., \& {Kawai}, N. 2006, GRB Coordinates Network,
  4517, 1

\bibitem[{{Yoshida} {et~al.}(2009{\natexlab{a}}){Yoshida}, {Kuroda},
  {Yanagisawa}, {Shimizu}, {Nagayama}, {Toda}, \& {Kawai}}]{Yoshida2009b}
{Yoshida}, M., {Kuroda}, D., {Yanagisawa}, K., {et~al.} 2009{\natexlab{a}}, GRB
  Coordinates Network, 10258, 1

\bibitem[{{Yoshida} {et~al.}(2008{\natexlab{a}}){Yoshida}, {Yanagisawa},
  {Kuroda}, {Shimizu}, {Nagayama}, {Toda}, \& {Kawai}}]{Yoshida2008a}
{Yoshida}, M., {Yanagisawa}, K., {Kuroda}, D., {et~al.} 2008{\natexlab{a}}, GRB
  Coordinates Network, 7863, 1

\bibitem[{{Yoshida} {et~al.}(2008{\natexlab{b}}){Yoshida}, {Yanagisawa},
  {Kuroda}, {Shimizu}, {Nagayama}, {Toda}, \& {Kawai}}]{Yoshida2008b}
{Yoshida}, M., {Yanagisawa}, K., {Kuroda}, D., {et~al.} 2008{\natexlab{b}}, GRB
  Coordinates Network, 8097, 1

\bibitem[{{Yoshida} {et~al.}(2009{\natexlab{b}}){Yoshida}, {Yanagisawa},
  {Kuroda}, {Shimizu}, {Nagayama}, {Toda}, \& {Kawai}}]{Yoshida2009a}
{Yoshida}, M., {Yanagisawa}, K., {Kuroda}, D., {et~al.} 2009{\natexlab{b}}, GRB
  Coordinates Network, 9218, 1

\bibitem[{{Yoshii} {et~al.}(2013){Yoshii}, {Ito}, {Saito}, {Yano}, {Usui},
  {Tachibana}, {Kurita}, {Yatsu}, \& {Kawai}}]{Yoshii2013}
{Yoshii}, T., {Ito}, K., {Saito}, Y., {et~al.} 2013, GRB Coordinates Network,
  15143, 1

\bibitem[{{Yoshii} {et~al.}(2014){Yoshii}, {Saito}, {Kurita}, {Tachibana},
  {Ito}, {Usui}, {Tanigawa}, {Yano}, {Yatsu}, \& {Kawai}}]{Yoshii2014}
{Yoshii}, T., {Saito}, Y., {Kurita}, S., {et~al.} 2014, GRB Coordinates
  Network, 15778, 1

\bibitem[{{Yost} {et~al.}(2006){Yost}, {Schaefer}, \& {Yuan}}]{Yost2006}
{Yost}, S.~A., {Schaefer}, B.~E., \& {Yuan}, F. 2006, GRB Coordinates Network,
  5824, 1

\bibitem[{{Yuan}(2009)}]{Yuan2009b}
{Yuan}, F. 2009, GRB Coordinates Network, 9224, 1

\bibitem[{{Yuan} \& {Rujopakarn}(2008)}]{Yuan2008}
{Yuan}, F. \& {Rujopakarn}, W. 2008, GRB Coordinates Network, 8536, 1

\bibitem[{{Yuan} \& {Rujopakarn}(2009)}]{Yuan2009a}
{Yuan}, F. \& {Rujopakarn}, W. 2009, GRB Coordinates Network, 9150, 1

\bibitem[{{Yuan} {et~al.}(2010){Yuan}, {Schady}, {Racusin}, {Willingale},
  {Kr{\"u}hler}, {O'Brien}, {Greiner}, {Oates}, {Rykoff}, {Aharonian},
  {Akerlof}, {Ashley}, {Barthelmy}, {Filgas}, {Flewelling}, {Gehrels},
  {G{\"o}{\v g}{\"u}{\c s}}, {G{\"u}ver}, {Horns}, {K{\i}z{\i}lo{\v g}lu},
  {Krimm}, {McKay}, {{\"O}zel}, {Phillips}, {Quimby}, {Rowell}, {Rujopakarn},
  {Schaefer}, {Vestrand}, {Wheeler}, \& {Wren}}]{Yuan2010}
{Yuan}, F., {Schady}, P., {Racusin}, J.~L., {et~al.} 2010, \apj, 711, 870

\bibitem[{{Zafar} {et~al.}(2012){Zafar}, {Watson}, {El{\'{\i}}asd{\'o}ttir},
  {Fynbo}, {Kr{\"u}hler}, {Schady}, {Leloudas}, {Jakobsson}, {Th{\"o}ne},
  {Perley}, {Morgan}, {Bloom}, \& {Greiner}}]{Zafar2012}
{Zafar}, T., {Watson}, D., {El{\'{\i}}asd{\'o}ttir}, {\'A}., {et~al.} 2012,
  \apj, 753, 82

\bibitem[{{Zhang} \& {Kobayashi}(2005)}]{Zhang2005}
{Zhang}, B. \& {Kobayashi}, S. 2005, \apj, 628, 315

\bibitem[{{Zhao} {et~al.}(2011){Zhao}, {Bai}, \& {Mao}}]{Zhao2011}
{Zhao}, X.~H., {Bai}, J.~M., \& {Mao}, J. 2011, GRB Coordinates Network, 11733,
  1

\bibitem[{{Zhao} {et~al.}(2013){Zhao}, {Mao}, \& {Bai}}]{Zhao2013}
{Zhao}, X.-H., {Mao}, J., \& {Bai}, J.-M. 2013, GRB Coordinates Network, 14418,
  1

\bibitem[{{Zhao} {et~al.}(2012){Zhao}, {Mao}, {Xu}, \& {Bai}}]{Zhao2012}
{Zhao}, X.-H., {Mao}, J., {Xu}, D., \& {Bai}, J.-M. 2012, GRB Coordinates
  Network, 13122, 1

\bibitem[{{Zheng} {et~al.}(2014){Zheng}, {Filippenko}, {Morgan}, \&
  {Cenko}}]{Zheng2014}
{Zheng}, W., {Filippenko}, A.~V., {Morgan}, A., \& {Cenko}, S.~B. 2014, GRB
  Coordinates Network, 16137, 1

\bibitem[{{Zheng} {et~al.}(2012){Zheng}, {Shen}, {Sakamoto}, {Beardmore}, {De
  Pasquale}, {Wu}, {Gorosabel}, {Urata}, {Sugita}, {Zhang}, {Pozanenko},
  {Nissinen}, {Sahu}, {Im}, {Ukwatta}, {Andreev}, {Klunko}, {Volnova},
  {Akerlof}, {Anto}, {Barthelmy}, {Breeveld}, {Carsenty},
  {Castillo-Carri{\'o}n}, {Castro-Tirado}, {Chester}, {Chuang}, {Cunniffe}, {De
  Ugarte Postigo}, {Duffard}, {Flewelling}, {Gehrels}, {G{\"u}ver}, {Guziy},
  {Hentunen}, {Huang}, {Jel{\'{\i}}nek}, {Koch}, {Kub{\'a}nek}, {Kuin},
  {McKay}, {Mottola}, {Oates}, {O'Brien}, {Ohno}, {Page}, {Pandey}, {P{\'e}rez
  del Pulgar}, {Rujopakarn}, {Rykoff}, {Salmi}, {S{\'a}nchez-Ram{\'{\i}}rez},
  {Schaefer}, {Sergeev}, {Sonbas}, {Sota}, {Tello}, {Yamaoka}, {Yost}, \&
  {Yuan}}]{Zheng2012}
{Zheng}, W., {Shen}, R.~F., {Sakamoto}, T., {et~al.} 2012, \apj, 751, 90

\bibitem[{{Zheng} {et~al.}(2009){Zheng}, {Yuan}, \& {Pandey}}]{Zheng2009}
{Zheng}, W., {Yuan}, F., \& {Pandey}, S.~B. 2009, GRB Coordinates Network,
  10284, 1

\bibitem[{{Zimmerman} \& {Tyagi}(2006)}]{Zimmerman2006}
{Zimmerman}, N. \& {Tyagi}, S. 2006, GRB Coordinates Network, 4440, 1

\end{thebibliography}

\end{document}